\documentclass[preprint,12pt]{elsarticle}

\usepackage{amssymb}
\usepackage{amsmath}

\usepackage{orcidlink}
 \usepackage[english]{babel}
 \usepackage{pstricks}
 \usepackage{verbatim}
 \usepackage{graphicx}
 \usepackage{psfrag}
 \usepackage{fancybox}
 \usepackage{amsmath}
 \usepackage{amsfonts}
 \usepackage{amssymb}
 \usepackage{listings}
 \usepackage{enumerate}
 \usepackage{url}

\csname PSTplotLoaded\endcsname
\let\PSTplotLoaded 
\ifx\PSTricksLoaded \else\input pstricks.tex\fi
\ifx\PSTXKeyLoaded \else \input pst-xkey.tex\fi
\ifx\PSTtoolsloaded \else\input pst-tools.tex\fi
\ifx\PSTtoolsloaded \else\input pstricks-add.tex\fi
\ifx\PSTFPloaded \else   \input pst-fp.tex  \fi
\ifx\MultidoLoaded \else \input multido.tex \fi
\def\fileversion{1.92}
\def\filedate{2019/05/16}
\edef\TheAtCode{\the\catcode`\@}
\catcode`\@=11
\pst@addfams{pst-plot}
\def\pst@linetype{2}%
\SpecialCoor

\newdimen\pstRadUnit
\newdimen\pstRadUnitInv
\begingroup
\catcode`\{=13
\catcode`\}=13
\catcode`\(=13
\catcode`\)=13
\catcode`\,=13
\catcode`\!=1
\catcode`\*=2
\catcode`\ =13
\catcode`\_=13
\catcode`\^^M=13
\gdef\pst@datadelimiters!
\catcode`\{=13%
\catcode`\}=13%
\catcode`\(=13%
\catcode`\)=13%
\catcode`\,=13%
\catcode`\ =13%
\catcode`\^^M=13%
\def,##1!%
\ifcat\noexpand,\noexpand##1%
\expandafter##1%
\else\space%
D\space##1%
\fi*%
\let(,\let),\let{,\let},\let ,\let^^M,\let_\@empty*
\endgroup
\define@key[psset]{pst-plot}{ignoreLines}[0]{\def\psk@ignoreLines{#1}}
\psset[pst-plot]{ignoreLines=0}
\newcount\pst@linecnt
\begingroup
\catcode`\,=13
\catcode`\_=13
\gdef\savedata@#1[#2]{%
  \xdef\pst@tempg{#2_}%
  \endgroup
  \let#1\pst@tempg
  \global\let\pst@tempg\relax
  \ignorespaces}
\gdef\readdata@{%
  \read1 to \pst@tempA
  \ifnum\pst@linecnt=\psk@nStep
    \global\pst@linecnt=0
    \expandafter\readdata@@\pst@tempA_\@nil
  \fi
  \global\advance\pst@linecnt by 1
  \ifeof1\else\expandafter\readdata@\fi}
\gdef\pst@@readfile#1#2\@nil{\addto@pscode{,#1#2}}%
\gdef\readdata@@#1#2\@nil{\xdef\pst@tempg{\pst@tempg,#1#2}}%
\endgroup
\def\readdata{\@ifnextchar[{\readdata@i}{\readdata@i[]}}
\def\readdata@i[#1]#2#3{%
  \openin1=#3
  \begingroup
  \ifx#1\relax#1\else\psset{#1}\fi
  \def\pst@tempg{}%
  \ifeof1
    \@pstrickserr{Data file `#3' not found.}\@ehpa
  \else
    \pst@datadelimiters
    \catcode`\[=1
    \catcode`\]=2
    \pst@cnta=0
    \loop \ifnum\the\pst@cnta<\psk@ignoreLines
      \advance\pst@cnta by 1\relax
      \read1 to \pst@tempA
    \repeat
    \psDEBUG[pst-plot]{>>> ignored \the\pst@cnta\space data lines}%
    \global\pst@linecnt=\psk@nStep
    \readdata@
  \fi
  \endgroup
  \global\let#2\pst@tempg%
  \global\let\pst@tempg\relax%
\ignorespaces}
\def\pst@readfile#1{{\let\readdata@@\pst@@readfile\readdata\pst@tempg{#1}}}
\def\pst@altreadfile#1{%
  \openin1=#1
  \ifeof1
    \@pstrickserr{Data file `#1' not found.}\@ehpa
  \else
    \catcode`\{=10
    \catcode`\}=10
    \catcode`\(=10
    \catcode`\)=10
    \catcode`\,=10
    \catcode`\^^M=10
    \catcode`\[=1
    \catcode`\]=2
    \pst@@altreadfile
  \fi}
\def\pst@@altreadfile{%
  \read1 to \pst@tempg
  \expandafter\pst@@@altreadfile\pst@tempg\@empty\@nil
  \ifeof1\else\expandafter\pst@@@altreadfile\fi}
\def\pst@@@altreadfile#1#2\@nil{\addto@pscode{#1#2}}%
\def\savedata#1{\begingroup\pst@datadelimiters\savedata@{#1}}
\newread\RCD@file
\def\psreadDataColumn{\@ifnextchar[\psreadDataColumn@i{\psreadDataColumn@i[]}}
\def\psreadDataColumn@i[#1]{%
  \psset{#1}%
  \psreadDataColumn@ii
}
\def\psreadDataColumn@ii#1#2#3#4{
  \immediate\openin\RCD@file=#4\relax
  \global\let#3=\@empty
  \pst@cnta=0
  \loop \ifnum\the\pst@cnta<\psk@ignoreLines
      \advance\pst@cnta by 1\relax
      \read\RCD@file to \@tempa
  \repeat
  \loop
    \read\RCD@file to \@tempa
    \ifeof\RCD@file\else
      \edef\@tempa{\@tempa#2}%
      \def\reserved@b{}%
      \@tempswafalse
      \@tempcnta=#1\relax
    \expandafter\@tfor\expandafter\reserved@a
      \expandafter:\expandafter=\@tempa\do{
      \if\reserved@a#2\relax
        \advance\@tempcnta \m@ne
        \ifnum \@tempcnta=\z@
          \expandafter\g@addto@macro\expandafter#3\expandafter{\reserved@b\space}%
          \@tempswatrue
        \fi
        \def\reserved@b{}
      \else
        \edef\reserved@b{\reserved@b\reserved@a}
      \fi
      \if@tempswa\@break@tfor\fi
    }%
  \repeat
  \immediate\closein\RCD@file
}

\def\beginplot@line{\begin@OpenObj}
\def\endplot@line{\psline@ii}
\def\beginplot@polygon{\begin@ClosedObj}
\def\endplot@polygon{\pspolygon@ii}
\def\beginplot@curve{\begin@OpenObj}
\def\endplot@curve{\pscurve@ii}
\def\beginplot@ecurve{\begin@OpenObj}
\def\endplot@ecurve{\psecurve@ii}
\def\beginplot@ccurve{\begin@ClosedObj}
\def\endplot@ccurve{\psccurve@ii}
\def\beginplot@dots{\begin@SpecialObj}
\def\endplot@dots{\psdots@ii}
\define@key[psset]{pst-plot}{Hue}[180]{\pst@getint{#1}\pst@HueAngle}
\psset[pst-plot]{Hue=180}
\def\beginplot@colordots{\begin@SpecialObj}
\def\endplot@colordots{%
  \addto@pscode{%
    \psk@dotsize
    \@nameuse{psds@\psk@dotstyle}
    newpath
    /MaxValue 0 def
    /m n 2 mul def
    n { 
      dup MaxValue gt { dup /MaxValue ED } if
      m 2 roll
    } repeat
    n { dup MaxValue div  
      \pst@number\psyunit div abs 
      \pst@HueAngle\space 360 div exch dup sethsbcolor 
      transform floor .5 add exch floor
      .5 add exch itransform Dot stroke } repeat }%
  \end@SpecialObj%
}
\def\beginplot@bubble{\begin@SpecialObj}
\def\endplot@bubble{%
  \addto@pscode{%
    newpath
    n { dup 
      \pst@number\psyunit div abs 
      transform floor .5 add exch floor
      .5 add exch itransform 
      0 360 arc \psk@fill 
      stroke } repeat }%
  \end@SpecialObj%
}
\def\beginplot@bezier{\begin@OpenObj}
\def\endplot@bezier{\psbezier@ii}
\def\beginplot@cbezier{\begin@ClosedObj}
\def\endplot@cbezier{\pscbezier@ii}
\def\beginplot@cspline{\begin@OpenObj}
\def\endplot@cspline{\pscspline@ii}
\let\beginplot@LineToYAxis\beginplot@line  
\def\endplot@LineToYAxis{\psLineToYAxis@ii}
\let\beginqp@LineToYAxis\beginqp@line
\let\doqp@LineToYAxis\doqp@line
\let\endqp@LineToYAxis\endqp@line
\let\testqp@LineToYAxis\testqp@line
\let\beginplot@LineToXAxis\beginplot@line
\def\endplot@LineToXAxis{\psLineToXAxis@ii}
\let\beginqp@LineToXAxis\beginqp@line
\let\doqp@LineToXAxis\doqp@line
\let\endqp@LineToXAxis\endqp@line
\let\testqp@LineToXAxis\testqp@line
\newif\ifPst@interrupt \Pst@interruptfalse
\define@key[psset]{pst-plot}{barwidth}[0.25cm]{\pst@getlength{#1}\Add@barwidth}
\psset[pst-plot]{barwidth=0.25cm}
\define@key[psset]{pst-plot}{interrupt}[]{\expandafter\pst@interrupt#1,,,\@nil}
\def\pst@interrupt#1,#2,#3,#4\@nil{%
  \ifx\relax#1\relax \Pst@interruptfalse
  \else
    \Pst@interrupttrue
    \def\pst@interrupt@YMax{#1 }%
    \def\pst@interrupt@YMaxSep{#2 }%
    \def\pst@interrupt@YMaxDiff{#3 }%
  \fi
}
\def\psbar@ii{\addto@pscode{false \tx@NArray \psbar@iii}}

\def\psbar@iii{%
  \ifPst@interrupt
    /YMax \pst@interrupt@YMax \strip@pt\psyunit\space mul def
    /YMaxSep \pst@interrupt@YMaxSep \strip@pt\psyunit\space mul def
    /YMaxDiff \pst@interrupt@YMaxDiff \strip@pt\psyunit\space mul def
    /Tilde { 
      /Op ED 
      /DX ED
      currentpoint 2 copy
      /Y ED /X ED   
      X DX add Y YMaxSep 2 div Op   
      X DX dup add add Y           
      curveto
      currentpoint 2 copy pop /X ED 
      X DX add Y YMaxSep 2 div neg Op  
      X DX dup add add Y    
      curveto      
    } def  
    newpath
    n { 
      /Yval exch def /Xval exch def 
      Xval \number\Add@barwidth 0.5 mul sub 0 moveto 
      Yval YMax le {  
        0 Yval rlineto \number\Add@barwidth 0 rlineto 
        0 Yval neg rlineto \number\Add@barwidth neg 0 rlineto
      }{
        0 YMax rlineto 
        \number\Add@barwidth 4 div 
        { add } Tilde
        0 YMax neg rlineto 
        \number\Add@barwidth neg 0 rlineto
        closepath
        Xval \number\Add@barwidth 0.5 mul sub YMax YMaxSep add moveto 
        0 Yval YMax sub YMaxSep sub YMaxDiff sub rlineto 
        \number\Add@barwidth 0 rlineto 
        0 Yval YMax YMaxSep add sub YMaxDiff sub neg rlineto 
        \number\Add@barwidth 4 div neg 
        { sub } Tilde
      } ifelse
    } repeat
  \else
    newpath
    n { 
      /Yval exch def /Xval exch def 
      Xval \number\Add@barwidth 0.5 mul sub 0 moveto 
      0 Yval rlineto \number\Add@barwidth 0 rlineto 
      0 Yval neg rlineto \number\Add@barwidth neg 0 rlineto
    } repeat
  \fi
}%
\def\beginplot@bar{\begin@SpecialObj}
\def\endplot@bar{%
  \psbar@ii\psk@fillstyle\ifpsshadow\pst@closedshadow\fi%
  \pst@stroke
  \end@SpecialObj}
\def\psybar@ii{\addto@pscode{false \tx@NArray \psybar@iii}}
\def\psybar@iii{%
  newpath
  n { 
    /Yval exch def /Xval exch def 
    0 Yval \number\Add@barwidth 0.5 mul sub moveto 
    Xval 0 rlineto 0 \number\Add@barwidth rlineto 
    Xval neg 0 rlineto 0 \number\Add@barwidth neg rlineto
  } repeat
}%
\def\beginplot@ybar{\begin@SpecialObj}
\def\endplot@ybar{%
  \psybar@ii\psk@fillstyle\ifpsshadow\pst@closedshadow\fi%
  \pst@stroke
  \end@SpecialObj}
\def\psLSM@ii{\addto@pscode{ false \tx@NArray \psLSM@iii }}
\def\psLSM@iii{%
  /xiSquare 0 def				
  /xi 0 def					
  /fi 0 def					
  /xifi 0 def					
  exch dup dup /xEnd ED /xStart ED exch
  n { 						
    /Yval ED /Xval ED 				
    /xi xi Xval add def				
    /xiSquare xiSquare Xval dup mul add def	
    /xifi xifi Xval Yval mul add def		
    /fi fi Yval add def				
    Xval xStart lt { /xStart Xval def } if	
    Xval xEnd gt { /xEnd Xval def } if		
  } repeat
  /u xiSquare fi mul xi xifi mul sub n xiSquare mul xi dup mul sub div def
  /v n xifi mul xi fi mul sub n xiSquare mul xi dup mul sub div def
  \Pst@Debug\space 0 gt { 			
    /NimbusSanL-Regu findfont 12 scalefont setfont	
    0 -50 moveto (y=) show 			
    v \pst@number\psyunit \pst@number\psxunit div div 20 string cvs show ( x+) show		
    u \pst@number\psyunit div 20 string cvs show } if
  newpath
  (\psk@xStart) length 0 gt 			
    { \psk@xStart\space \pst@number\psxunit mul }
    { xStart } ifelse 
  dup v mul u add 				
  moveto		 			
  (\psk@xEnd) length 0 gt 			
    { \psk@xEnd\space \pst@number\psxunit mul }
    { xEnd } ifelse 
  dup v mul u add 				
  lineto					
}%
\def\beginplot@LSM{\begin@SpecialObj}
\def\endplot@LSM{%
  \psLSM@ii\psk@fillstyle\ifpsshadow\pst@closedshadow\fi%
  \pst@stroke
  \end@SpecialObj}
\define@key[psset]{pst-plot}{IQLfactor}{\pst@checknum{#1}\pst@IQLfactor}
\psset[pst-plot]{IQLfactor=1.5}
\define@key[psset]{pst-plot}{postAction}[]{\def\psk@postAction{%
  \ifx\relax#1\relax\else\pst@number\psyunit div #1 \pst@number\psyunit mul \fi }}
\psset[pst-plot]{postAction=}
\define@key[psset]{pst-plot}{mediancolor}[black]{\pst@getcolor{#1}\median@linecolor}
\psset[pst-plot]{mediancolor=black}
\define@boolkey[psset]{pst-plot}[Pst@]{markMedian}[true]{}
\psset[pst-plot]{markMedian=false}

\def\psBoxplot@ii{%
  \addto@pscode{
    /Barwidth \number\Add@barwidth 2 div def  
    /Endwidth Barwidth \psk@arrowlength\space mul def  
   NArray bubblesort
   /NArray ED 				
   [ NArray { yUnit mul } forall ] /NArray ED 
   NArray 0 get /MinVal ED		
   NArray m 1 sub get /MaxVal ED	
   m 2 div cvi /M ED 			
   NArray length 2 mod 0 eq {		
     M 1 sub NArray exch get 		
     NArray M get          		
     add 2 div /Median ED  		
   }{
     NArray M get /Median ED  		
   } ifelse
   m 4 mod 0 eq {	  		
     m 4 div cvi dup 1 sub NArray exch get
     exch NArray exch get
     add 2 div floor /LowerQuartil ED
   }{ 
     NArray M 2 div cvi get /LowerQuartil ED 
   } ifelse				
   m 0.75 mul dup dup cvi sub 0 eq {	
     cvi dup 1 sub NArray exch get exch NArray exch get
     add 2 div floor /UpperQuartil ED
   }{					
     NArray m 0.75 mul floor cvi get /UpperQuartil ED
   } ifelse 
   /IQL UpperQuartil LowerQuartil sub \pst@IQLfactor\space mul def
   0 1 m 1 sub { 
     dup /Index ED
     NArray exch get LowerQuartil sub abs IQL sub 0 gt { 
       \psk@dotsize
       \@nameuse{psds@\psk@dotstyle}
       0 NArray Index get \psk@postAction
       Dot
       NArray Index LowerQuartil UpperQuartil LowerQuartil sub \pst@IQLfactor\space mul sub 
       dup /MinVal ED put 
       NArray Index 1 add get /MinVal ED 
    } { exit } ifelse
   } for
   m 1 sub -1 0 {	
     dup /Index ED
     NArray exch get UpperQuartil sub abs IQL sub 0 gt { 
       \psk@dotsize
       \@nameuse{psds@\psk@dotstyle}
       0 NArray Index get \psk@postAction\space
       Dot
       NArray Index UpperQuartil LowerQuartil sub \pst@IQLfactor\space mul UpperQuartil add 
       dup /MaxVal ED put 
       NArray Index 1 sub get /MaxVal ED 
     }{ exit } ifelse
   } for
   Endwidth neg MaxVal \psk@postAction moveto			
   Endwidth dup add 0 rlineto 
   0 MaxVal \psk@postAction moveto 
   0 UpperQuartil \psk@postAction lineto			
   MinVal \psk@postAction MaxVal \psk@postAction lt {
     0 LowerQuartil \psk@postAction moveto			
     0 MinVal \psk@postAction lineto 
     Endwidth neg MinVal \psk@postAction moveto 
     Endwidth dup add 0 rlineto 
   } if
   gsave
   \pst@number\pslinewidth SLW
   \pst@usecolor\pslinecolor
   \tx@setStrokeTransparency 
   \@nameuse{psls@\pslinestyle}
   stroke
   grestore
   newpath
   Barwidth neg LowerQuartil \psk@postAction moveto	
   Barwidth neg UpperQuartil \psk@postAction lineto
   Barwidth dup add 0 rlineto
   Barwidth LowerQuartil \psk@postAction lineto
   closepath
   \pst@usecolor\psfillcolor
   gsave \pst@usecolor\psfillcolor \tx@setTransparency fill grestore
   \@nameuse{psls@solid}
   \ifPst@markMedian
     \pst@number\pslabelsep neg Median moveto currentpoint 
     /YMedian ED /XMedian ED 
      Barwidth neg Median \psk@postAction lineto  
   \else
      Barwidth neg Median \psk@postAction moveto  
   \fi
   Barwidth dup add 0 rlineto 
   \pst@number\pslinewidth SLW
   \pst@usecolor\median@linecolor
   \tx@setStrokeTransparency
   stroke
  }
}%
\def\beginplot@Boxplot{\init@pscode}
\def\endplot@Boxplot{%
  \psBoxplot@ii\psk@fillstyle\ifpsshadow\pst@closedshadow\fi%
  \pst@stroke
  \end@SpecialObj}
\def\psBoxplot{\def\pst@par{}\pst@object{psBoxplot}}
\def\psBoxplot@i#1{%
  \leavevmode
  \pst@killglue
  \begingroup
  \addbefore@par{barwidth=40pt,arrowlength=0.75}%
  \addto@par{plotstyle=Boxplot}%
  \use@par
  \@nameuse{beginplot@\psplotstyle}%
  \addto@pscode{
    /D {} def
    [ #1 ] /NArray ED 
    NArray aload length /m ED
    /xUnit \pst@number\psxunit def
    /yUnit \pst@number\psyunit def
  }%
  \@nameuse{endplot@\psplotstyle}%
  \ignorespaces%
}
\define@key[psset]{pst-plot}{plotstyle}[line]{%
  \@ifundefined{beginplot@#1}%
    {\@pstrickserr{Plot style `#1' not defined}\@eha}%
    {\def\psplotstyle{#1}}}
\psset[pst-plot]{plotstyle=line}
\define@key[psset]{pst-plot}{plotpoints}[50]{%
  \pst@cntg=#1\relax
  \ifnum\pst@cntg<2
    \@pstrickserr{plotpoints parameter must be at least 2}\@ehpa
  \else
    \advance\pst@cntg-1
    \edef\psk@plotpoints{\the\pst@cntg\space}%
  \fi}
\psset[pst-plot]{plotpoints=50}
\define@key[psset]{pst-plot}{yMaxValue}[1.e30]{\def\psk@yMaxValue{#1 }\def\psk@yMinValue{#1 neg }}
\psset{yMaxValue=1.e30}
\define@key[psset]{pst-plot}{yMinValue}[-1.e30]{\def\psk@yMinValue{#1 }}
\psset{yMinValue=-1.e30}
\def\beginqp@line{\pst@oplineto}
\def\doqp@line{ 
  dup
  \psk@yMaxValue \pst@number\psyunit mul gt 
    { moveto }
    { dup \psk@yMinValue \pst@number\psyunit mul lt 
      { moveto }
      { L } ifelse 
    } ifelse
}
\def\endqp@line{%
  \ifPst@variableLW \addto@pscode{ \pst@flattenpath }\fi%
  \end@OpenObj}%

\def\testqp@line{%
  \ifdim\pslinearc>\z@\else
    \ifshowpoints\else
      \ifx\psk@arrowA\@empty
        \ifx\psk@arrowB\@empty
          \@psttrue
        \fi
      \fi
    \fi
  \fi}
\def\beginqp@polygon{moveto }
\def\doqp@polygon{ 
      dup
      \psk@yMaxValue \pst@number\psyunit mul gt 
      { moveto }{ 
          dup
          \psk@yMinValue \pst@number\psyunit mul lt 
          { moveto }{ L } ifelse 
      } ifelse
}
\def\endqp@polygon{%
  \addto@pscode{closepath}%
  \end@ClosedObj}
\def\testqp@polygon{%
  \ifdim\pslinearc>\z@\else
    \ifshowpoints\else
      \@psttrue
    \fi
  \fi}
\def\beginqp@dots{%
  \psk@dotsize
  \@nameuse{psds@\psk@dotstyle}
  Dot }
\def\doqp@dots{Dot }
\def\endqp@dots{\end@SpecialObj}
\def\testqp@dots{\@psttrue}
\def\beginqp@bezier{/n 0 def \pst@oplineto}
\def\doqp@bezier{/n n 1 add def n 3 mod 0 eq { 
    dup \psk@yMaxValue\space \pst@number\psyunit mul gt 
    { moveto pop pop pop pop}
    { dup \psk@yMinValue\space \pst@number\psyunit mul lt 
      { moveto pop pop pop pop}{ curveto } ifelse 
    } ifelse 
  } if
}
\def\endqp@bezier{%
  \addto@pscode{n 3 mod { pop pop } repeat}
  \end@OpenObj}%
\def\testqp@bezier{%
  \ifshowpoints\else
    \ifx\psk@arrowA\@empty
      \ifx\psk@arrowB\@empty
        \@psttrue
      \fi
    \fi
  \fi}
\def\beginqp@cbezier{/n 0 def moveto }
\def\doqp@cbezier{\doqp@bezier}
\def\endqp@cbezier{%
  \addto@pscode{n 3 mod { pop pop } repeat closepath}
  \end@ClosedObj}%
\def\testqp@cbezier{\ifshowpoints\else\@psttrue\fi}
\def\tx@LineToYAxis{LineToYAxis }
\def\psLineToYAxis@ii{%
\addto@pscode{\pst@cp \psline@iii \psk@Ox\space \pst@number\psxunit mul \tx@LineToYAxis}%
\end@OpenObj}
\def\tx@LineToXAxis{LineToXAxis }
\def\psLineToXAxis@ii{%
\addto@pscode{\pst@cp \psline@iii \psk@Oy\space \pst@number\psyunit mul \tx@LineToXAxis}%
\end@OpenObj}
\define@key[psset]{pst-plot}{PSfont}[NimbusRomNo9L-Regu]{\def\psk@PSfont{/#1 }}
\define@key[psset]{pst-plot}{valuewidth}[10]{\pst@getint{#1}\psk@valuewidth }
\define@key[psset]{pst-plot}{fontscale}[10]{\pst@checknum{#1}\psk@fontscale }
\define@key[psset]{pst-plot}{decimals}[-1]{\pst@getint{#1}\psk@decimals }
\psset[pst-plot]{PSfont=NimbusRomNo9L-Regu,fontscale=10,valuewidth=10,decimals=-1}
\newdimen\psxlabelsep
\newdimen\psylabelsep
\define@key[psset]{pst-plot}{xlabelsep}[5pt]{\pssetlength\psxlabelsep{#1}}
\define@key[psset]{pst-plot}{ylabelsep}[5pt]{\pssetlength\psylabelsep{#1}}
\psset[pst-plot]{xlabelsep=5pt,ylabelsep=5pt}

\newif\ifPst@valuesStar\Pst@valuesStarfalse
\newif\ifPst@xvalues\Pst@xvaluesfalse
\def\psvalues@ii{\addto@pscode{ false \tx@NArray \psvalues@iii }}
\def\psvalues@iii{
  \psk@PSfont findfont \psk@fontscale scalefont setfont 
  newpath 
  n { /yO ED /xO ED
      gsave
      \ifPst@xvalues
        xO \pst@number\psxunit div
      \else
        yO \pst@number\psyunit div
      \fi
      \psk@decimals 0 eq { cvi } if
      \psk@decimals 0 gt { 10 \psk@decimals exp dup 3 1 roll mul cvi exch div } if
      \psk@valuewidth string cvs /Str ED
      \ifPst@valuesStar
      Str stringwidth pop /yS \psk@fontscale def /xS ED 
      gsave newpath 
        xO \ifPst@xvalues \pst@number\pslabelsep add \fi 
        yO \ifPst@xvalues \psk@fontscale 4 div sub \else \pst@number\pslabelsep add \fi 
        moveto \ifx\psk@rot\@empty\else\psk@rot rotate \fi
        xS 0 rlineto 0 yS rlineto xS neg 0 rlineto 0 yS neg rlineto 
        closepath  1 setgray fill stroke 
      grestore 
      \fi
      xO \ifPst@xvalues \pst@number\pslabelsep add \fi
      yO \ifPst@xvalues \psk@fontscale 4 div sub \else \pst@number\pslabelsep add \fi 
      moveto \ifx\psk@rot\@empty\else\psk@rot rotate \fi 
      Str show 
      grestore } repeat 
}%
\def\beginplot@values{\Pst@valuesStarfalse\begin@SpecialObj}
\expandafter\def\csname beginplot@values*\endcsname{\Pst@valuesStartrue\begin@SpecialObj}
\def\beginplot@xvalues{\Pst@valuesStarfalse\begin@SpecialObj}
\expandafter\def\csname beginplot@xvalues*\endcsname{\Pst@valuesStartrue\begin@SpecialObj}
\def\endplot@values{%
  \Pst@xvaluesfalse%
  \psvalues@ii%
  \pst@stroke
  \end@SpecialObj}
\@namedef{endplot@values*}{\endplot@values}
\def\endplot@xvalues{%
  \Pst@xvaluestrue%
  \psvalues@ii%
  \pst@stroke
  \end@SpecialObj}
\@namedef{endplot@xvalues*}{\endplot@xvalues}
\def\psdataplot{\def\pst@par{}\pst@object{dataplot}}
\def\dataplot{\def\pst@par{}\pst@object{dataplot}}
\def\dataplot@i#1{%
  \pst@killglue
  \begingroup
    \use@par
    \@pstfalse
    \@nameuse{testqp@\psplotstyle}%
    \if@pst
      \dataplot@ii{\addto@pscode{#1}}%
    \else
      \listplot@ii{\addto@pscode{#1}}%
    \fi
  \endgroup
  \ignorespaces}
\def\dataplot@ii#1{%
  \@nameuse{beginplot@\psplotstyle}%
    \addto@pscode{%
      /Dx { \pst@number\psxunit mul /D { Dy } def } def
      /Dy { \pst@number\psyunit mul Do /D { Dx } def } def
      /D { /D { Dx } def } def
      /Do {
        \@nameuse{beginqp@\psplotstyle}%
        /Do { \@nameuse{doqp@\psplotstyle}} def
      } def}%
    #1
    \addto@pscode{ D }%
  \@nameuse{endqp@\psplotstyle}}
\def\psfileplot{\def\pst@par{}\pst@object{fileplot}}
\def\fileplot{\def\pst@par{}\pst@object{fileplot}}
\def\fileplot@i#1{%
  \pst@killglue%
  \begingroup%
    \use@par%
    \@pstfalse%
    \@nameuse{testqp@\psplotstyle}%
    \if@pst\dataplot@ii{\pst@readfile{#1}}\else\listplot@ii{\pst@altreadfile{#1}}\fi%
  \endgroup%
  \ignorespaces}
\def\pslistplot{\pst@object{listplot}}
\def\listplot{\pst@object{listplot}}
\def\listplot@i#1{\listplot@ii{\addto@pscode{#1}}}
\def\listplot@ii#1{%
  \@nameuse{beginplot@\psplotstyle}%
  \addto@pscode{/D {} def mark}%
  #1%
  \addto@pscode{
    \tx@PreparePoints
    \pst@number\psxunit
    \pst@number\psyunit
    \tx@ScalePoints
  }%
  \@nameuse{endplot@\psplotstyle}%
}
\define@boolkey[psset]{pst-plot}[Psk@]{xyValues}[true]{}
\define@boolkey[psset]{pst-plot}[Pst@]{ChangeOrder}[true]{}
\psset[pst-plot]{xyValues,ChangeOrder=false}
\pst@def{PreparePoints}<{%
  counttomark /m exch def
  /maxYValues \psk@plotNoMax\space def
  /YValuePos \psk@plotNo\space def
  /XValuePos \psk@plotNoX\space def
  /maxYvalue \ifx\empty\psk@plotYMax false \else true \fi def
  maxYvalue { /YMaxValue \psk@plotYMax\space def } if
  \ifPsk@xyValues\else 
    /mm m def
    /M m 1 add def
    m { mm exch M 2 roll /M M 1 add def /mm mm 1 sub def } repeat
    /m m dup add def
  \fi
  \ifPst@ChangeOrder
    /m0 m def
    m maxYValues 1 add div 1 sub cvi {
      m0 maxYValues 1 add roll /m0 m0 maxYValues 1 add sub def
    } repeat
  \fi
  /n m maxYValues 1 add div cvi def
  XValuePos 1 gt {
    n {
      maxYValues 1 add XValuePos neg roll    
      dup /XValue ED
      maxYValues 1 add XValuePos 1 sub  roll 
      pop XValue                             
      maxYValues 1 add 1 roll                
      m maxYValues 1 add  roll               
    } repeat
  } if 
  maxYValues 1 gt {
    n {
      maxYValues YValuePos 1 sub neg roll 
      maxYValues 1 sub { pop } repeat     
      /m m maxYValues 1 sub sub def
      m 2 roll
    } repeat
  } if 
  /xMax -99999 def /yMax -99999 def
  /xP 0 def /yP 0 def
  m copy
  n {
    /y exch def /x exch def
    xMax x lt { /xMax x def } if
    yMax y lt {/yMax y def } if
    xP x gt { /xP x def } if
    yP y gt { /yP y def } if
  } repeat
  \psk@xStep\space 0 gt \psk@yStep\space 0 gt or (\psk@xStart) length 0 gt or
  (\psk@yStart) length 0 gt or (\psk@xEnd) length 0 gt or (\psk@yEnd) length 0 gt or {
    (\psk@xStart) length 0 gt {\psk@xStart\space }{ xP } ifelse /xStart exch def
    (\psk@yStart) length 0 gt {\psk@yStart\space }{ yP } ifelse /yStart exch def
    (\psk@xEnd) length 0 gt { \psk@xEnd\space }{ xMax } ifelse /xEnd exch def
    (\psk@yEnd) length 0 gt { \psk@yEnd\space }{ yMax } ifelse /yEnd exch def
    n {
      m -2 roll
      2 copy /yVal exch def /xVal exch def
      xVal xP ge
      yVal yP ge and
      xVal xEnd le and
      yVal yEnd le and
      xVal xStart ge and
      yVal yStart ge and {
        /xP xP \psk@xStep\space add def
        /yP yP \psk@yStep\space add def
      }{%
        pop pop
        /m m 2 sub def
      } ifelse
    } repeat
  }{%
    /ncount 1 def
    (\psk@nEnd) length 0 gt { \psk@nEnd\space }{ m } ifelse 
    /nEnd exch def
    n {
      m -2 roll
      \psk@nStep\space 1 gt { ncount \psk@nStart\space sub \psk@nStep\space mod 0 eq }{ true } ifelse
        ncount nEnd le and
        ncount \psk@nStart\space ge and not {
          pop pop
          /m m 2 sub def
        } if
        /ncount ncount 1 add def
      } repeat
  } ifelse
  maxYvalue { 
    /m0 m def
    n {
      dup YMaxValue gt { pop pop /m m 2 sub def } if
      m -2 roll 
    } repeat
  } if
}>
\define@boolkey[psset]{pst-plot}[Pst@]{polarplot}[true]{}
\define@boolkey[psset]{pst-plot}[Pst@]{VarStep}[true]{}
\define@key[psset]{pst-plot}{PlotDerivative}{\def\psk@PlotDerivative{#1}}%
\define@key[psset]{pst-plot}{VarStepEpsilon}{\def\psk@VarStepEpsilon{#1}}%
\define@key[psset]{pst-plot}{method}{\def\psk@method{#1}}
\def\@rkiv{rk4}
\def\@varrkiv{varrkiv}
\def\@adams{adams}
\def\@default{default}
\psset[pst-plot]{VarStep=false,PlotDerivative=none,VarStepEpsilon=default,polarplot=false,method={}}
\def\psplotinit#1{\xdef\psplot@init{#1 }}
\def\psplot@init{}
\def\psplot{\def\pst@par{}\pst@object{psplot}}
\def\psplot@i#1#2{\@ifnextchar[{\psplot@x{#1}{#2}}{\psplot@x{#1}{#2}[]}}
\def\psplot@x#1#2[#3]#4{%
  \pst@killglue
  \begingroup
    \use@par
    \@nameuse{beginplot@\psplotstyle}%
    \ifPst@polarplot
      \addto@pscode{
        \psplot@init
        #3 
        /x #1 def
        /x1 #2 def
        /dx x1 x sub \psk@plotpoints div def
        /F@pstplot \ifPst@algebraic (#4)
                    \ifx\psk@PlotDerivative\@none\else
                      \psk@PlotDerivative\space { (x) tx@Derive begin Derive end } repeat
                    \fi\space
                    tx@AlgToPs begin AlgToPs end cvx
                 \else { #4 } \fi  def
        \ifPst@VarStep
          /StillZero 0 def /LastNonZeroStep dx def
          /F2@pstplot tx@Derive begin (#4) (x) Derive (x) Derive end
                     \ifx\psk@PlotDerivative\@none\else
                       \psk@PlotDerivative\space { (x) tx@Derive begin Derive end } repeat
                     \fi\space
                    tx@AlgToPs begin AlgToPs end cvx def
          /epsilon12 \ifx\psk@VarStepEpsilon\@default tx@Derive begin F2@pstplot end dx 3 exp abs mul abs
                    \else\psk@VarStepEpsilon\space 12 mul \fi def
          /ComputeStep {
            dup 1e-4 lt
            { pop StillZero 2 ge { LastNonZeroStep 2 mul } { LastNonZeroStep } ifelse /StillZero StillZero 1 add def }
            { epsilon12 exch div 1 3 div exp /StillZero 0 def }
            ifelse } bind def
        \fi
        /xy {
          F@pstplot x \ifPst@algebraic RadtoDeg \fi PtoC
          \pst@number\psyunit mul exch
          \pst@number\psxunit mul exch
        } def}%
    \else
    \addto@pscode{
      \psplot@init
      #3 
      /x #1 def
      /x1 #2 def
      /dx x1 x sub \psk@plotpoints div def
      /F@pstplot \ifPst@algebraic (#4)
                    \ifx\psk@PlotDerivative\@none\else
                      \psk@PlotDerivative\space { (x) tx@Derive begin Derive end } repeat
                    \fi\space
                    tx@AlgToPs begin AlgToPs end cvx
                 \else { #4 } \fi  def
      \ifPst@VarStep
         /StillZero 0 def /LastNonZeroStep dx def
         /F2@pstplot tx@Derive begin (#4) (x) Derive (x) Derive end
                     \ifx\psk@PlotDerivative\@none\else
                       \psk@PlotDerivative\space { (x) tx@Derive begin Derive end } repeat
                     \fi\space
                    tx@AlgToPs begin AlgToPs end cvx def
         /epsilon12 \ifx\psk@VarStepEpsilon\@default tx@Derive begin F2@pstplot end dx 3 exp abs mul abs
                    \else\psk@VarStepEpsilon\space 12 mul \fi def
         /ComputeStep {
           dup 1e-4 lt
           { pop StillZero 2 ge { LastNonZeroStep 2 mul } { LastNonZeroStep } ifelse /StillZero StillZero 1 add def }
           { epsilon12 exch div 1 3 div exp /StillZero 0 def }
           ifelse } bind def
      \fi
      /xy { x \pst@number\psxunit mul F@pstplot \pst@number\psyunit mul
      } def}%
    \fi
    \gdef\psplot@init{}%
    \ifx\pslinestyle\psls@@symbol
      \psplot@iii
    \else
      \@pstfalse
      \@nameuse{testqp@\psplotstyle}%
      \if@pst\psplot@ii\else\psplot@iii\fi
    \fi
  \endgroup
  \ignorespaces}
\def\psplot@ii{%
  \ifPst@VarStep%
    \addto@pscode{%
      mark xy \@nameuse{beginqp@\psplotstyle}
      { F2@pstplot abs ComputeStep
        x 2 copy add dup x1 gt {pop x1} if /x exch def F2@pstplot abs ComputeStep
        /x 3 -1 roll def 2 copy gt { exch } if pop
        /x x 3 -1 roll add dup x1 gt {pop x1} if def
        xy \@nameuse{doqp@\psplotstyle}
        x x1 eq { exit } if} loop}%
  \else
    \pst@killglue%
    \addto@pscode{
      /ps@Exit false def
      xy \@nameuse{beginqp@\psplotstyle}
      \ifx\psk@method\@varrkiv\else\psk@plotpoints 1 sub \fi {
        /x x dx add \ifx\psk@method\@varrkiv  dup x1 gt { pop x1 } if \fi def
        xy \@nameuse{doqp@\psplotstyle}
        \ifx\psk@method\@varrkiv  x x1 eq { exit } if \fi
      } 
      ps@Exit { exit } if
      \ifx\psk@method\@varrkiv loop \else repeat \fi
      ps@Exit not {
        /x x1 def
        xy \@nameuse{doqp@\psplotstyle}
      } if }%
  \fi%
  \@nameuse{endqp@\psplotstyle}}
\def\psplot@iii{%
  \ifPst@VarStep%
    \addto@pscode{
      /n 2 def
      mark
      { xy n 2 roll F2@pstplot abs
        ComputeStep x 2 copy add dup x1 gt {pop x1} if
        /x exch def F2@pstplot abs ComputeStep
        /x 3 -1 roll def 2 copy gt { exch } if pop
        /x x 3 -1 roll dup /LastNonZeroStep exch def add dup x1 gt {pop x1} if def /n n 2 add def
        x x1 eq { exit } if } loop
      xy 
      n 2 roll}%
  \else\pst@killglue%
    \addto@pscode{
      mark
      /n 2 def
      \ifx\psk@method\@varrkiv\else\psk@plotpoints\fi {
        xy
        n 2 roll
        /n n 2 add def
        /x x dx add \ifx\psk@method\@varrkiv  dup x1 gt { pop x1 } if \fi def
        \ifx\psk@method\@varrkiv  x x1 eq { exit } if \fi
      } \ifx\psk@method\@varrkiv loop\else repeat \fi \space
      /x x1 def
      xy 
      2 copy \tx@UserCoor 2 array astore /FinalState ED
      n 2 roll}%
  \fi%
  \@nameuse{endplot@\psplotstyle}}
\def\psparametricplot{\pst@object{parametricplot}}
\def\parametricplot{\pst@object{parametricplot}}
\def\parametricplot@i#1#2{\@ifnextchar[{\parametricplot@x{#1}{#2}}{\parametricplot@x{#1}{#2}[]}}
\def\parametricplot@x#1#2[#3]{\@ifnextchar[{\parametricplot@xi{#1}{#2}[#3]}{\parametricplot@xi{#1}{#2}[#3][]}}
\def\parametricplot@xi#1#2[#3][#4]#5{%
  \pst@killglue%
  \begingroup%
    \use@par%
    \@nameuse{beginplot@\psplotstyle}%
    \addto@pscode{%
      #3 
      \psplot@init
      /t #1 def
      /t1 #2 def
      /dt t1 t sub \psk@plotpoints div def
      /F@pstplot \ifPst@algebraic (#5)
                    \ifx\psk@PlotDerivative\@none\else
                      \psk@PlotDerivative\space { (t) tx@Derive begin Derive end } repeat
                    \fi\space
                    tx@AlgToPs begin AlgToPs end cvx
                 \else { #5 } \fi  def
      \ifPst@VarStep
         /StillZero 0 def /LastNonZeroStep dt def
         /F2@pstplot tx@Derive begin (#5) (t) Derive (t) Derive end
                     \ifx\psk@PlotDerivative\@none\else
                       \psk@PlotDerivative\space { (t) tx@Derive begin Derive end } repeat
                     \fi\space
                    tx@AlgToPs begin AlgToPs end cvx def
         /epsilon12 \ifx\psk@VarStepEpsilon\@default
                       tx@Derive begin F2@pstplot end Pyth
                       dt 3 exp abs mul
                    \else\psk@VarStepEpsilon\space 12 mul \fi def
         /ComputeStep {
           dup 1e-4 lt
           { pop StillZero 2 ge { LastNonZeroStep 2 mul } { LastNonZeroStep } ifelse /StillZero StillZero 1 add def }
           { epsilon12 exch div 1 3 div exp /StillZero 0 def }
           ifelse } bind def
      \fi
      /xy {
        \ifPst@algebraic F@pstplot \else #5 \fi
        \pst@number\psyunit mul exch
        \pst@number\psxunit mul exch
      } def
      }%
    \gdef\psplot@init{}%
    \@pstfalse
    \@nameuse{testqp@\psplotstyle}%
    \if@pst\parametricplot@ii{#4}\else\parametricplot@iii{#4}\fi
  \endgroup%
  \ignorespaces}
\def\parametricplot@ii#1{
  \ifPst@VarStep%
    \addto@pscode{%
      mark xy \@nameuse{beginqp@\psplotstyle}
      { F2@pstplot Pyth ComputeStep
        t 2 copy add dup t1 gt {pop t1} if /t exch def F2@pstplot Pyth ComputeStep
        /t 3 -1 roll def 2 copy gt { exch } if pop
        /t t 3 -1 roll add dup t1 gt {pop t1} if def
        xy \@nameuse{doqp@\psplotstyle}
        t t1 eq { exit } if } loop}%
  \else\pst@killglue%
    \addto@pscode{%
      /ps@Exit false def
      xy \@nameuse{beginqp@\psplotstyle}
      \psk@plotpoints 1 sub {
        /t t dt add def
        xy \@nameuse{doqp@\psplotstyle}
        ps@Exit { exit } if 
      } repeat
      ps@Exit not {
        /t t1 def
        xy \@nameuse{doqp@\psplotstyle}
      } if 
    }%
  \fi%
  \addto@pscode{ #1 }%
  \@nameuse{endqp@\psplotstyle}}
\def\parametricplot@iii#1{%
  \ifPst@VarStep%
    \addto@pscode{%
      /n 2 def
      mark
      { xy n 2 roll F2@pstplot Pyth
        ComputeStep t 2 copy add dup t1 gt {pop t1} if
        /t exch def F2@pstplot Pyth ComputeStep
        /t 3 -1 roll def 2 copy gt { exch } if pop
        /t t 3 -1 roll dup /LastNonZeroStep exch def add dup t1 gt {pop t1} if def /n n 2 add def
        t t1 eq { exit } if } loop
      xy 
      2 copy \tx@UserCoor 2 array astore /FinalState ED
      n 2 roll}%
  \else\pst@killglue%
    \addto@pscode{
      mark
      /n 2 def
      \psk@plotpoints {
        xy
        n 2 roll
        /n n 2 add def
        /t t dt add def
      } repeat
      /t t1 def
      xy
      n 2 roll}%
  \fi%
  \addto@pscode{ #1 }%
  \@nameuse{endplot@\psplotstyle}}
\newdimen\psk@subticksize\psk@subticksize=\z@
\newdimen\pst@xticksizeA
\newdimen\pst@xticksizeB
\newdimen\pst@xticksizeC
\newdimen\pst@yticksizeA
\newdimen\pst@yticksizeB
\newdimen\pst@yticksizeC
\define@key[psset]{pst-plot}{ticks}[all]{\pst@expandafter\psset@@ticks{#1}\@nil\psk@ticks}
\def\psset@@ticks#1#2\@nil#3{%
  \ifx#1a\let#3\z@\else
    \ifx#1x\let#3\@ne\else
      \ifx#1y\let#3\tw@\else
        \ifx#1n\let#3\thr@@\else
          \@pstrickserr{Bad argument: `#1#2'}\@ehpa
  \fi\fi\fi\fi}
\psset[pst-plot]{ticks=all}
\define@key[psset]{pst-plot}{labels}[all]{\pst@expandafter\psset@@ticks{#1}\@nil\psk@labels}
\psset[pst-plot]{labels=all}
\define@key[psset]{pst-plot}{Ox}[0]{\def\psk@Ox{#1}}
\psset[pst-plot]{Ox=0}
\define@key[psset]{pst-plot}{Dx}[1]{\def\psk@Dx{#1}}
\psset[pst-plot]{Dx=1}
\define@key[psset]{pst-plot}{dx}[0]{%
  \pssetxlength\pst@dimg{#1}%
  \edef\psk@dx{\number\pst@dimg}}
\psset[pst-plot]{dx=0}
\define@key[psset]{pst-plot}{Oy}[0]{\def\psk@Oy{#1}}
\psset[pst-plot]{Oy=0}
\define@key[psset]{pst-plot}{Dy}[1]{\def\psk@Dy{#1}}
\psset[pst-plot]{Dy=1}
\define@key[psset]{pst-plot}{dy}[0]{%
  \pssetylength\pst@dimg{#1}%
  \edef\psk@dy{\number\pst@dimg}}
\psset[pst-plot]{dy=0}
\define@boolkey[psset]{pst-plot}[]{showXorigin}[true]{}
\define@boolkey[psset]{pst-plot}[]{showYorigin}[true]{}
\define@boolkey[psset]{pst-plot}[]{showorigin}[true]{%
  \ifshoworigin
    \showXorigintrue\showYorigintrue
  \else
    \showXoriginfalse\showYoriginfalse
  \fi
}
\psset[pst-plot]{showorigin=true}
\long\def\psrotatebox#1#2{%
  \leavevmode
  \Grot@setangle{#1}%
  \setbox\z@\hbox{{#2}}%
  \Grot@x\z@
  \Grot@y\z@
  \Grot@box}
\def\Grot@setangle#1{\edef\Grot@angle{#1}}
\def\Grot@Px#1#2#3{%
        #1\Grot@cos#2%
        \advance#1-\Grot@sin#3}
\def\Grot@Py#1#2#3{%
        #1\Grot@sin#2%
        \advance#1\Grot@cos#3}
\def\Grot@box{%
  \begingroup
  \CalculateSin\Grot@angle
  \CalculateCos\Grot@angle
  \edef\Grot@sin{\UseSin\Grot@angle}%
  \edef\Grot@cos{\UseCos\Grot@angle}%
  \Grot@r\wd\z@  \advance\Grot@r-\Grot@x
  \Grot@l\z@     \advance\Grot@l-\Grot@x
  \Grot@h\ht\z@  \advance\Grot@h-\Grot@y
  \Grot@d-\dp\z@ \advance\Grot@d-\Grot@y
  \ifdim\Grot@sin\p@>\z@
    \ifdim\Grot@cos\p@>\z@
      \Grot@Py\Grot@height \Grot@r\Grot@h
      \Grot@Px\Grot@right  \Grot@r\Grot@d
      \Grot@Px\Grot@left   \Grot@l\Grot@h
      \Grot@Py\Grot@depth  \Grot@l\Grot@d
    \else
      \Grot@Py\Grot@height \Grot@r\Grot@d
      \Grot@Px\Grot@right  \Grot@l\Grot@d
      \Grot@Px\Grot@left   \Grot@r\Grot@h
      \Grot@Py\Grot@depth  \Grot@l\Grot@h
    \fi
  \else
    \ifdim\Grot@cos\p@<\z@
      \Grot@Py\Grot@height \Grot@l\Grot@d
      \Grot@Px\Grot@right  \Grot@l\Grot@h
      \Grot@Px\Grot@left   \Grot@r\Grot@d
      \Grot@Py\Grot@depth  \Grot@r\Grot@h
    \else
      \Grot@Py\Grot@height \Grot@l\Grot@h
      \Grot@Px\Grot@right  \Grot@r\Grot@h
      \Grot@Px\Grot@left   \Grot@l\Grot@d
      \Grot@Py\Grot@depth  \Grot@r\Grot@d
    \fi
  \fi
  \advance\Grot@height\Grot@y
  \advance\Grot@depth\Grot@y
  \Grot@Px\dimen@  \Grot@x\Grot@y
  \Grot@Py\dimen@ii \Grot@x\Grot@y
  \dimen@-\dimen@     \advance\dimen@-\Grot@left
  \dimen@ii-\dimen@ii \advance\dimen@ii\Grot@y
  \setbox\z@\hbox{%
    \kern\dimen@
    \raise\dimen@ii\hbox{\Grot@start\box\z@\Grot@end}}%
  \ht\z@\Grot@height
  \dp\z@-\Grot@depth
  \advance\Grot@right-\Grot@left\wd\z@\Grot@right
  \leavevmode\box\z@
  \endgroup}
\define@key[psset]{pst-plot}{labelFontSize}[{}]{\def\psk@xlabelFontSize{#1}\def\psk@ylabelFontSize{#1}}%
\define@key[psset]{pst-plot}{xlabelFontSize}[{}]{\def\psk@xlabelFontSize{#1}}%
\define@key[psset]{pst-plot}{ylabelFontSize}[{}]{\def\psk@ylabelFontSize{#1}}%
\define@boolkey[psset]{pst-plot}[Pst@]{mathLabel}[true]{%
  \ifPst@mathLabel%
    \Pst@xmathLabeltrue \Pst@ymathLabeltrue
    \def\pshlabel##1{$\psk@xlabelFontSize##1$}%
    \def\psvlabel##1{$\psk@ylabelFontSize##1$}%
  \else%
    \Pst@xmathLabelfalse \Pst@ymathLabelfalse
    \def\pshlabel##1{\psk@xlabelFontSize##1}%
    \def\psvlabel##1{\psk@ylabelFontSize##1}%
  \fi}
\define@boolkey[psset]{pst-plot}[Pst@]{xmathLabel}[true]{%
  \ifPst@xmathLabel%
    \def\pshlabel##1{$\psk@xlabelFontSize##1$}\else\def\pshlabel##1{\psk@xlabelFontSize##1}\fi}
\define@boolkey[psset]{pst-plot}[Pst@]{ymathLabel}[true]{%
  \ifPst@ymathLabel%
    \def\psvlabel##1{$\psk@ylabelFontSize##1$}\else\def\psvlabel##1{\psk@ylabelFontSize##1}\fi}
\psset[pst-plot]{labelFontSize={},mathLabel}
\define@boolkey[psset]{pst-plot}[Pst@]{xAxis}[true]{}
\define@boolkey[psset]{pst-plot}[Pst@]{yAxis}[true]{}
\define@boolkey[psset]{pst-plot}[Pst@]{xyAxes}[true]{%
    \@nameuse{Pst@xAxis#1}\@nameuse{Pst@yAxis#1}}%
\psset[pst-plot]{xAxis,yAxis}%
\define@key[psset]{pst-plot}{xlabelPos}[b]{\pst@expandafter\psset@@xlabelPos#1\@nil}
\define@key[psset]{pst-plot}{ylabelPos}[l]{\pst@expandafter\psset@@ylabelPos#1\@nil}
\def\psset@@xlabelPos#1#2\@nil{%
  \ifx#1t\relax
    \let\psk@xlabelPos\tw@
    \pst@xticksizeC=\pst@xticksizeB
  \else
    \ifx#1a\relax
      \let\psk@xlabelPos\@ne 
      \pst@xticksizeC=\z@
    \else
      \def\psk@xlabelPos{\z@}
      \pst@xticksizeC=\pst@xticksizeA
  \fi\fi
}
\def\psset@@ylabelPos#1#2\@nil{%
  \ifx#1r\relax
    \def\psk@ylabelPos{\tw@}
    \pst@yticksizeC=\pst@yticksizeB
  \else
    \ifx#1a\relax
      \def\psk@ylabelPos{\@ne}
      \pst@yticksizeC=\z@
    \else 
      \def\psk@ylabelPos{\z@}
      \pst@yticksizeC=\pst@yticksizeA
  \fi\fi
}
\psset[pst-plot]{xlabelPos=b, ylabelPos=l}%
\define@key[psset]{pst-plot}{xyDecimals}[{}]{\def\psk@xDecimals{#1}\def\psk@yDecimals{#1}}
\define@key[psset]{pst-plot}{xDecimals}[{}]{\def\psk@xDecimals{#1}}
\define@key[psset]{pst-plot}{yDecimals}[{}]{\def\psk@yDecimals{#1}}
\psset[pst-plot]{xyDecimals={}}%
\define@key[psset]{pst-plot}{xlogBase}[{}]{\def\psk@xlogBase{#1}}
\define@key[psset]{pst-plot}{ylogBase}[{}]{\def\psk@ylogBase{#1}}
\define@key[psset]{pst-plot}{xylogBase}[{}]{\def\psk@xlogBase{#1}\def\psk@ylogBase{#1}}%
\psset[pst-plot]{xylogBase={}}%
\define@key[psset]{pst-plot}{trigLabelBase}[0]{\pst@getint{#1}{\psk@xtrigLabelBase}\let\psk@ytrigLabelBase\psk@xtrigLabelBase}
\define@key[psset]{pst-plot}{xtrigLabelBase}[0]{%
  \pst@getint{#1}{\psk@xtrigLabelBase}\psset{xtrigLabels}}
\define@key[psset]{pst-plot}{ytrigLabelBase}[0]{%
  \pst@getint{#1}{\psk@ytrigLabelBase}\psset{ytrigLabels}}
\psset[pst-plot]{trigLabelBase=0}
\define@key[psset]{pst-plot}{fractionLabelBase}[0]{\pst@getint{#1}{\psk@xfractionLabelBase}\let\psk@yfractionLabelBase\psk@xfractionLabelBase}
\define@key[psset]{pst-plot}{xfractionLabelBase}[0]{%
  \pst@getint{#1}{\psk@xfractionLabelBase}\psset{xfractionLabels}}
\define@key[psset]{pst-plot}{yfractionLabelBase}[0]{%
  \pst@getint{#1}{\psk@yfractionLabelBase}\psset{yfractionLabels}}
\psset[pst-plot]{fractionLabelBase=0}
\def\setDefaulthLabels{%
  \ifPst@xmathLabel\def\pshlabel##1{$\psk@xlabelFontSize##1$}\else\def\pshlabel##1{\psk@xlabelFontSize##1}\fi
  \def\pst@@@hlabel##1{%
      \edef\@xyDecimals{\psk@xDecimals}%
      \ifnum\psk@labels<\tw@\relax
        \ifx\psk@xlogBase\@empty
          \pshlabel{\psk@xlabelFontSize\expandafter\@LabelComma##1..\@nil\psk@xlabelFactor}%
        \else
          \ifPst@xmathLabel
            \pshlabel{\psk@xlabelFontSize\psk@xlogBase^{\expandafter\@stripDecimals##1..\@nil}}%
          \else
            \pshlabel{\psk@xlabelFontSize\psk@xlogBase\textsuperscript{\expandafter\@stripDecimals##1..\@nil}}%
          \fi
        \fi
      \fi
    }%
    \ifPst@xmathLabel\def\pshlabel##1{$\psk@xlabelFontSize##1$}\else\def\pshlabel##1{\psk@xlabelFontSize##1}\fi
}
\def\setTrighLabels{%
    \def\pst@@@hlabel##1{\pshlabel{##1}}%
    \def\pshlabel##1{%
      \ifnum\psk@xtrigLabelBase<2
        \def\de@nominator{\@ne}\else\def\de@nominator{\psk@xtrigLabelBase}\fi
      \def\pst@tempA{##1}%
      \pst@abs{\pst@tempA}\pst@cntm 
      \pst@mod{\pst@cntm}{\de@nominator}\pst@cntp 
      \ifnum\@ne>\pst@cntp                  
        \pst@cnto=\pst@cntm \divide\pst@cnto by \de@nominator  
	\ifPst@xmathLabel
          $\psk@xlabelFontSize
  	  \ifnum\pst@tempA<0 -\fi
          \ifnum\pst@cnto=\@ne                
            \pi                 	      
          \else
            \ifnum\pst@cnto=\z@ 0\else
            \the\pst@cnto\pi 	              
          \fi\fi$%
	\else%
          \psk@xlabelFontSize
  	  \ifnum\pst@tempA<0 -\fi
          \ifnum\pst@cnto=\@ne
            $\pi$
          \else%
            \the\pst@cnto$\pi$
          \fi%
	\fi%
      \else%
	\ifPst@xmathLabel%
          $\psk@xlabelFontSize%
          \ifnum\pst@cntp=\@ne
            \if\pst@cntm=\@ne%
              \frac{\pi}{\de@nominator}
            \else\ifnum\pst@tempA=-1 \frac{-\pi}{\de@nominator}%
              \else \ifnum\pst@tempA=1 \frac{\pi}{\de@nominator}%
                \else\frac{\pst@tempA\pi}{\de@nominator}
            \fi\fi\fi%
          \else%
            \ifnum\pst@tempA=1 \frac{\pi}{\de@nominator}%
            \else\ifnum\pst@tempA=\de@nominator \pi%
              \else\frac{\pst@tempA\pi}{\de@nominator}%
          \fi\fi\fi$%
	\else%
          \psk@xlabelFontSize%
          \ifnum\pst@cntp=\@ne
            \if\pst@cntm=\@ne%
              $\frac{\pi}{\de@nominator}$
            \else\ifnum\pst@tempA=-1 $\frac{-\pi}{\de@nominator}$%
              \else \ifnum\pst@tempA=1 $\frac{\pi}{\de@nominator}$%
                \else$\frac{\pst@tempA\pi}{\de@nominator}$
            \fi\fi\fi
          \else
            \ifnum\pst@tempA=1 $\frac{\pi}{\de@nominator}$%
            \else\ifnum\pst@tempA=\de@nominator $\pi$%
              \else$\frac{\pst@tempA\pi}{\de@nominator}$%
          \fi\fi\fi
	\fi
      \fi
    }%
}
\def\setDefaultvLabels{%
  \ifPst@ymathLabel\def\psvlabel##1{$\psk@ylabelFontSize##1$}\else\def\psvlabel##1{\psk@ylabelFontSize##1}\fi
    \def\pst@@@vlabel##1{%
      \edef\@xyDecimals{\psk@yDecimals}%
      \ifodd\psk@labels 
      \else%
        \ifx\psk@ylogBase\@empty
          \psvlabel{\expandafter\@LabelComma##1..\@nil\psk@ylabelFactor}%
        \else%
          \ifPst@ymathLabel%
            \psvlabel{\psk@ylogBase^{\expandafter\@stripDecimals##1..\@nil }}%
	  \else
            \psvlabel{\psk@ylogBase\textsuperscript{\expandafter\@stripDecimals##1..\@nil }}%
          \fi%
        \fi%
      \fi%
    }%
}%
\def\setTrigvLabels{%
  \def\pst@@@vlabel##1{\psvlabel{##1}}%
    \def\psvlabel##1{%
      \ifnum\psk@ytrigLabelBase<2 \def\de@nominator{\@ne}\else\def\de@nominator{\psk@ytrigLabelBase}\fi
      \def\pst@tempA{##1} 
      \pst@abs{\pst@tempA}\pst@cntm 
      \pst@mod{\pst@cntm}{\de@nominator}\pst@cntp 
      \ifnum\@ne>\pst@cntp                  
        \pst@cnto=\pst@cntm \divide\pst@cnto by \de@nominator  
	\ifPst@ymathLabel%
          $\psk@ylabelFontSize
  	  \ifnum\pst@tempA<0 -\fi
          \ifnum\pst@cnto=\@ne                
            \pi                 	      
          \else
            \the\pst@cnto\pi 	              
          \fi$%
	\else%
          \psk@ylabelFontSize%
  	  \ifnum\pst@tempA<0 -\fi
          \ifnum\pst@cnto=\@ne
            $\pi$
          \else
            \the\pst@cnto$\pi$
          \fi
	\fi
      \else
	\ifPst@ymathLabel%
          $\psk@ylabelFontSize
          \ifnum\pst@cntp=\@ne
            \if\pst@cntm=\@ne%
              \frac{\pi}{\de@nominator}
            \else\ifnum\pst@tempA=-1 \frac{-\pi}{\de@nominator}%
              \else \ifnum\pst@tempA=1 \frac{\pi}{\de@nominator}%
                \else\frac{\pst@tempA\pi}{\de@nominator}
            \fi\fi\fi%
          \else%
            \ifnum\pst@tempA=1 \frac{\pi}{\de@nominator}%
            \else\ifnum\pst@tempA=\de@nominator \pi%
              \else\frac{\pst@tempA\pi}{\de@nominator}%
          \fi\fi\fi$%
	\else
          \psk@ylabelFontSize
          \ifnum\pst@cntp=\@ne
            \if\pst@cntm=\@ne
              $\frac{\pi}{\de@nominator}$
            \else\ifnum\pst@tempA=-1 $\frac{-\pi}{\de@nominator}$%
              \else \ifnum\pst@tempA=1 $\frac{\pi}{\de@nominator}$%
                \else$\frac{\pst@tempA\pi}{\de@nominator}$
            \fi\fi\fi
          \else
            \ifnum\pst@tempA=1 $\frac{\pi}{\de@nominator}$%
            \else\ifnum\pst@tempA=\de@nominator $\pi$%
              \else$\frac{\pst@tempA\pi}{\de@nominator}$%
          \fi\fi\fi
	\fi
      \fi
    }%
}
\def\setFractionvLabels{%
  \def\pst@@@vlabel##1{\psvlabel{##1}}%
  \def\psvlabel##1{%
      \ifnum\psk@yfractionLabelBase<2 \def\de@nominator{\@ne}\else\def\de@nominator{\psk@yfractionLabelBase}\fi
      \def\pst@tempA{##1}%
      \pst@abs{\pst@tempA}\pst@cntm 
      \pst@mod{\pst@cntm}{\de@nominator}\pst@cntp 
      \ifnum\@ne>\pst@cntp                  
        \pst@cnto=\pst@cntm \divide\pst@cnto by \de@nominator  
	\ifPst@ymathLabel$\psk@ylabelFontSize\ifnum\pst@tempA<0 -\fi\the\pst@cnto\psk@ylabelFactor$%
	\else             \psk@ylabelFontSize\ifnum\pst@tempA<0 -\fi\the\pst@cnto\psk@ylabelFactor
	\fi
      \else
	\ifPst@ymathLabel
          $\psk@ylabelFontSize
          \ifnum\pst@cntp=\@ne                
            \if\pst@cntm=\@ne
              \frac{1}{\de@nominator}\psk@ylabelFactor
            \else\ifnum\pst@tempA=-1 \frac{-1}{\de@nominator}\psk@ylabelFactor%
              \else \ifnum\pst@tempA=1 \frac{1}{\de@nominator}\psk@ylabelFactor%
                \else\frac{\pst@tempA}{\de@nominator}\psk@ylabelFactor
            \fi\fi\fi
          \else
            \ifnum\pst@tempA=1 \frac{1}{\de@nominator}\psk@ylabelFactor%
            \else\ifnum\pst@tempA=\de@nominator 1\psk@xlabelFactor \else\frac{\pst@tempA}{\de@nominator}\psk@ylabelFactor%
          \fi\fi\fi$
	\else
          \psk@ylabelFontSize
          \ifnum\pst@cntp=\@ne
            \if\pst@cntm=\@ne
              $\frac{1}{\de@nominator}\psk@ylabelFactor$
            \else\ifnum\pst@tempA=-1 $\frac{-1}{\de@nominator}\psk@ylabelFactor$%
              \else \ifnum\pst@tempA=1 $\frac{1}{\de@nominator}\psk@ylabelFactor$%
                \else$\frac{\pst@tempA}{\de@nominator}\psk@ylabelFactor$
            \fi\fi\fi%
          \else%
            \ifnum\pst@tempA=1 $\frac{1}{\de@nominator}\psk@ylabelFactor$%
            \else\ifnum\pst@tempA=\de@nominator 1\psk@ylabelFactor
              \else$\frac{\pst@tempA}{\de@nominator}\psk@ylabelFactor$
          \fi\fi\fi
	\fi
      \fi
    }%
}
\def\setFractionhLabels{%
  \def\pst@@@hlabel##1{\pshlabel{##1}}%
  \def\pshlabel##1{%
      \ifnum\psk@xfractionLabelBase<2 \def\de@nominator{\@ne}\else\def\de@nominator{\psk@xfractionLabelBase}\fi
      \def\pst@tempA{##1}%
      \pst@abs{\pst@tempA}\pst@cntm 
      \pst@mod{\pst@cntm}{\de@nominator}\pst@cntp
      \ifnum\@ne>\pst@cntp                  
        \pst@cnto=\pst@cntm \divide\pst@cnto by \de@nominator  
	\ifPst@xmathLabel$\psk@xlabelFontSize\ifnum\pst@tempA<0 -\fi\the\pst@cnto\psk@xlabelFactor$%
	\else             \psk@xlabelFontSize\ifnum\pst@tempA<0 -\fi\the\pst@cnto\psk@xlabelFactor
	\fi
      \else
	\ifPst@xmathLabel
          $\psk@xlabelFontSize
          \ifnum\pst@cntp=\@ne
            \if\pst@cntm=\@ne \frac{1}{\de@nominator}\psk@xlabelFactor
            \else\ifnum\pst@tempA=-1 \frac{-1}{\de@nominator}\psk@xlabelFactor%
              \else\ifnum\pst@tempA=1 \frac{1}{\de@nominator}\psk@xlabelFactor%
                \else\frac{\pst@tempA}{\de@nominator}\psk@xlabelFactor
            \fi\fi\fi%
          \else%
            \ifnum\pst@tempA=1 \frac{1}{\de@nominator}\psk@xlabelFactor%
            \else\ifnum\pst@tempA=\de@nominator 1\psk@xlabelFactor\else\frac{\pst@tempA}{\de@nominator}\psk@xlabelFactor%
          \fi\fi\fi$
	\else
          \psk@xlabelFontSize
          \ifnum\pst@cntp=\@ne
            \if\pst@cntm=\@ne $\frac{1}{\de@nominator}\psk@xlabelFactor$
            \else\ifnum\pst@tempA=-1 $\frac{-1}{\de@nominator}\psk@xlabelFactor$%
              \else \ifnum\pst@tempA=1 $\frac{1}{\de@nominator}\psk@xlabelFactor$%
                \else$\frac{\pst@tempA}{\de@nominator}\psk@xlabelFactor$
            \fi\fi\fi
          \else
            \ifnum\pst@tempA=1 $\frac{1}{\de@nominator}\psk@xlabelFactor$%
            \else\ifnum\pst@tempA=\de@nominator 1\psk@xlabelFactor%
              \else$\frac{\pst@tempA}{\de@nominator}\psk@xlabelFactor$
          \fi\fi\fi
	\fi
      \fi
    }%
}
\define@boolkey[psset]{pst-plot}[Pst@]{xtrigLabels}[true]{%
  \ifPst@xtrigLabels \setTrighLabels \else \setDefaulthLabels\fi}
\define@boolkey[psset]{pst-plot}[Pst@]{ytrigLabels}[true]{%
  \ifPst@ytrigLabels \setTrigvLabels \else \setDefaultvLabels \fi}
\define@boolkey[psset]{pst-plot}[Pst@]{trigLabels}[true]{%
  \ifPst@trigLabels\psset[pst-plot]{xtrigLabels,ytrigLabels=false}
  \else            \psset[pst-plot]{xtrigLabels=false,ytrigLabels=false}%
  \fi}
\psset[pst-plot]{trigLabels=false}
\define@boolkey[psset]{pst-plot}[Pst@]{xfractionLabels}[true]{%
  \ifPst@xfractionLabels \setFractionhLabels \else \setDefaulthLabels \fi}
\define@boolkey[psset]{pst-plot}[Pst@]{yfractionLabels}[true]{%
  \ifPst@yfractionLabels \setFractionvLabels \else \setDefaultvLabels \fi}
\define@boolkey[psset]{pst-plot}[Pst@]{fractionLabels}[true]{%
  \ifPst@fractionLabels \setFractionhLabels\setFractionvLabels\Pst@xfractionLabelstrue\Pst@yfractionLabelstrue\fi}
\psset[pst-plot]{fractionLabels=false}
\def\psk@logLines{3}
\define@key[psset]{pst-plot}{logLines}[none]{\pst@expandafter\psset@@logLines#1\@nil\psk@logLines}
\def\psset@@logLines#1#2\@nil#3{%
  \ifx#1a\relax
    \let#3\z@
    \Pst@maxxTickstrue\Pst@maxyTickstrue
    \set@xticksize{0 4pt}\set@yticksize{0 4pt}%
    \def\psk@xsubticksize{1}\def\psk@ysubticksize{1}%
  \else
    \ifx#1x\relax
      \let#3\@ne
      \Pst@maxxTickstrue\Pst@maxyTicksfalse
      \set@xticksize{0 4pt}\def\psk@xsubticksize{1}%
    \else
      \ifx#1y\relax
        \let#3\tw@
	\Pst@maxyTickstrue\Pst@maxxTicksfalse
	\set@yticksize{0 4pt}\def\psk@ysubticksize{1}%
      \else
        \ifx#1n\let#3\thr@@\else
          \@pstrickserr{Bad argument: `#1#2'}\@ehpa
  \fi\fi\fi\fi}
\psset[pst-plot]{logLines=none}%
\define@key[psset]{pst-plot}{ylabelFactor}[\relax]{\def\psk@ylabelFactor{#1}}
\define@key[psset]{pst-plot}{xlabelFactor}[\relax]{\def\psk@xlabelFactor{#1}}
\define@boolkey[psset]{pst-plot}[Pst@]{showOriginTick}[true]{}%
\psset[pst-plot]{xlabelFactor=\relax,ylabelFactor=\relax,showOriginTick}%

\def\psxTick{\pst@object{psxTick}}
\def\psxTick@i{\@ifnextchar({\psxTick@ii{0}}\psxTick@ii}
\def\psxTick@ii#1(#2)#3{{%
  \pst@killglue
  \addbefore@par{arrows=-,linewidth=\psk@xtickwidth\pslinewidth}
  \ifPst@xtrigLabels\addto@par{xtrigLabels=false}\fi 
  \use@par
  \edef\temp@coor{(!#2 \pst@number\pst@xticksizeB \pst@number\psyunit div)(!#2 \pst@number\pst@xticksizeA \pst@number\psyunit div)}%
  \expandafter\psline\temp@coor
  \rput[t]{#1}(! \psk@origin 
                 #2 \pst@number\psxlabelsep \pst@number\pst@xticksizeB add
                 \pst@number\psyunit div neg ){\pshlabel{#3\vphantom{1}}}%
  }\ignorespaces}
\def\psyTick{\pst@object{psyTick}}
\def\psyTick@i{\@ifnextchar({\psyTick@ii{0}}\psyTick@ii}
\def\psyTick@ii#1(#2)#3{{%
  \pst@killglue
  \addbefore@par{arrows=-,linewidth=\psk@ytickwidth\pslinewidth}
  \ifPst@ytrigLabels \setDefaultvLabels \fi
  \use@par
  \edef\temp@coor{(!\pst@number\pst@yticksizeB \pst@number\psxunit div #2)(!\pst@number\pst@yticksizeA \pst@number\psxunit div #2)}%
  \expandafter\psline\temp@coor
    \rput[r]{#1}(!\psk@origin
                  \pst@number\pst@yticksizeB \pst@number\psylabelsep add
                  \pst@number\psxunit div neg #2){\psvlabel{#3}}}\ignorespaces}
\define@boolkey[psset]{pst-plot}[Pst@]{markPoint}[true]{}%
\psset[pst-plot]{markPoint}
\def\psCoordinates{\pst@object{psCoordinates}}
\def\psCoordinates@i(#1){%
  \pst@killglue%
  \begingroup
  \addbefore@par{showpoints=false,markPoint}
  \use@par
  \psline(#1|0,0)(#1)
  \psline(#1)(0,0|#1)%
  \ifPst@markPoint\psdot(#1)\fi%
  \endgroup
  \ignorespaces
}
\def\stripDecimals#1{\expandafter\@stripDecimals#1..\@nil}
\def\@stripDecimals#1.#2.#3\@nil{%
  \def\pst@dummy{#1}%
  \ifx\pst@dummy\@empty\the\@zero\else#1\fi
}
\newcount\@digitcounter\@digitcounter=0\relax
\def\@inc@digitcounter{\global\advance\@digitcounter by 1\relax}
\def\@get@digitcounter{\the\@digitcounter\relax}
\def\@Reset@digitcounter{\global\@digitcounter=0\relax}
\def\@zeroFill{%
  \ifnum \@xyDecimals>\@get@digitcounter
    \bgroup
      0\@inc@digitcounter\@zeroFill
    \egroup
  \fi
}
\def\@process@digits#1#2;{%
  \ifx *#1\@zeroFill\else#1\@inc@digitcounter 
  \ifnum\@xyDecimals>\@get@digitcounter\expandafter\@process@digits#2;\fi\fi%
}
\def\@writeDecimals#1{%
  \ifx\@xyDecimals\@empty
    \def\@tempa{#1}
    \ifx\@tempa\@empty
    \else\ifmmode\expandafter\mathord\expandafter{\psk@decimalSeparator}\else\psk@decimalSeparator\fi#1\fi%
  \else
    \ifnum\@xyDecimals>\@zero
      \ifmmode\expandafter\mathord\expandafter{\psk@decimalSeparator}\else\psk@decimalSeparator\fi%
        \@Reset@digitcounter
        \expandafter\@process@digits#1*;%
      \fi%
  \fi%
}
\def\@LabelComma#1.#2.#3\@nil{%
  \def\pst@tempA{#1}%
  \ifx\pst@tempA\@empty\the\@zero\else#1\fi
  \def\pst@tempA{#2}%
  \ifx\pst@tempA\@empty\@writeDecimals{}\else\@writeDecimals{#2}\fi
}
\def\set@xticksize#1{%
  \pst@expandafter\pst@getydimdim{#1} {} {}\@nil
  \ifdim\pst@dimm>\pst@dimn
    \pst@xticksizeA=\the\pst@dimn%
    \pst@xticksizeB=\the\pst@dimm%
  \else%
    \pst@xticksizeA=\the\pst@dimm%
    \pst@xticksizeB=\the\pst@dimn
  \fi%
  \edef\psk@xticksize{\pst@number\pst@xticksizeA \pst@number\pst@xticksizeB}%
  \ifnum\psk@xlabelPos<\z@\relax
    \pst@xticksizeC=\pst@dimn
  \else
    \pst@xticksizeC=\pst@dimm
  \fi
}
\def\set@yticksize#1{%
  \pst@expandafter\pst@getxdimdim{#1} {} {}\@nil
  \ifdim\pst@dimm>\pst@dimn\relax
    \pst@yticksizeA=\the\pst@dimn%
    \pst@yticksizeB=\the\pst@dimm%
  \else%
    \pst@yticksizeA=\the\pst@dimm%
    \pst@yticksizeB=\the\pst@dimn
  \fi%
  \edef\psk@yticksize{\pst@number\pst@yticksizeA \pst@number\pst@yticksizeB}%
  \ifnum\psk@ylabelPos<\z@	
    \pst@yticksizeC=\pst@dimn%
  \else%
      \pst@yticksizeC=\pst@dimo
  \fi%
}
\newif\ifPst@maxxTicks
\newif\ifPst@maxyTicks
\define@key[psset]{pst-plot}{ticksize}[-4pt 4pt]{%
  \def\pst@tempA{max}%
  \def\pst@tempB{#1}%
  \ifx\pst@tempA\pst@tempB%
    \Pst@maxxTickstrue\Pst@maxyTickstrue%
    \set@xticksize{0 4pt}\set@yticksize{0 4pt}%
  \else%
    \Pst@maxxTicksfalse\Pst@maxyTicksfalse%
    \set@xticksize{#1}\set@yticksize{#1}%
  \fi}
\define@key[psset]{pst-plot}{xticksize}{%
  \def\pst@tempA{max}%
  \def\pst@tempB{#1}%
  \ifx\pst@tempA\pst@tempB
    \Pst@maxxTickstrue\set@xticksize{0 4pt}%
  \else\set@xticksize{#1}\Pst@maxxTicksfalse\fi}
\define@key[psset]{pst-plot}{yticksize}{%
  \def\pst@tempA{max}%
  \def\pst@tempB{#1}%
  \ifx\pst@tempA\pst@tempB%
    \Pst@maxyTickstrue\set@yticksize{0 4pt}%
  \else\set@yticksize{#1}\Pst@maxyTicksfalse\fi}%
\psset[pst-plot]{ticksize=-4pt 4pt}
\define@key[psset]{pst-plot}{tickstyle}[full]{\pst@expandafter\psset@@tickstyle{#1}\@nil}
\def\psset@@tickstyle#1#2\@nil{%
  \ifx#1f\let\psk@tickstyle\z@\else			
    \ifx#1t\let\psk@tickstyle\@ne			
      \edef\psk@xticksize{0 \pst@number\pst@xticksizeB}%
      \edef\psk@yticksize{0 \pst@number\pst@yticksizeB}%
    \else\ifx#1b\let\psk@tickstyle\m@ne			
      \edef\psk@xticksize{\pst@number\pst@xticksizeA 0}%
      \edef\psk@yticksize{\pst@number\pst@yticksizeA 0}%
      \else\ifx#1i\let\psk@tickstyle\tw@
        \else\@pstrickserr{Bad tick style: `#1#2'}\@ehpa
  \fi\fi\fi\fi}
\psset[pst-plot]{tickstyle=full}
\define@key[psset]{pst-plot}{subticks}[1]{\def\psk@xsubticks{#1}\def\psk@ysubticks{#1}}
\define@key[psset]{pst-plot}{xsubticks}[1]{\def\psk@xsubticks{#1}}
\define@key[psset]{pst-plot}{ysubticks}[1]{\def\psk@ysubticks{#1}}
\define@key[psset]{pst-plot}{subticksize}[0.75]{\def\psk@xsubticksize{#1}\def\psk@ysubticksize{#1}}
\define@key[psset]{pst-plot}{xsubticksize}[0.75]{\def\psk@xsubticksize{#1}}
\define@key[psset]{pst-plot}{ysubticksize}[0.75]{\def\psk@ysubticksize{#1}}
\define@key[psset]{pst-plot}{tickwidth}[0.5\pslinewidth]{%
  \pst@getlength{#1}\psk@xtickwidth%
  \pst@getlength{#1}\psk@ytickwidth}
\define@key[psset]{pst-plot}{xtickwidth}[0.5\pslinewidth]{\pst@getlength{#1}\psk@xtickwidth}
\define@key[psset]{pst-plot}{ytickwidth}[0.5\pslinewidth]{\pst@getlength{#1}\psk@ytickwidth}
\define@key[psset]{pst-plot}{subtickwidth}[0.25\pslinewidth]{%
  \pst@getlength{#1}\psk@xsubtickwidth%
  \pst@getlength{#1}\psk@ysubtickwidth}
\define@key[psset]{pst-plot}{xsubtickwidth}[0.25\pslinewidth]{\pst@getlength{#1}\psk@xsubtickwidth}
\define@key[psset]{pst-plot}{ysubtickwidth}[0.25\pslinewidth]{\pst@getlength{#1}\psk@ysubtickwidth}
\define@key[psset]{pst-plot}{labelOffset}[0pt]{%
  \pst@getlength{#1}\psk@xlabelOffset%
  \pst@getlength{#1}\psk@ylabelOffset}
\define@key[psset]{pst-plot}{xlabelOffset}[0pt]{\pst@getlength{#1}\psk@xlabelOffset}
\define@key[psset]{pst-plot}{ylabelOffset}[0pt]{\pst@getlength{#1}\psk@ylabelOffset}
\define@key[psset]{pst-plot}{frameOffset}[0pt]{\pst@getlength{#1}\psk@frameOffset}
\define@key[psset]{pst-plot}{tickcolor}[black]{%
    \pst@getcolor{#1}\psk@xtickcolor%
    \pst@getcolor{#1}\psk@ytickcolor}
\define@key[psset]{pst-plot}{xtickcolor}[black]{\pst@getcolor{#1}\psk@xtickcolor}
\define@key[psset]{pst-plot}{ytickcolor}[black]{\pst@getcolor{#1}\psk@ytickcolor}
\define@key[psset]{pst-plot}{subtickcolor}[gray]{%
  \pst@getcolor{#1}\psk@xsubtickcolor%
  \pst@getcolor{#1}\psk@ysubtickcolor}
\define@key[psset]{pst-plot}{xsubtickcolor}[gray]{\pst@getcolor{#1}\psk@xsubtickcolor}
\define@key[psset]{pst-plot}{ysubtickcolor}[gray]{\pst@getcolor{#1}\psk@ysubtickcolor}
\define@key[psset]{pst-plot}{xticklinestyle}[solid]{%
  \@ifundefined{psls@#1}%
    {\@pstrickserr{Line style `#1' not defined}\@eha}%
    {\def\psxticklinestyle{#1}}}
\define@key[psset]{pst-plot}{xsubticklinestyle}[solid]{%
  \@ifundefined{psls@#1}%
    {\@pstrickserr{Line style `#1' not defined}\@eha}%
    {\def\psxsubticklinestyle{#1}}}
\define@key[psset]{pst-plot}{yticklinestyle}[solid]{%
  \@ifundefined{psls@#1}%
    {\@pstrickserr{Line style `#1' not defined}\@eha}%
    {\def\psyticklinestyle{#1}}}
\define@key[psset]{pst-plot}{ysubticklinestyle}[solid]{%
  \@ifundefined{psls@#1}%
    {\@pstrickserr{Line style `#1' not defined}\@eha}%
    {\def\psysubticklinestyle{#1}}}
\define@key[psset]{pst-plot}{ticklinestyle}[solid]{%
  \@ifundefined{psls@#1}%
    {\@pstrickserr{Line style `#1' not defined}\@eha}%
    {\def\psxticklinestyle{#1}\def\psyticklinestyle{#1}}}
\define@key[psset]{pst-plot}{subticklinestyle}[solid]{%
  \@ifundefined{psls@#1}%
    {\@pstrickserr{Line style `#1' not defined}\@eha}%
    {\def\psxsubticklinestyle{#1}\def\psysubticklinestyle{#1}}}
\psset[pst-plot]{subticksize=0.75,subticks=1,tickcolor=black,ticklinestyle=solid,
  subticklinestyle=solid,subtickcolor=gray,tickwidth=0.5\pslinewidth,
  subtickwidth=0.25\pslinewidth,labelOffset=0pt,frameOffset=0pt}
\define@key[psset]{pst-plot}{nStep}[1]{\def\psk@nStep{#1}}
\define@key[psset]{pst-plot}{nStart}[0]{\def\psk@nStart{#1}}
\define@key[psset]{pst-plot}{nEnd}[{}]{\def\psk@nEnd{#1}}
\define@key[psset]{pst-plot}{xStep}[0]{\def\psk@xStep{#1}}
\define@key[psset]{pst-plot}{yStep}[0]{\def\psk@yStep{#1}}
\define@key[psset]{pst-plot}{xStart}[{}]{\def\psk@xStart{#1}}
\define@key[psset]{pst-plot}{xEnd}[{}]{\def\psk@xEnd{#1}}
\define@key[psset]{pst-plot}{yStart}[{}]{\def\psk@yStart{#1}}
\define@key[psset]{pst-plot}{yEnd}[{}]{\def\psk@yEnd{#1}}
\define@key[psset]{pst-plot}{plotNoX}[1]{\def\psk@plotNoX{#1}}
\define@key[psset]{pst-plot}{plotNo}[1]{\def\psk@plotNo{#1}}
\define@key[psset]{pst-plot}{plotNoMax}[1]{\def\psk@plotNoMax{#1}}
\define@key[psset]{pst-plot}{plotYMax}[1]{\def\psk@plotYMax{#1}}
\psset[pst-plot]{nStep=1, nStart=0, nEnd={},%
  xStep=0, yStep=0, xStart={}, xEnd={},  yStart={}, yEnd={}, 
  plotNo=1,plotNoMax=1,plotNoX=1,plotYMax={}}%
\def\pstScalePoints(#1,#2)#3#4{%
  \def\pstXScale{#1 }%
  \def\pstYScale{#2 }%
  \def\pstXPSScale{#3 }%
  \def\pstYPSScale{#4 }%
  \pst@def{ScalePoints}<%
    /yVal ED /xVal ED
    /yPSOp { #4 yVal mul #2 mul } def
    /xPSOp { #3 xVal mul #1 mul } def
    counttomark dup dup cvi eq not { exch pop } if
    /m exch def /n m 2 div cvi def
    n {
      \ifPst@polarplot exch cvi 360 mod PtoC \fi  
      yPSOp m 1 roll xPSOp m 1 roll 
      /m m 2 sub
      def } repeat>%
}
\pstScalePoints(1,1){}{}
\def\psxs@none{\let\psk@arrowA\@empty\let\psk@arrowB\@empty\psxs@axes}
\def\psxs@axes{{%
  \ifPst@xAxis\psxs@@axes\pst@dima\pst@dimb\pst@dimc\pst@dimd{}{x}\fi%
  \ifPst@yAxis\psxs@@axes\pst@dima\pst@dimb\pst@dimc\pst@dimd{exch}{y}\fi%
}}
\newif\ifSpecialLabelsDone
\def\psaxes{\pst@object{psaxes}}
\def\psaxes@i{%
  \let\pst@par@save\pst@par
  \pst@getarrows\psaxes@ii}
\def\psaxes@ii(#1){\@ifnextchar({\psaxes@iii(#1)}{\psaxes@iv(0,0)(0,0)(#1)}}
\def\psaxes@iii(#1)(#2){\@ifnextchar({\psaxes@iv(#1)(#2)}{\psaxes@iv(#1)(#1)(#2)}}
\def\psaxes@iv(#1)(#2)(#3){\@ifnextchar[{\psaxes@v(#1)(#2)(#3)}{\psaxes@vii(#1)(#2)(#3)}}%
\def\psaxes@v(#1)(#2)(#3)[#4]{\@ifnextchar[{\psaxes@vi(#1)(#2)(#3)[#4]}{\psaxes@vi(#1)(#2)(#3)[#4][]}}%
\def\psaxes@vi(#1)(#2)(#3)[#4,#5][#6,#7]{%
  \psaxes@vii(#1)(#2)(#3)%
  \let\pst@par\pst@par@save
  \begingroup
  \SpecialCoor
  \use@par
  \ifshowgrid\psgrid[style=gridstyleA]\fi
  \uput{\psxlabelsep}[#5](#3|#1){#4}\uput{\psylabelsep}[#7](#1|#3){#6}%
  \endgroup
  \ignorespaces
}
\def\psaxes@vii(#1,#2)(#3,#4)(#5,#6){%
  \pst@killglue
  \begingroup
  \ifdim\pst@dimc<\z@\relax 
    \ifdim\pst@dimd<\z@\relax 
      \addbefore@par{xlabelPos=t,ylabelPos=r}%
  \fi\fi
  \use@par
  \pssetxlength\pst@dimc{#5}
  \pssetylength\pst@dimd{#6}
    \pssetxlength\pst@dimg{#1}
    \pssetylength\pst@dimh{#2}
    \pssetxlength\pst@dima{#3}
    \pssetylength\pst@dimb{#4}
    \pst@dima=\dimexpr\pst@dima-\pst@dimg\relax
    \pst@dimb=\dimexpr\pst@dimb-\pst@dimh\relax
    \pst@dimc=\dimexpr\pst@dimc-\pst@dimg\relax
    \pst@dimd=\dimexpr\pst@dimd-\pst@dimh\relax
   \setbox\pst@hbox=\hbox\bgroup
    \ifshowgrid\psgrid[style=gridstyleA]\fi
    \@nameuse{psxs@\psk@axesstyle}
    \ifPst@xAxis
      \SpecialLabelsDonefalse
      \begingroup
      \ifnum\psk@dx=\z@
        \pst@dimg=\psk@Dx\psxunit
        \ifdim\pst@dimg<\p@ 
          \pst@cnta=\psk@Dx
          \edef\psk@Dx{\the\numexpr-1*\pst@cnta}%
        \fi
        \edef\psk@dx{\number\pst@dimg}%
      \fi
      \pst@hlabels{\pst@dimc}{\psk@arrowB}{#3}{#5}
      \ifPst@yAxis\showoriginfalse\fi
      \pst@hlabels{\pst@dima}{\psk@arrowA}{#3}{#5}
      \endgroup
    \fi
    \ifPst@yAxis
      \SpecialLabelsDonefalse
      \begingroup
      \ifdim\pst@dima=\z@ \else\ifPst@xtrigLabels\showoriginfalse\fi\fi
      \ifnum\psk@dy=\z@
        \pst@dimg=\psk@Dy\psyunit
        \ifdim\pst@dimg<\p@ 
          \pst@cnta=\psk@Dy
          \edef\psk@Dy{\the\numexpr-1*\pst@cnta}%
        \fi
        \edef\psk@dy{\number\pst@dimg}%
      \fi
      \pst@vlabels{\pst@dimb}{\psk@arrowA}{#4}{#6}%
      \ifPst@xAxis\ifdim\pst@dima<\z@ \showoriginfalse\fi\fi 
      \pst@vlabels{\pst@dimd}{\psk@arrowB}{#4}{#6}%
      \endgroup
    \fi
  \egroup%
  \pssetxlength\pst@dimg{#1}%
  \pssetylength\pst@dimh{#2}%
  \leavevmode
  \psput@cartesian\pst@hbox
  \endgroup
  \ignorespaces
}
\newif\ifis@yAxis%
\def\psxs@@axes#1#2#3#4#5#6{
  \pst@killglue
  \begin@SpecialObj
    \ifx#6x\relax
      \is@yAxisfalse
      \ifnum\psk@dx=\z@
        \pst@dimg=\psk@Dx\psxunit
        \def\psk@dx{\number\pst@dimg}%
      \fi
    \else
      \is@yAxistrue
      \ifnum\psk@dy=\z@
        \pst@dimg=\psk@Dy\psyunit
        \def\psk@dy{\number\pst@dimg}%
      \fi
    \fi
    \let\pst@linetype\pst@arrowtype
    \def\pst@axes{axes}%
    \pst@addarrowdef
    \addto@pscode{
      /showOrigin \ifPst@showOriginTick true \else false \fi def 	
      \ifis@yAxis 0 \pst@number#4 \else \pst@number#3 0 \fi
      \ifis@yAxis 0 \pst@number#2 \else \pst@number#1 0 \fi
      ArrowA
      CP 4 2 roll
      ArrowB 
      2 copy
      /yEnd exch def /xEnd exch def
      \ifx\psk@axesstyle\@none   
        pop pop 
      \else
        gsave                              		
        L                                  		
        \@nameuse{psls@\pslinestyle}                 	
        stroke                                       	
        grestore
      \fi
      /yStart exch def
      /xStart exch def
      \number\psk@ticks\space dup 2 mod 0 eq \ifis@yAxis true \else false \fi and 
      exch 2 lt \ifis@yAxis false \else true \fi and or {
      /viceversa 
        \ifis@yAxis\pst@number#2 \pst@number#4 \else\pst@number#1 \pst@number#3 \fi
         gt { true }{ false } ifelse def           
      /epsilon 0.01 def                            
      /minTickline \ifis@yAxis \pst@number#1 \else \pst@number#2 \fi def
      /maxTickline \ifis@yAxis \pst@number#3 \else \pst@number#4 \fi def
      /dT \ifis@yAxis \psk@dy \else \psk@dx \fi\space abs  
        65536 div viceversa { neg } if def                 
      /DT \ifis@yAxis \psk@Dy \else \psk@Dx \fi\space abs viceversa { neg } if def  
      /subTNo \ifis@yAxis\psk@ysubticks\else\psk@xsubticks\fi \space def
      subTNo 0 gt { /dsubT dT subTNo div def}{ /dsubT 0 def } ifelse  
      \ifis@yAxis \psk@yticksize \else \psk@xticksize \fi
      /tickend exch def /tickstart exch def
      /Twidth \ifis@yAxis \psk@ytickwidth \else \psk@xtickwidth \fi\space def
      /subTwidth \ifis@yAxis \psk@ysubtickwidth \else \psk@xsubtickwidth \fi\space def
      /STsize \ifis@yAxis \psk@ysubticksize \else \psk@xsubticksize \fi\space def
      /TColor {
        \ifis@yAxis\pst@usecolor\psk@ytickcolor
        \else\pst@usecolor\psk@xtickcolor\fi\space } def
      /subTColor {
        \ifis@yAxis\pst@usecolor\psk@ysubtickcolor
        \else\pst@usecolor\psk@xsubtickcolor\fi\space } def
      /MinValue { \ifis@yAxis yStart \else xStart \fi
        \ifx\psk@arrowA\@empty\else 
          \psk@arrowsize\space CLW mul add \psk@arrowlength\space mul 
           viceversa { sub epsilon add }{ add epsilon sub } ifelse \fi } def
      /MaxValue { \ifis@yAxis yEnd \else xEnd \fi 
        \ifx\psk@arrowB\@empty\else
          \psk@arrowsize\space CLW mul add \psk@arrowlength\space mul 
           viceversa { add epsilon sub }{ sub epsilon add } ifelse \fi } def
      /logLines {
        \ifnum\psk@logLines=\z@ true \else         
          \ifnum\psk@logLines<\tw@                 
            \ifis@yAxis false \else true \fi       
          \else
            \ifnum\psk@logLines<\thr@@             
              \ifis@yAxis true \else false \fi     
            \else 
              false                                
            \fi
          \fi
        \fi
      } def
      /LSstroke {                                  
        \ifis@yAxis \@nameuse{psls@\psyticklinestyle}
        \else       \@nameuse{psls@\psxticklinestyle}\fi 
        stroke} def
      /subLSstroke {                               
        \ifis@yAxis \@nameuse{psls@\psysubticklinestyle}
        \else       \@nameuse{psls@\psxsubticklinestyle}\fi 
        stroke} def
      0 dT MaxValue 1 add {                        
        /cntTick exch def                          
        logLines {                                 
          gsave
          1 1 DT {
           1 sub /OffSet exch def
          -10 subTNo 1 add div dup 10 add exch dup -0.1 mul 1 add {                   
            /dx exch def                           
            /x dx log OffSet add \ifis@yAxis\pst@number\psyunit\else\pst@number\psxunit\fi\space mul cntTick add def       %
            x abs MaxValue abs le {                
	      \ifis@yAxis
	        \ifPst@maxyTicks true \else false \fi
	      \else
	        \ifPst@maxxTicks true \else false \fi
	      \fi
                { x minTickline #5 moveto
                  x maxTickline #5 lineto }
                { x tickstart STsize mul #5 moveto
                  x tickend STsize mul #5 lineto } ifelse
            } if
          } for } for
          subTwidth SLW subTColor                  
          subLSstroke
          grestore                                 
          stroke
          /dsubT 0 def                             
        } if 					   
        dsubT abs 0 gt {                           
          gsave                                    
          /cntsubTick cntTick dsubT add def
          subTNo 1 sub {
            cntsubTick abs MaxValue abs le {       
    	    \ifis@yAxis
              \ifPst@maxyTicks true \else false \fi
    	    \else
              \ifPst@maxxTicks true \else false \fi
    	    \fi
              { cntsubTick minTickline STsize mul #5 moveto
                cntsubTick maxTickline STsize mul #5 lineto }
              { cntsubTick tickstart STsize mul #5 moveto
                cntsubTick tickend STsize mul #5 lineto } ifelse
            }{ exit }  ifelse
            /cntsubTick cntsubTick dsubT add def
          } repeat 
          subTwidth SLW subTColor               
          subLSstroke
          grestore                              
        } if
        showOrigin {
          gsave
          \ifis@yAxis
            \ifPst@maxyTicks true \else false \fi
          \else
            \ifPst@maxxTicks true \else false \fi
          \fi
            { cntTick minTickline #5 moveto
              cntTick maxTickline #5 lineto }
            { cntTick tickstart #5 moveto        
              cntTick tickend #5 lineto } ifelse 
          Twidth SLW TColor                      
          LSstroke
          grestore
        }{ /showOrigin true def } ifelse         
      } for
      /showOrigin \ifPst@showOriginTick true \else false \fi def 
      /dT dT neg def                               
      /dsubT dsubT neg def
      0 dT MinValue epsilon viceversa { add }{ sub } ifelse {
        /cntTick exch def
        logLines {                                 
          gsave
          1 1 DT cvi {
            1 sub /OffSet exch def
          -10 subTNo 1 add div dup 10 add exch dup -0.1 mul 1 add {                   
            /dx exch def                           
            /x dx log OffSet add \ifis@yAxis\pst@number\psyunit\else\pst@number\psxunit\fi\space mul cntTick add def
            x abs MinValue abs le {                
	      \ifis@yAxis
	        \ifPst@maxyTicks true \else false \fi
	      \else
	        \ifPst@maxxTicks true \else false \fi
	      \fi
                { x minTickline #5 moveto
                  x maxTickline #5 lineto }
                { x tickstart STsize mul #5 moveto
                  x tickend STsize mul #5 lineto } ifelse
            } if
          } for } for
          /dsubT 0 def 
          subTwidth SLW subTColor                  
          subLSstroke
          grestore
        }                                          
        dsubT abs 0 gt {                           
          gsave                                    
          /cntsubTick cntTick dsubT add def
          subTNo 1 sub {
            cntsubTick abs MinValue abs le {       
              cntsubTick tickstart STsize mul #5 moveto
              cntsubTick tickend STsize mul #5 lineto
            }{ exit } ifelse
            /cntsubTick cntsubTick dsubT add def
          } repeat 
          subTwidth SLW subTColor                  
          subLSstroke
          grestore                                 
        } if
        showOrigin {
          gsave
          cntTick tickstart #5 moveto         	
          cntTick tickend #5 lineto    	       	
          Twidth SLW TColor                         
          LSstroke
          grestore
        }{ /showOrigin true def } ifelse         
      } for
    } if
   }
  \end@SpecialObj%
  \ifx\psk@axesstyle\@none\else
    \ifPst@yAxis\psline[linecolor=\pslinecolor](0,#2)(0,#4)\fi
    \ifPst@xAxis\psline[linecolor=\pslinecolor](#1,0)(#3,0)\fi
  \fi
  \ignorespaces
}%
\def\psxs@frame{%
  \psset{axesstyle=none}%
  \begin@SpecialObj%
    \addto@pscode{					
      \pst@number\pst@dima \psk@frameOffset sub \pst@number\pst@dimb \psk@frameOffset sub moveto 	
      \pst@number\pst@dimc \psk@frameOffset add \pst@number\pst@dimb \psk@frameOffset sub L	
      \pst@number\pst@dimc \psk@frameOffset add \pst@number\pst@dimd \psk@frameOffset add L 	
      \pst@number\pst@dima \psk@frameOffset sub \pst@number\pst@dimd \psk@frameOffset add L 	
      closepath 
      }%
    \pst@stroke%
    \psk@fillstyle%
  \end@SpecialObj%
  \let\psk@arrowA\@empty%
  \let\psk@arrowB\@empty%
  \pst@xticksizeC=\z@\pst@yticksizeC=\z@  
  \ifPst@xAxis\psxs@@axes\pst@dima\pst@dimb\pst@dimc\pst@dimd{}{x}\fi
  \ifPst@yAxis\psxs@@axes\pst@dima\pst@dimb\pst@dimc\pst@dimd{ exch }{y}\fi
  \ifnum\psk@tickstyle=\tw@	
    \psDEBUG[psxs@frame]{psk@tickstyle=2 (inner)}%
    \psDEBUG[psxs@frame]{pst@dima=\pst@number\pst@dima}%
    \psDEBUG[psxs@frame]{pst@dimb=\pst@number\pst@dimb}%
    \psDEBUG[psxs@frame]{pst@dimc=\pst@number\pst@dimc}%
    \psDEBUG[psxs@frame]{pst@dimd=\pst@number\pst@dimd}%
    \ifPst@xAxis\psxs@@axes\pst@dima\pst@dimb\pst@dimc\pst@dimd{ neg \pst@number\pst@dimd add }{x}\fi
    \ifPst@yAxis\psxs@@axes\pst@dima\pst@dimb\pst@dimc\pst@dimd{ neg \pst@number\pst@dimc add exch }{y}\fi
  \fi%
}
\def\psxs@polar{
  \pst@killglue
  \begingroup
  \edef\pst@dimC{\strip@pt\pst@dimc}
  \pstFPDiv\pstR@dius{\pst@dimC}{\strip@pt\psxunit}
  \edef\pst@dimD{\strip@pt\pst@dimd}
  \pstFPDiv\psk@EndAngle{\pst@dimD}{\strip@pt\psyunit}
  \ifnum\psk@EndAngle=0 \def\psk@EndAngle{360}\fi
  \use@keep@par
  \pstFPDiv\pstN@lpha{\psk@EndAngle}{\psk@Dy}
  \pstFPdiv\pstd@lpha{\psk@Dy}{\psk@ysubticks}
  \pstFPdiv\pstdR@dius{1}{\psk@xsubticks}
  \pst@cntm=\psk@xsubticks\advance\pst@cntm by \m@ne
  \multido{\iA=\psk@Dx+\psk@Dx,\rB=\pstdR@dius+\psk@Dx,\iB=0+1}{\pstR@dius}{%
    \multido{\rA=\rB+\pstdR@dius}{\the\pst@cntm}{%
      \psarc[linestyle=\psxsubticklinestyle,
         linecolor=\psk@xsubtickcolor,linewidth=\psk@xsubtickwidth pt](0,0){\rA}{0}{\psk@EndAngle}}    
    \psarc[linestyle=\psxticklinestyle,linecolor=\psk@xtickcolor,
		linewidth=\psk@xtickwidth pt](0,0){\iA}{0}{\psk@EndAngle}%
    \ifnum\psk@labels<2\relax
      \uput[-45](\iB,0){\pshlabel{\iB}}\uput[45](0,\iB){\pshlabel{\iB}}%
    \fi%
  }%
  \pst@cntm=\psk@ysubticks\advance\pst@cntm by \m@ne
  \multido{\iA=\psk@Oy+\psk@Dy,\rB=\pstd@lpha+\psk@Dy}{\pstN@lpha}{%
    \multido{\rA=\rB+\pstd@lpha}{\the\pst@cntm}{\psline[linestyle=\psysubticklinestyle,
      linecolor=\psk@ysubtickcolor,linewidth=\psk@ysubtickwidth pt](\pstR@dius;\rA)} 
    \psline[linestyle=\psyticklinestyle,
      linecolor=\psk@ytickcolor,linewidth=\psk@ytickwidth pt](\pstR@dius;\iA)%
    \ifodd\psk@labels\else
      \uput[\iA](\pstR@dius;\iA){\psvlabel{\iA\psk@ylabelFactor}}%
    \fi%
  }%
  \ifnum\psk@EndAngle<360 \psline[linestyle=\psyticklinestyle,
      linecolor=\psk@ytickcolor,linewidth=\psk@ytickwidth pt](\pstR@dius;0)\fi
  \endgroup\ignorespaces%
  \Pst@xAxisfalse\Pst@yAxisfalse%
}
\def\@polar{polar}
\define@key[psset]{pst-plot}{axesstyle}[axes]{%
  \@ifundefined{psxs@#1}%
    {\@pstrickserr{Axes style `#1' not defined}\@eha}%
    {\def\psk@axesstyle{#1}%
     \ifx\psk@axesstyle\@polar\psset{Dy=30}\fi}}
\psset[pst-plot]{axesstyle=axes}
\define@key[psset]{pst-plot}{xLabels}[]{\def\psk@xLabels{#1}}
\define@key[psset]{pst-plot}{xLabelsRot}[0]{\pst@getangle{#1}\pst@xLabelsRot}
\psset[pst-plot]{xLabels=,xLabelsRot=0}
\define@key[psset]{pst-plot}{yLabels}[]{\def\psk@yLabels{#1}}
\define@key[psset]{pst-plot}{yLabelsRot}[0]{\pst@getangle{#1}\pst@yLabelsRot}
\psset[pst-plot]{yLabels=,yLabelsRot=0}
\def\pst@hlabels#1#2#3#4{%
  \ifSpecialLabelsDone
  \else
    \kern\psk@xlabelOffset pt            
    \ifx\empty\psk@xLabels
      \ifdim#1=\z@
      \else                   
        \ifx#2\empty
        \else
          \advance#1\ifdim#1>\z@-\fi7\pslinewidth
        \fi
        \pst@cnta=#1\relax                
        \divide\pst@cnta\psk@dx\relax     
        \ifnum\pst@cnta=\z@
        \else
          \pst@dimb=\psk@dx sp            
          \ifnum\psk@labels<\tw@ \ifPst@xAxis\pst@@hlabels\fi\fi
          \showoriginfalse
        \fi
      \fi
   \else
     \ifnum\psk@xlabelPos=\tw@ \def\pst@tempC{90}\else\def\pst@tempC{-90}\fi
       \pstFPsub\pst@pmtempa{#4}{#3}%
       \pstFPDiv\pst@pmtempb{\pst@pmtempa}{\psk@Dx}%
       \pstFPadd\pst@pmtempc{\pst@pmtempb}{-1}%
       \pstFPadd\pst@pmtempd{\pst@pmtempb}{1}%
       \ifdim\pst@pmtempb pt < \z@ 
         \def\pst@pmtempe{\pst@int{\pst@pmtempc}}%
       \else
         \def\pst@pmtempe{\pst@int{\pst@pmtempd}}%
       \fi
       \multido{\nA=0+1,\rA=#3+\psk@Dx}{\pst@pmtempe}{%
         \ifdim \nA pt < \z@ \def\nB{-\nA} \else \def\nB{\nA} \fi
         \uput{\psxlabelsep}[\pst@tempC]{\pst@xLabelsRot}(\rA,0){%
              \strut\expandafter\pshlabel\expandafter{\psPutXLabel{\nB}}}}%
       \SpecialLabelsDonetrue
    \fi
  \fi
}
\def\pst@@hlabels{%
  \setbox\z@=\vbox{
    \ifcase\psk@xlabelPos
      \vskip-\pst@xticksizeA\vskip\psxlabelsep\or
      \vskip-1ex\vskip-\pslabelsep\or
      \vskip-\pst@xticksizeB\vskip-\psxlabelsep\vskip-1ex
    \fi
    \ifnum\pst@cnta<\z@ \pst@dimb=-\pst@dimb\fi
    \hbox to \z@{%
      \ifshoworigin\hbox to \z@{\hss\pst@@@hlabel{\psk@Ox}\hss}\fi
      \mmultido{\nA=\psk@Ox+\psk@Dx}{\pst@cnta}{%
        \hskip\pst@dimb \hbox to \z@{\hss
          \ifdim\nA pt=\z@\relax\ifshoworigin\pst@@@hlabel{0}\fi
          \else\expandafter\pst@@@hlabel{\nA}%
          \fi
        \hss}%
      }\hss
    }%
  }\ht\z@\z@ \dp\z@\z@ \box\z@}
\def\pst@vlabels#1#2#3#4{%
  \ifSpecialLabelsDone\else
      \ifx\empty\psk@yLabels
        \ifdim#1=\z@\else
          \ifx#2\empty\else\ifdim#1>\z@ \advance#1 by -7\pslinewidth\else\advance#1 by 7\pslinewidth\fi\fi
          \pst@cnta=#1\relax           
          \divide\pst@cnta\psk@dy\relax
          \ifnum\pst@cnta=\z@\else
            \pst@dima=\psk@dy sp
            \ifodd\number\psk@labels\else\ifPst@yAxis\pst@@vlabels\fi
          \fi
          \showoriginfalse
        \fi
      \fi
    \else
	\pstFPsub\pst@pmtempa{#4}{#3}%
	\pstFPDiv\pst@pmtempb{\pst@pmtempa}{\psk@Dy}%
	\pstFPadd\pst@pmtempc{\pst@pmtempb}{-1}%
	\pstFPadd\pst@pmtempd{\pst@pmtempb}{1}%
	\ifdim\pst@pmtempb pt < \z@ \def\pst@pmtempe{\pst@int{\pst@pmtempc}}\else\def\pst@pmtempe{\pst@int{\pst@pmtempd}}\fi
	\multido{\nA=0+1,\rA=#3+\psk@Dy}{\pst@pmtempe}{%
	  \ifdim \nA pt < \z@ \def\nB{-\nA}\else \def\nB{\nA}\fi
	  \ifnum\psk@ylabelPos=0
            \uput{\psylabelsep}[180]{\pst@yLabelsRot}(0,\rA){%
              \strut\expandafter\psvlabel\expandafter{\psPutYLabel{\nB}}}%
          \else
            \uput{\psylabelsep}[0]{\pst@yLabelsRot}(0,\rA){%
              \strut\expandafter\psvlabel\expandafter{\psPutYLabel{\nB}}}%
          \fi
        }%
      \SpecialLabelsDonetrue
    \fi
  \fi
}
\def\pst@@vlabels{%
  \vbox to\z@{%
   \vbox to -\psk@ylabelOffset pt{}
    \ifnum\pst@cnta>\z@ \pst@dima=-\pst@dima\fi
    \offinterlineskip
    \ifshoworigin
      \vbox to \z@{\vss\hbox to\z@{%
        \ifcase\psk@ylabelPos
	  \hss\pst@@@vlabel{\psk@Oy}\hskip\psylabelsep\hskip-\pst@yticksizeA\or%
	  \hskip\pslabelsep\hss\pst@@@vlabel{\psk@Oy}\hss\or
	  \hskip\pst@yticksizeB\hskip\psylabelsep\pst@@@vlabel{\psk@Oy}%
	\fi}\vss}%
    \fi
    \mmultido{\nA=\psk@Oy+\psk@Dy}{\pst@cnta}{%
      \vbox to\pst@dima{\vss}%
      \vbox to \z@{%
        \vss\hbox to\z@{%
        \ifcase\psk@ylabelPos 
	  \hss\ifdim\nA pt=\z@ \ifshoworigin\pst@@@vlabel{0}\fi\else\pst@@@vlabel{\nA}\fi
	    \hskip\psylabelsep\hskip-\pst@yticksizeA\or
	  \hss\ifdim\nA pt=\z@ \ifshoworigin\pst@@@vlabel{0}\fi\else\pst@@@vlabel{\nA}\fi
	  \ifdim\psylabelsep=\z@\hss\else\kern-\psylabelsep\fi\or
	  \hskip\pst@yticksizeB\hskip\psylabelsep
	  \ifdim\nA pt=\z@ \ifshoworigin\pst@@@vlabel{0}\fi\else\pst@@@vlabel{\nA}\fi
	\fi}\vss}%
    }\vss}%
}
\define@key[psset]{pst-plot}{xAxisLabel}[x]{\def\psk@xAxisLabel{#1}}
\define@key[psset]{pst-plot}{yAxisLabel}[y]{\def\psk@yAxisLabel{#1}}
\psset[pst-plot]{xAxisLabel=x,yAxisLabel=y}
\define@key[psset]{pst-plot}{xAxisLabelPos}[{}]{\def\psk@xAxisLabelPos{#1}}
\define@key[psset]{pst-plot}{yAxisLabelPos}[{}]{\def\psk@yAxisLabelPos{#1}}
\psset[pst-plot]{yAxisLabelPos={},xAxisLabelPos={}}
\newdimen\psk@llx
\newdimen\psk@lly
\newdimen\psk@urx
\newdimen\psk@ury
\define@key[psset]{pst-plot}{llx}[\z@]{\pssetxlength\psk@llx{#1}}
\define@key[psset]{pst-plot}{lly}[\z@]{\pssetylength\psk@lly{#1}}
\define@key[psset]{pst-plot}{urx}[\z@]{\pssetxlength\psk@urx{#1}}
\define@key[psset]{pst-plot}{ury}[\z@]{\pssetylength\psk@ury{#1}}
\psset[pst-plot]{llx=\z@, lly=\z@, urx=\z@, ury=\z@}
\define@boolkey[psset]{pst-plot}[Pst@]{psgrid}[true]{}
\define@key[psset]{pst-plot}{gridpara}[{}]{\def\psk@gridpara{#1}}
\define@key[psset]{pst-plot}{gridcoor}[\relax]{\def\psk@gridcoor{#1}}
\psset[pst-plot]{psgrid=false,gridpara={gridlabels=0pt,gridcolor=red!30,subgridcolor=green!30,subgridwidth=0.5\pslinewidth,
  subgriddiv=5},gridcoor=\relax}
\define@key[psset]{pst-plot}{axespos}[bottom]{\pst@expandafter\psset@@axespos{#1}\@nil}
\def\psset@@axespos#1#2\@nil{%
  \ifx#1b\let\psk@axespos\z@\else		
    \ifx#1t\let\psk@axespos\@ne			
      \else\@pstrickserr{Bad axes position: `#1#2'}\@ehpa
  \fi\fi}
\psset[pst-plot]{axespos=b}
\newdimen\pst@xunit
\newdimen\pst@yunit
\def\pslegend{\@ifnextchar[\pslegend@i{\pslegend@i[rt]}}
\def\pslegend@i[#1]{\@ifnextchar({\pslegend@ii[#1]}{\pslegend@ii[#1](\pst@number\pslabelsep,\pst@number\pslabelsep)}}
\def\pslegend@ii[#1](#2,#3)#4{%
  \gdef\pslegend@ref{#1}%
  \xdef\pslegend@sepx{#2 }%
  \xdef\pslegend@sepy{#3 }%
  \gdef\pslegend@text{#4}}
\newpsstyle{legendstyle}{fillstyle=solid,fillcolor=white,linewidth=0.5pt}
\def\pslegend@iii[#1](#2){\rput[#1](#2){\psframebox[style=legendstyle]{%
  \footnotesize\tabcolsep=2pt%
  \tabular[t]{@{}ll@{}}\pslegend@text\endtabular}}\global\let\pslegend@text\relax}
\let\pslegend@text\relax
\def\psgraph{\pst@object{psgraph}}
\def\psgraph@i{%
  \let\psgraph@para\pst@par
  \let\psk@save@arrowA\psk@arrowA
  \let\psk@save@arrowB\psk@arrowB
  \pst@getarrows\psgraph@ii}
\def\psgraph@ii(#1,#2){\catcode`\!=12\relax
  \@ifnextchar({\psgraph@iii(#1,#2)}{\psgraph@iv(0,0)(#1,#2)}}
\def\psgraph@iii(#1,#2)(#3,#4){\@ifnextchar({\psgraph@v(#1,#2)(#3,#4)}{\psgraph@iv(#1,#2)(#3,#4)}}
\def\psgraph@iv(#1,#2)(#3,#4)#5#6{
  \pst@killglue%
  \begingroup
  \use@keep@par
  \pstFPsub\pst@tempA{#3}{#1}%
  \pst@dimm=#5
  \pst@dimo=\pst@tempA pt
  \pstFPdiv\pst@@dx{\strip@pt\pst@dimm}{\pst@tempA}%
  \pst@xunit=\pst@@dx\p@
  \ifx!#6\let\pst@yunit=\pst@xunit\else
    \pst@dimm=#6
    \pstFPsub\pst@tempA{#4}{#2}%
    \pstFPdiv\pst@@dy{\strip@pt\pst@dimm}{\pst@tempA}%
    \pst@yunit=\pst@@dy\p@
  \fi
  \pst@dimm=#1\pst@xunit\advance\pst@dimm by \psk@llx
  \pst@dimn=#2\pst@yunit\advance\pst@dimn by \psk@lly
  \pst@dimo=#3\pst@xunit\advance\pst@dimo by \psk@urx
  \pst@dimp=#4\pst@yunit\advance\pst@dimp by \psk@ury
  \if@star\pspicture*(\pst@dimm,\pst@dimn)(\pst@dimo,\pst@dimp)\else
  \pspicture(\pst@dimm,\pst@dimn)(\pst@dimo,\pst@dimp)\fi
  \let\psxunit\pst@xunit \let\psyunit\pst@yunit
  \ifdim\pst@xunit=\pst@yunit\relax\psset{runit=\pst@xunit}\fi%
  \bgroup
    \use@par
  \ifPst@psgrid
     \expandafter\psset\expandafter{\psk@gridpara}%
      \rput[lb](0,0){\expandafter\psgrid\psk@gridcoor}  
  \fi
    \ifnum\psk@axespos=0
      \expandafter\psaxes\expandafter[\psgraph@para](#1,#2)(#3,#4)%
    \else
      \xdef\psgraph@coor{(#1,#2)(#3,#4)(#5,#6)}%
    \fi
  \egroup
  \psgraph@vi(#1,#2)(#1,#2)(#3,#4)%
}
\def\psgraph@v(#1,#2)(#3,#4)(#5,#6)#7#8{
  \pst@killglue%
  \let\psgraph@para\pst@par
  \begingroup%
  \use@keep@par
  \pstFPsub\pst@tempA{#5}{#3}%
  \pst@dimm=#7%
  \pst@dimo=\pst@tempA pt%
  \pstFPdiv\pst@@dx{\strip@pt\pst@dimm}\pst@tempA%
  \pst@xunit=\pst@@dx\p@%
  \ifx!#8\let\pst@yunit=\pst@xunit\else
    \pst@dimm=#8%
    \pstFPsub\pst@tempA{#6}{#4}%
    \pstFPdiv\pst@@dy{\strip@pt\pst@dimm}\pst@tempA%
    \pst@yunit=\pst@@dy\p@%
  \fi%
  \pst@dima=#3\pst@xunit \advance\pst@dima by \psk@llx%
  \pst@dimb=#4\pst@yunit \advance\pst@dimb by \psk@lly%
  \pst@dimc=#5\pst@xunit \advance\pst@dimc by \psk@urx%
  \pst@dimd=#6\pst@yunit \advance\pst@dimd by \psk@ury%
  \if@star\pspicture*(\pst@dima,\pst@dimb)(\pst@dimc,\pst@dimd)\else%
          \pspicture(\pst@dima,\pst@dimb)(\pst@dimc,\pst@dimd)\fi%
  \psset{xunit=\pst@xunit,yunit=\pst@yunit}
  \ifdim\pst@xunit=\pst@yunit \psset{runit=\pst@xunit}\fi%
  \bgroup%
    \use@par%
  \ifPst@psgrid
     \expandafter\psset\expandafter{\psk@gridpara}%
      \rput[lb](0,0){\expandafter\psgrid\psk@gridcoor}
  \fi%
    \ifnum\psk@axespos=0
      \psaxes(#1,#2)(#3,#4)(#5,#6)%
    \else
      \xdef\psgraph@coor{(#1,#2)(#3,#4)(#5,#6)}%
    \fi
  \egroup
  \psgraph@vi(#1,#2)(#3,#4)(#5,#6)%
}
\def\setxLabelC@@r#1,#2(#3,#4)(#5){%
  \pst@getcoor{#5}\pst@tempB%
  \ifx c#1 
    \pssetylength\pst@dimm{#2}%
    \rput(! #4 #3 add 2 div \pst@number\pst@dimm \pst@tempB\space exch pop add 
      \pst@number\psyunit div ){\psk@xAxisLabel}%
  \else%
    \pst@getcoor{\psk@xAxisLabelPos}\pst@tempA%
    \rput(! \pst@tempA\space \pst@tempB\space exch pop add \tx@UserCoor ){\psk@xAxisLabel}%
  \fi}
\def\setyLabelC@@r#1,#2(#3,#4)(#5){%
  \pst@getcoor{#5}\pst@tempB%
  \ifx c#2
    \pssetxlength\pst@dimm{#1}%
    \rput{90}(! \pst@number\pst@dimm \pst@tempB\space pop add \pst@number\psxunit div #4 #3 add 2 div ){\psk@yAxisLabel}%
  \else%
    \pst@getcoor{\psk@yAxisLabelPos}\pst@tempA%
    \rput{90}(! \pst@tempB\space pop \pst@tempA\space 3 1 roll add exch \tx@UserCoor ){\psk@yAxisLabel}%
  \fi}
\def\psgraph@vi(#1,#2)(#3,#4)(#5,#6){%
  \ifx\psk@xAxisLabel\@empty\else%
    \ifx\psk@xAxisLabelPos\@empty\uput[0](#5,#2){\psk@xAxisLabel}%
    \else\expandafter\setxLabelC@@r\psk@xAxisLabelPos(#3,#5)(#1,#2)\fi%
  \fi%
  \ifx\psk@yAxisLabel\@empty\else%
    \ifx\psk@yAxisLabelPos\@empty\uput[90](#1,#6){\psk@yAxisLabel}%
    \else\expandafter\setyLabelC@@r\psk@yAxisLabelPos(#4,#6)(#1,#2)\fi%
  \fi%
  \def\lt@@{lt}\def\lb@@{lb}\def\rb@@{rb}%
  \ifx\pslegend@ref\lb@@    \gdef\pslegend@coor{#3 \pslegend@sepx \pst@number\psxunit div add 
                                                   \pslegend@sepy \pst@number\psyunit div}%
  \else%
    \ifx\pslegend@ref\lt@@  \gdef\pslegend@coor{#3 \pslegend@sepx \pst@number\psxunit div add 
                                                #6 \pslegend@sepy \pst@number\psyunit div sub}%
    \else%
      \ifx\pslegend@ref\rb@@\gdef\pslegend@coor{#5 \pslegend@sepx \pst@number\psxunit div sub 
                                                   \pslegend@sepy \pst@number\psyunit div}%
      \else                 \gdef\pslegend@coor{#5 \pslegend@sepx \pst@number\psxunit div sub 
                                                #6 \pslegend@sepy \pst@number\psyunit div sub}%
      \fi%
    \fi%
  \fi%
  \xdef\psgraphLLx{#3}\xdef\psgraphLLy{#4}\xdef\psgraphURx{#5}\xdef\psgraphURy{#6}%
  \global\let\psk@arrowA\psk@save@arrowA
  \global\let\psk@arrowB\psk@save@arrowB
  \ignorespaces
}
\def\endpsgraph{%
  \ifx\relax\pslegend@text\relax \else\pslegend@iii[\pslegend@ref](!\pslegend@coor)\fi
  \expandafter\psset\expandafter{\psgraph@para}%
  \ifnum\psk@axespos>0
    \expandafter\psaxes\psgraph@coor
  \fi
  \endpspicture
  \endgroup\ignorespaces}
\@namedef{psgraph*}{\psgraph*}
\@namedef{endpsgraph*}{\endpsgraph}
\def\psPutXLabel#1{%
  \global\pst@cnto=0\relax
  \global\pst@cntp=#1\relax
  \expandafter\get@Label\psk@xLabels,,\@nil%
}
\def\psPutYLabel#1{%
  \global\pst@cnto=0\relax
  \global\pst@cntp=#1\relax
  \expandafter\get@Label\psk@yLabels,,\@nil%
}
\def\get@Label#1,#2,#3\@nil{%
    \ifnum\the\pst@cnto<\the\pst@cntp
      \global\advance\pst@cnto by \@ne 
      \ifx\relax#3\relax\else\expandafter\get@Label#2,#3\@nil\fi%
    \else #1\fi%
}
\def\psVectorfield{\pst@object{psVectorfield}}
\def\psVectorfield@i(#1,#2)(#3,#4)#5{{%
  \addbefore@par{Dx=0.1,Dy=0.1,Ox=3,arrows=->,linewidth=0.2pt}%
  \begin@SpecialObj
  \SpecialCoor
  \pstFPsub\pst@tempA{#3}{#1}%
  \pstFPsub\pst@tempB{#4}{#2}%
  \pstFPDiv{\pst@tempC}{\pst@tempA}{\psk@Dx}%
  \pstFPDiv{\pst@tempD}{\pst@tempB}{\psk@Dy}%
  \pstVerb{ /yStrich \ifPst@algebraic (#5) tx@AlgToPs begin AlgToPs end cvx
                \else { #5 } \fi def }%
  \multido{\rX=#1+\psk@Dx}{\numexpr\pst@tempC+1}{%
    \multido{\rY=#2+\psk@Dy}{\numexpr\pst@tempD+1}{%
       \psline%
         (! /x \rX\space def 
            /y \rY\space def 
            /yTemp yStrich \psk@Dx\space \psk@Ox\space div mul def 
            \rX\space \psk@Dx\space \psk@Ox\space div sub \rY\space yTemp sub)%
         (! /x \rX\space def 
            /y \rY\space def 
            /yTemp yStrich \psk@Dx\space \psk@Ox\space div mul def 
            \rX\space \psk@Dx\space \psk@Ox\space div add \rY\space yTemp add)%
   }}%
  \end@SpecialObj
}\ignorespaces}  
\def\psFixpoint{\pst@object{psFixpoint}}
\def\psFixpoint@i#1#2#3{
  \pst@killglue%
  \begingroup%
  \use@par%
  \@nameuse{beginplot@\psplotstyle}%
  \addto@pscode{
    \psplot@init
      /x #1 def
      /F@pstplot \ifPst@algebraic (#2) tx@AlgToPs begin AlgToPs end cvx
                 \else { #2 } \fi  def
      /xy { x \pst@number\psxunit mul F@pstplot dup /x ED \pst@number\psyunit mul } def 
  }%
  \gdef\psplot@init{}%
  \@pstfalse%
  \@nameuse{testqp@\psplotstyle}%
  \addto@pscode{
      mark
      x \pst@number\psxunit mul 0
      /n 2 def
      #3 {
        xy 
        dup dup 
        /n n 4 add def
      } repeat 
  }%
  \@nameuse{endplot@\psplotstyle}%
  \endgroup%
  \ignorespaces}
\define@boolkey[psset]{pst-plot}[Pst@]{showDerivation}[true]{}
\psset{showDerivation}
\def\psNewton{\pst@object{psNewton}}
\def\psNewton@i#1#2{\@ifnextchar[{\psNewton@ii{#1}{#2}}{\psNewton@iii{#1}{#2}}}
\def\psNewton@ii#1#2[#3]#4{
  \pst@killglue%
  \begingroup%
  \addbefore@par{showDerivation}%
  \use@par%
  \@nameuse{beginplot@\psplotstyle}%
  \addto@pscode{
    \psplot@init
      /x #1 def
      /F@pstplot \ifPst@algebraic (#2) tx@AlgToPs begin AlgToPs end cvx \else { #2 } \fi  def
      /F@pstplotDerive \ifPst@algebraic (#3) tx@AlgToPs begin AlgToPs end cvx \else { #3 } \fi  def
      /newxVal { 
        F@pstplotDerive 
        div neg 
      } def
  }%
  \gdef\psplot@init{}%
  \@pstfalse%
  \@nameuse{testqp@\psplotstyle}%
  \addto@pscode{
      mark
      x 0 \tx@ScreenCoor 
      /n 2 def
      #4 {
        F@pstplot /yVal ED
        x yVal \tx@ScreenCoor
        /n n 2 add def
        yVal newxVal x add /x ED
        x 0 \tx@ScreenCoor 
        \ifPst@showDerivation /n n 4 add def \else moveto /n n 2 add def\fi
      } repeat 
      pstack
  }%
  \@nameuse{endplot@\psplotstyle}%
  \endgroup%
  \ignorespaces}
\def\psNewton@iii#1#2#3{
  \pst@killglue%
  \begingroup%
  \addbefore@par{VarStepEpsilon=0.01,showDerivation}%
  \use@par%
  \@nameuse{beginplot@\psplotstyle}%
  \addto@pscode{
    \psplot@init
      /epsilon \psk@VarStepEpsilon\space def
      /x #1 def
      /F@pstplot \ifPst@algebraic (#2) tx@AlgToPs begin AlgToPs end cvx \else { #2 } \fi  def
      /newxVal { 
        /saveX x def
        saveX epsilon add /x ED F@pstplot saveX epsilon sub /x ED F@pstplot sub epsilon dup add div 
        div neg 
        /x saveX def
      } def
  }%
  \gdef\psplot@init{}%
  \@pstfalse%
  \@nameuse{testqp@\psplotstyle}%
  \addto@pscode{
      mark
      x 0 \tx@ScreenCoor 
      /n 2 def
      #3 {
        F@pstplot /yVal ED
        x yVal \tx@ScreenCoor
        yVal newxVal x add /x ED
        x 0 \tx@ScreenCoor 
        \ifPst@showDerivation /n n 4 add def \else moveto /n n 2 add def\fi
      } repeat 
  }%
  \@nameuse{endplot@\psplotstyle}%
  \endgroup%
  \ignorespaces}
\def\psResetPlotValues{%
  \psset{method={}}%
}%
\catcode`\@=\TheAtCode\relax
 
\csname PSTnodesLoaded\endcsname
\let\PSTnodesLoaded 
\ifx\PSTricksLoaded \else\input pstricks.tex \fi\relax
\ifx\PSTXKeyLoaded \else \input pst-xkey \fi
\def\fileversion{1.42}
\def\filedate{2019/03/03}
\edef\TheAtCode{\the\catcode`\@}
\catcode`\@=11
\pstheader{pst-node.pro}
\pst@addfams{pst-node}
\SpecialCoor
\define@boolkey[psset]{pst-node}[Pst@]{trueAngle}[true]{}
\psset[pst-node]{trueAngle=false}
\define@boolkey[psset]{pst-node}[Pst@]{storeNodeInfo}[true]{}
\psset[pst-node]{storeNodeInfo=false}
\def\pst@nodedict{tx@NodeDict begin }
\def\pst@zapspace#1 #2{%
#1%
\ifx#2\@empty\else\expandafter\pst@zapspace\fi
#2}
\def\pst@getnode#1#2{\pst@expandafter\pst@@getnode{#1},,\@nil#2}
\def\pst@@getnode#1,#2,#3\@nil#4{%
  \ifx\@empty#3\@empty
    \edef#4{/N@\pst@zapspace#1 \@empty\space}%
  \else
    \pst@cntg=#1\relax
    \pst@cnth=#2\relax
    \edef#4{/N@M-\ifnum\psmatrixcnt=\z@ 1\else\the\psmatrixcnt\fi
    -\the\pst@cntg-\the\pst@cnth\space}%
  \fi}
\def\tx@NewNode{/NodeScale {\ifx\pstnodescale\@undefined  \else\pstnodescale \fi} def NewNode }
\def\psopenNodeFile{%
  \pst@Verb{ 
    (\jobname.nodes) (w) file /NodeFile exch def 
  }}
\def\pscloseNodeFile{\pstVerb{ tx@NodeDict begin NodeFile closefile end }}
\define@boolkey[psset]{pst-node}[Pst@]{showNode}[true]{\ifPst@showNode\psopenNodeFile\fi}
\define@boolkey[psset]{pst-node}[Pst@]{markNode}[true]{}
\define@boolkey[psset]{pst-node}[Pst@]{saveNodeCoors}[true]{}
\define@key[psset]{pst-node}{NodeCoorPrefix}[]{\def\psk@NodeCoorPrefix{#1}}
\psset[pst-node]{saveNodeCoors=false,showNode=false,markNode=false,NodeCoorPrefix=}
\def\pst@newnode#1#2#3#4{%
\pst@killglue
\leavevmode
\pst@getnode{#1}\pst@thenode
\pst@Verb{
  \ifPst@saveNodeCoors
    \ifx\relax#3\relax 0 0 \else gsave \pst@dict STV CP T end #3 \tx@UserCoor grestore \fi 
    \if$\psk@NodeCoorPrefix$
      /N-#1.y exch def
      /N-#1.x exch def
    \else
      /\psk@NodeCoorPrefix#1y exch def
      /\psk@NodeCoorPrefix#1x exch def
    \fi
  \fi
  \pst@nodedict
  {#3}
  \ifx\psk@name\relax false \else \psk@name true \fi
  \pst@thenode
  #2
  {#4}
  \ifPst@showNode 
  exch dup /NodeType ED 
  exch
   NodeType 10 eq {  
    5 copy 
    cvlit aload pop
    20 string cvs (; )   6 2 roll 
    20 string cvs (; )   7 2 roll 
    20 string cvs (; )   8 2 roll 
    20 string cvs (; )   9 2 roll 
    cvlit dup length 2 eq 
      { aload pop exch 
        20 string cvs (; ) 11 2 roll 
        20 string cvs (, ) 12 2 roll  
        (\string\n)                   
        13 array astore concatstringarray 
      }
      { 255 string cvs (; ) 10 2 roll 
        (\string\n)                   
        11 array astore concatstringarray 
      } ifelse 
    NodeFile exch writestring 
  } if
  NodeType 14 eq {  
    5 copy 
    /@@temp ED 
    @@temp  
    4 -1 roll cvlit pop
    ( OvalNodePos ) (; )  5 2 roll
    20 string cvs (; )   6 2 roll 
    20 string cvs (; )   7 2 roll 
    20 string cvs (; )   8 2 roll 
    Y 20 string cvs (; ) 10 2 roll
    X 20 string cvs (, ) 12 2 roll
    (\string\n)                   
    13 array astore concatstringarray 
    tx@NodeDict begin NodeFile exch writestring end
  } if
  \fi
  \tx@NewNode
  end 
}%
\global\let\psk@name\relax%
\pstree@nodehook%
\global\let\pstree@nodehook\relax}
\let\pstree@nodehook\relax
\define@boolkey[psset]{pst-node}[Pst@]{nodealign}[true]{}
\psset[pst-node]{nodealign=false}

\def\pst@nodealign{%
\pst@dimg=\ht\pst@hbox
\advance\pst@dimg by -\dp\pst@hbox
\divide\pst@dimg by \tw@
\lower\pst@dimg}
\def\tx@InitPnode{InitPnode }
\def\pnode{\@ifnextchar[{\pnode@i}{\pnode@iii}}
\def\pnode@i[#1]{\@ifnextchar({\pnode@ii[#1]}{\pnode@ii[#1](0,0)}}
\def\pnode@ii[#1](#2)#3{%
  \pst@getcoor{#1}\pst@tempA%
  \pst@getcoor{#2}\pst@tempB%
  \pst@newnode{#3}{10}{\pst@tempA \pst@tempB 3 -1 roll add 3 1 roll add exch }{\tx@InitPnode}%
  \ifPst@showNode\psdot(#3)\uput[\ifx\psk@rot\@empty0\else\psk@rot\fi]{0}(#3){#3}\fi
  \ignorespaces}
\def\pnode@iii{\@ifnextchar({\pnode@}{\pnode@(0,0)}}
\def\pnode@(#1)#2{%
  \pst@@getcoor{#1}%
  \pst@newnode{#2}{10}{\pst@coor}{\tx@InitPnode}%
  \ifPst@showNode\psdot(#2)\uput[\ifx\psk@rot\@empty0\else\psk@rot\fi]{0}(#2){#2}\fi
  \ignorespaces}
\def\pnodes{\@ifnextchar[{\pnodes@i}{\pnodes@i[0,0]}}
\def\pnodes@i[#1]{\@ifnextchar({\psnodes@ii[#1]}{\pnodes@ii}}
\def\psnodes@ii[#1](#2)#3{%
  \pnode[#1](#2){#3}%
  \@ifnextchar({\psnodes@ii[#1]}{}%
}
\def\tx@InitCnode{InitCnode }
\def\cnode{\pst@object{cnode}}
\def\cnode@i{\@ifnextchar({\cnode@ii}{\cnode@ii(0,0)}}
\def\cnode@ii(#1)#2#3{%
  \leavevmode
  \hbox{%
    \use@par
    \pst@@getcoor{#1}%
    \pssetlength\pst@dimc{#2}%
    \pst@dimg=\psk@dimen\pslinewidth
    \advance\pst@dimc-\pst@dimg
    \advance\pst@dimc.5\pslinewidth
    \ifPst@nodealign
      \kern\pst@dimc
      \vrule width\z@ height \pst@dimc depth \pst@dimc
    \fi
    \pscircle@do(#1){#2}%
    \pst@newnode{#3}{11}{\pst@coor \pst@number\pst@dimc}{\tx@InitCnode}%
    \ifPst@nodealign\kern\pst@dimc\fi%
  }%
  \ignorespaces}
\def\Cnode{\pst@object{Cnode}}
\def\Cnode@i{\@ifnextchar({\Cnode@ii}{\Cnode@ii(0,0)}}
\def\Cnode@ii(#1)#2{\cnode@ii(#1){\psk@radius}{#2}}%
\def\cnodeput{\pst@object{cnodeput}}
\def\cnodeput@i{\@ifnextchar({\cnodeput@iii}{\cnodeput@ii}}
\def\cnodeput@ii#1{%
  \addto@par{rot={#1}}%
  \@ifnextchar({\cnodeput@iii}{\cnodeput@iii(\z@,\z@)}%
}
\def\cnodeput@iii(#1)#2{%
  \pst@killglue
  \@fixedradiusfalse
  \def\pst@nodehook{\cnodeput@iv{#2}}%
  \pst@makebox{\cput@v{#1}}%
}
\def\cnodeput@iv#1{%
  \pst@newnode{#1}{11}{\pscirclebox@iv \pst@number\pslinewidth add}{\tx@InitCnode}%
  \global\let\pst@nodehook\relax
  \ignorespaces
}
\def\Cnodeput{\pst@object{Cnodeput}}
\def\Cnodeput@i{\@ifnextchar({\Cnodeput@iii}{\Cnodeput@ii}}
\def\Cnodeput@ii#1{%
  \addto@par{rot={#1}}%
  \@ifnextchar({\Cnodeput@iii}{\Cnodeput@iii(\z@,\z@)}}
\def\Cnodeput@iii(#1)#2{%
  \pst@killglue
  \@fixedradiustrue
  \def\pst@nodehook{\Cnodeput@iv{#2}}%
  \pst@makebox{\cput@v{#1}}%
}
\def\Cnodeput@iv#1{%
  \pst@newnode{#1}{11}{%
    \pst@number{\wd\pst@hbox} 2 div \pst@number\pst@dima 
    \pst@number\pst@dimb \pst@number\pslinewidth \psk@dimen .5 sub mul sub }
       {\tx@InitCnode}%
  \global\let\pst@nodehook\relax}
\def\circlenode{\pst@object{circlenode}}
\def\circlenode@i#1{\pst@makebox{\circlenode@ii{#1}}}
\def\circlenode@ii#1{%
  \begingroup
  \pst@useboxpar
  \setbox\pst@hbox=\hbox{%
    \cnodeput@iv{#1}%
    \pscirclebox@iii
    \box\pst@hbox}%
  \ifPst@nodealign \psboxseptrue \fi
  \ifpsboxsep \pscirclebox@sep \fi
  \leavevmode
  \ifPst@nodealign\pst@nodealign\fi
  \box\pst@hbox
  \endgroup}
\def\Circlenode{\pst@object{Circlenode}}
\def\Circlenode@i#1{\pst@makebox{\Circlenode@ii{#1}}}
\def\Circlenode@ii#1{%
\begingroup
  \pst@useboxpar
  \pst@dima=\ht\pst@hbox
  \advance\pst@dima by -\dp\pst@hbox
  \divide\pst@dima by \tw@
  \pssetlength\pst@dimb\psk@radius
  \setbox\pst@hbox=\hbox{%
  \Cnodeput@iv{#1}%
  \pscircle(.5\wd\pst@hbox,\pst@dima){\pst@dimb}%
  \box\pst@hbox}%
  \ifPst@nodealign \psboxseptrue \fi
  \ifpsboxsep \psCirclebox@sep \fi
  \leavevmode
  \ifPst@nodealign\pst@nodealign\fi
  \box\pst@hbox
  \endgroup}
\def\tx@GetRnodePos{GetRnodePos }
\def\tx@InitRnode{InitRnode }
\def\psnode{\pst@object{psnode}}
\def\psnode@i{\@ifnextchar(\psnode@ii{\psnode@ii(0,0)}}
\def\psnode@ii(#1)#2#3{
  \rput(#1){\rnode{#2}{#3}}}
\def\rnode{\@ifnextchar[{\rnode@i}{\def\pst@par{}\rnode@ii}}
\def\rnode@i[#1]{\def\pst@par{ref=#1}\rnode@ii}
\def\rnode@ii#1{\pst@makebox{\rnode@iii\rnode@iv{#1}}}
\def\rnode@iii#1#2{%
\leavevmode
\begingroup
\pst@useboxpar
#1%
\ifPst@nodealign\lower\pst@dimb\fi
\hbox{%
\pst@newnode{#2}{16}{%
\pst@number{\ht\pst@hbox}%
\pst@number{\dp\pst@hbox}%
\pst@number{\wd\pst@hbox}%
\pst@number\pst@dima%
\pst@number\pst@dimb}%
{\tx@InitRnode}%
\box\pst@hbox}%
\endgroup}
\def\rnode@iv{%
\pst@dima=\psk@xref\wd\pst@hbox
\ifx\psk@yref\relax
\pst@dimb=\z@
\else
\pst@dimb=\ht\pst@hbox
\advance\pst@dimb\dp\pst@hbox
\pst@dimb=\psk@yref\pst@dimb
\advance\pst@dimb-\dp\pst@hbox
\fi}
\define@key[psset]{pst-node}{href}[0]{\pst@checknum{#1}\psk@href}
\psset[pst-node]{href=0}
\define@key[psset]{pst-node}{vref}[0.7ex]{\def\psk@vref{#1}}
\psset[pst-node]{vref=0.7ex}
\def\Rnode{\pst@object{Rnode}}
\def\Rnode@i#1{\pst@makebox{\rnode@iii\Rnode@ii{#1}}}
\def\Rnode@ii{%
\use@par
\pst@dima=\psk@href\wd\pst@hbox
\advance\pst@dima\wd\pst@hbox
\divide\pst@dima 2
\pssetlength\pst@dimb{\psk@vref}}
\def\tx@DiaNodePos{DiaNodePos }
\def\dianode{\pst@object{dianode}}
\def\dianode@i#1{\pst@makebox{\dianode@ii{#1}}}
\def\dianode@ii#1{%
\begingroup
\pst@useboxpar
\psdiabox@iii
\setbox\pst@hbox=\hbox{%
\pst@newnode{#1}{14}{}{%
/X \pst@number\pst@dima def
/Y \pst@number\pst@dimb def
/w \pst@number\pst@dimc 2 mul def
/h \pst@number\pst@dimd 2 mul def
/NodePos { \tx@DiaNodePos } def}%
\box\pst@hbox}%
\ifPst@nodealign\psboxseptrue\fi
\ifpsboxsep\psdiabox@sep\fi
\leavevmode
\ifPst@nodealign\lower\pst@dimb\fi
\box\pst@hbox
\endgroup}
\def\tx@TriNodePos{TriNodePos }
\def\tx@InitTriNode{InitTriNode }
\def\trinode{\pst@object{trinode}}
\def\trinode@i#1{\pst@makebox{\trinode@ii{#1}}}
\def\trinode@ii#1{%
  \begingroup%
  \pst@useboxpar%
  \pstribox@iii
  \setbox\pst@hbox=\hbox{%
    \pst@newnode{#1}{14}{}{
      \pst@number\pst@dimc
      \pst@number\pst@dimd
      \ifodd\psk@trimode
        exch
        \pst@number\pst@dima
      \else
        \pst@number\pst@dimb
      \fi
      \psk@trimode
      \pst@number{\wd\pst@hbox}
      \pst@number{\ht\pst@hbox}
      \pst@number{\dp\pst@hbox}
      \tx@InitTriNode
    }%
    \box\pst@hbox%
  }%
  \ifPst@nodealign\psboxseptrue\fi
  \ifpsboxsep\pstribox@sep\fi
  \leavevmode
  \ifPst@nodealign\lower\pst@tempa\fi
  \box\pst@hbox%
  \endgroup}
\def\tx@OvalNodePos{OvalNodePos }
\def\ovalnode{\pst@object{ovalnode}}
\def\ovalnode@i#1{\pst@makebox{\ovalnode@ii{#1}}}
\def\ovalnode@ii#1{%
\begingroup
\pst@useboxpar
\psovalbox@iii
\setbox\pst@hbox=\hbox{%
\pst@newnode{#1}{14}{}{%
/X \pst@number\pst@dima def
/Y \pst@number\pst@dimb def
/w \pst@number\pst@dimc def
/h \pst@number\pst@dimd def
/NodePos { \tx@OvalNodePos } def}%
\unhbox\pst@hbox}%
\ifPst@nodealign\psboxseptrue\fi
\ifpsboxsep\psovalbox@sep\fi
\leavevmode
\ifPst@nodealign\lower\pst@dimb\fi
\box\pst@hbox
\endgroup}
\def\dotnode{\pst@object{dotnode}}
\def\dotnode@i{\@ifnextchar({\dotnode@ii}{\dotnode@ii(\z@,\z@)}}
\def\dotnode@ii(#1)#2{%
  \leavevmode
  \hbox{%
    \use@par
    \pst@@getcoor{#1}%
    \pst@getdotsize
    \pstree@nodehook
    \ifPst@nodealign
      \pst@dima=\pst@dimg
      \kern\pst@dima
      \vrule width\z@ height \pst@dimh depth \pst@dimh
    \fi
    \pst@newnode{#2}{14}{}{
      \pst@coor
      /Y exch def /X exch def
      /w \pst@number\pst@dimg def
      /h \pst@number\pst@dimh def
      /NodePos { \tx@OvalNodePos } def}%
    \psdot@ii(#1)%
    \ifPst@nodealign\kern\pst@dima\fi}%
  \ifPst@markNode\uput[\ifx\psk@rot\@empty0\else\psk@rot\fi]{0}(#2){#2}\fi
  \ignorespaces}
\def\dotnodes{\pst@object{dotnodes}}
\def\dotnodes@i{\use@par\dotnodes@ii}
\def\dotnodes@ii(#1)#2{%
  \dotnode(#1){#2}%
  \@ifnextchar(\dotnodes@ii{\def\pst@par{}}}
\define@key[psset]{pst-node}{framesize}{\pst@expandafter\psset@@framesize{#1} \@nil}
\def\psset@@framesize#1 #2\@nil{%
  \pssetlength\pst@dimg{#1}%
  \divide\pst@dimg2
  \edef\psk@framewidth{\pst@number\pst@dimg}%
  \ifx\@empty#2\@empty
    \let\psk@frameheight\psk@framewidth
  \else
    \pssetlength\pst@dimg{#2}%
    \divide\pst@dimg2
    \edef\psk@frameheight{\pst@number\pst@dimg}%
  \fi}
\psset[pst-node]{framesize=10pt}
\def\fnode{\pst@object{fnode}}
\def\fnode@i{\@ifnextchar({\fnode@ii}{\fnode@ii(\z@,\z@)}}
\def\fnode@ii(#1)#2{%
  \leavevmode
  \pst@killglue
  \hbox{%
    \use@par%
    \begin@ClosedObj%
    \ifPst@nodealign
      \kern\psk@framewidth\p@
      \vrule width\z@ height \psk@frameheight\p@ depth \psk@frameheight\p@
      \edef\pst@coor{0 0 }%
    \else\pst@@getcoor{#1}\fi
    \pst@newnode{#2}{14}{}{
      \pst@coor
      /Y exch def /X exch def
      /d \psk@dimen .5 sub CLW mul neg def
      /r \psk@framewidth d add def
      /l r neg def
      /u \psk@frameheight d add def
      /d u neg def
      /NodePos { \tx@GetRnodePos } def}%
    \addto@pscode{
      /x2 \psk@framewidth CLW \psk@dimen mul sub def
      /y2 \psk@frameheight CLW \psk@dimen mul sub def
      \pst@coor 2 copy
      y2 sub /y1 ED
      x2 sub /x1 exch def
      y2 add /y2 exch def
      x2 add /x2 exch def
      \psk@cornersize
      1 index 0 eq { pop pop \tx@Rect } { \tx@OvalFrame } ifelse}%
    \def\pst@linetype{2}%
    \showpointsfalse%
    \end@ClosedObj%
    \ifPst@nodealign\kern\psk@framewidth\p@\fi}
  \ignorespaces}
\define@key[psset]{}{XnodesepA}{\pst@getlength{#1}\psk@nodesepA\def\psk@nodeseptypeA{2 }}
\define@key[psset]{}{XnodesepB}{\pst@getlength{#1}\psk@nodesepB\def\psk@nodeseptypeB{2 }}
\define@key[psset]{}{Xnodesep}{%
    \pst@getlength{#1}\psk@nodesepA
    \let\psk@nodesepB\psk@nodesepA
    \def\psk@nodeseptypeA{2 }%
    \def\psk@nodeseptypeB{2 }}
\define@key[psset]{}{YnodesepA}{\pst@getlength{#1}\psk@nodesepA\def\psk@nodeseptypeA{1 }}
\define@key[psset]{}{YnodesepB}{\pst@getlength{#1}\psk@nodesepB\def\psk@nodeseptypeB{1 }}
\define@key[psset]{}{Ynodesep}{%
    \pst@getlength{#1}\psk@nodesepA
    \let\psk@nodesepB\psk@nodesepA
    \def\psk@nodeseptypeA{1 }%
    \def\psk@nodeseptypeB{1 }}
\define@key[psset]{pst-node}{nodesepA}[0pt]{\pst@getlength{#1}\psk@nodesepA \def\psk@nodeseptypeA{0 }}
\define@key[psset]{pst-node}{nodesepB}[0pt]{\pst@getlength{#1}\psk@nodesepB \def\psk@nodeseptypeB{0 }}
\define@key[psset]{pst-node}{nodesep}[0pt]{%
  \pst@getlength{#1}\psk@nodesepA
  \let\psk@nodesepB\psk@nodesepA
  \def\psk@nodeseptypeA{0 }%
  \def\psk@nodeseptypeB{0 }}
\psset[pst-node]{nodesep=0pt}
\define@key[psset]{pst-node}{armA}[10pt]{\pst@getlength{#1}\psk@armA \def\psk@armtypeA{0 }}
\define@key[psset]{pst-node}{armB}[10pt]{\pst@getlength{#1}\psk@armB \def\psk@armtypeB{0 }}
\define@key[psset]{pst-node}{arm}[10pt]{%
  \pst@getlength{#1}\psk@armA
  \let\psk@armB\psk@armA
  \def\psk@armtypeA{0 }%
  \def\psk@armtypeB{0 }}
\psset[pst-node]{arm=10pt}
\define@key[psset]{pst-node}{XarmA}[]{\pst@getlength{#1}\psk@armA \def\psk@armtypeA{1 }}
\define@key[psset]{pst-node}{XarmB}[]{\pst@getlength{#1}\psk@armB \def\psk@armtypeB{1 }}
\define@key[psset]{pst-node}{Xarm}{%
  \pst@getlength{#1}\psk@armA
  \let\psk@armB\psk@armA
  \def\psk@armtypeA{1 }%
  \def\psk@armtypeB{1 }}
\define@key[psset]{pst-node}{YarmA}[]{\pst@getlength{#1}\psk@armA \def\psk@armtypeA{2 }}
\define@key[psset]{pst-node}{YarmB}[]{\pst@getlength{#1}\psk@armB \def\psk@armtypeB{2 }}
\define@key[psset]{pst-node}{Yarm}[]{%
  \pst@getlength{#1}\psk@armA
  \let\psk@armB\psk@armA
  \def\psk@armtypeA{2 }%
  \def\psk@armtypeB{2 }}
\define@key[psset]{pst-node}{offsetA}[0pt]{\pst@getlength{#1}\psk@offsetA}
\define@key[psset]{pst-node}{offsetB}[0pt]{\pst@getlength{#1}\psk@offsetB}
\define@key[psset]{pst-node}{offset}[0pt]{\pst@getlength{#1}\psk@offsetA\let\psk@offsetB\psk@offsetA}
\psset[pst-node]{offset=0pt}
\define@key[psset]{pst-node}{angleA}[0]{\pst@getangle{#1}\psk@angleA}
\define@key[psset]{pst-node}{angleB}[0]{\pst@getangle{#1}\psk@angleB}%
\define@key[psset]{pst-node}{angle}[0]{%
  \pst@getangle{#1}\psk@angleA
  \let\psk@angleB\psk@angleA}
\psset[pst-node]{angle=0}
\define@key[psset]{pst-node}{arcangleA}[8]{\pst@getangle{#1}\psk@arcangleA}
\define@key[psset]{pst-node}{arcangleB}[8]{\pst@getangle{#1}\psk@arcangleB}%
\define@key[psset]{pst-node}{arcangle}[8]{%
  \pst@getangle{#1}\psk@arcangleA
  \let\psk@arcangleB\psk@arcangleA}
\psset[pst-node]{arcangle=8}
\define@key[psset]{pst-node}{ncurvA}[0.67]{\pst@checknum{#1}\psk@ncurvA}
\define@key[psset]{pst-node}{ncurvB}[0.67]{\pst@checknum{#1}\psk@ncurvB}%
\define@key[psset]{pst-node}{ncurv}[0.67]{\pst@checknum{#1}\psk@ncurvA\let\psk@ncurvB\psk@ncurvA}
\psset[pst-node]{ncurv=0.67}
\def\tx@GetCenter{GetCenter }
\def\tx@XYPos{XYPos }
\def\tx@GetEdge{GetEdge }
\def\tx@AddOffset{AddOffset }
\def\tx@GetEdgeA{GetEdgeA }
\def\tx@GetEdgeB{GetEdgeB }
\def\tx@GetArmA{GetArmA }
\def\tx@GetArmB{GetArmB }
\def\check@arrow#1#2{%
  \check@@arrow#2-\@nil
  \if@pst\addto@par{arrows=#2}\def\next{#1}%
  \else\def\next{#1{#2}}\fi
  \next}
\def\check@@arrow#1-#2\@nil{%
\ifx\@nil#2\@nil\@pstfalse\else\@psttrue\fi}
\def\tx@InitNC{InitNC }
\def\nc@object#1#2#3#4#5{%
  \csname begin@#1Obj\endcsname
  \showpointsfalse
  \pst@getnode{#2}\pst@tempa
  \pst@getnode{#3}\pst@tempb
  \gdef\npos@default{#4 }%
  \addto@pscode{%
    /NCLW CLW def
    \pst@nodedict
    \psk@offsetA
    \psk@offsetB neg
    \psk@nodesepA
    \psk@nodesepB
    \psk@nodeseptypeA
    \psk@nodeseptypeB
    \pst@tempa
    \pst@tempb
    \tx@InitNC { #5 } if
    end }%
  \def\use@pscode{%
    \pst@Verb{gsave \tx@STV newpath \pst@code\space grestore}%
    \gdef\pst@code{}}%
  \csname end@#1Obj\endcsname
  \pst@shortput}
\def\npos@default{.5 }
\def\pc@object#1{%
  \@ifnextchar({\pc@@object#1}{\pst@getarrows{\pc@@object#1}}}
\def\pc@@object#1(#2)(#3){%
  \pnode(#2){@@A}\pnode(#3){@@B}%
  #1{@@A}{@@B}}
\def\tx@LPutLine{LPutLine }
\def\tx@LPutLines{LPutLines }
\def\tx@BezierMidpoint{BezierMidpoint }
\def\tx@HPosBegin{HPosBegin }
\def\tx@HPosEnd{HPosEnd }
\def\tx@HPutLine{HPutLine }
\def\tx@HPutLines{HPutLines }
\def\tx@VPosBegin{VPosBegin }
\def\tx@VPosEnd{VPosEnd }
\def\tx@VPutLine{VPutLine }
\def\tx@VPutLines{VPutLines }
\def\tx@HPutCurve{HPutCurve }
\def\tx@NCCoor{NCCoor }
\def\tx@NCLine{NCLine }
\def\ncline{\pst@object{ncline}}
\def\ncline@i{\check@arrow{\ncline@ii}}
\def\ncline@ii#1#2{\nc@object{Open}{#1}{#2}{.5}{\tx@NCLine}}
\def\pcline{\pst@object{pcline}}
\def\pcline@i{\pc@object\ncline@ii}
\def\ncLine{\pst@object{ncLine}}
\def\ncLine@i{\check@arrow{\ncLine@ii}}
\def\ncLine@ii#1#2{\nc@object{Open}{#1}{#2}{.5}%
{\tx@NCLine /LPutPos { xB yB xA yA \tx@LPutLine } def}}
\def\tx@NCLines{NCLines }
\def\nclines{\pst@object{nclines}}
\def\nclines@i{\check@arrow\nclines@ii}
\def\nclines@ii#1#2{%
\begingroup
\use@par
\def\pst@aftercoors{\nclines@iii{#1}{#2}}%
\def\pst@coors{}%
\pst@@getcoors}
\def\nclines@iii#1#2{%
\nc@object{Open}{#1}{#2}{.5}{%
tx@Dict begin \psline@iii pop end
mark \pst@coors \tx@NCLines}%
\endgroup
\ignorespaces}
\def\tx@NCCurve{NCCurve }
\def\nccurve{\pst@object{nccurve}}
\def\nccurve@i{\check@arrow{\nccurve@ii}}
\def\nccurve@ii#1#2{\nc@object{Open}{#1}{#2}{.5}{%
  /AngleA \psk@angleA\space def /AngleB \psk@angleB\space def
  \psk@ncurvB\space \psk@ncurvA\space
  \tx@NCCurve}}
\def\pccurve{\pst@object{pccurve}}
\def\pccurve@i{\pc@object\nccurve@ii}
\def\ncarc{\pst@object{ncarc}}
\def\ncarc@i{\check@arrow{\ncarc@ii}}
\def\ncarc@ii#1#2{\nc@object{Open}{#1}{#2}{.5}{%
  yB yA sub xB xA sub \tx@Atan dup
  \psk@arcangleA\space add /AngleA exch def
  \psk@arcangleB\space sub 180 add /AngleB exch def
  \psk@ncurvB\space \psk@ncurvA\space
  \tx@NCCurve}}
\def\pcarc{\pst@object{pcarc}}
\def\pcarc@i{\pc@object\ncarc@ii}
\def\tx@NCAngles{NCAngles }
\define@boolkey[psset]{pst-node}[Pst@]{pcRef}[true]{}
\psset[pst-node]{pcRef=false}
\def\ncangles{\pst@object{ncangles}}
\def\ncangles@i{\check@arrow{\ncangles@ii}}
\def\ncangles@ii#1#2{%
  \nc@object{Open}{#1}{#2}{1.5}{\ncangles@iii \tx@NCAngles}}
\def\ncangles@iii{
  tx@Dict begin \psline@iii pop end
  /AngleA \psk@angleA def
  /AngleB \psk@angleB def
  /ArmA \psk@armA \ifPst@pcRef 
    GetEdgeA yA yA1 sub dup mul xA xA1 sub dup mul add sqrt sub \fi def
  /ArmB \psk@armB def
  /ArmTypeA \psk@armtypeA def
  /ArmTypeB \psk@armtypeB def }
\def\pcangles{\pst@object{pcangles}}
\def\pcangles@i{\pc@object\ncangles@ii}
\def\tx@NCAngle{NCAngle }
\def\ncangle{\pst@object{ncangle}}
\def\ncangle@i{\check@arrow{\ncangle@ii}}
\def\ncangle@ii#1#2{%
\nc@object{Open}{#1}{#2}{1.5}{\ncangles@iii \tx@NCAngle}}
\def\pcangle{\pst@object{pcangle}}
\def\pcangle@i{\pc@object\ncangle@ii}
\def\tx@NCBar{NCBar }
\def\ncbar{\pst@object{ncbar}}
\def\ncbar@i{\check@arrow{\ncbar@ii}}
\def\ncbar@ii#1#2{\nc@object{Open}{#1}{#2}{1.5}{%
\ncangles@iii /AngleB \psk@angleA def \tx@NCBar}}
\def\pcbar{\pst@object{pcbar}}
\def\pcbar@i{\pc@object\ncbar@ii}
\define@key[psset]{pst-node}{lineAngle}[0]{%
  \ifdim#1pt=\z@\else\psset{armB=0.5}\fi
  \def\psk@lineAngle{#1}}%
\psset[pst-node]{lineAngle=0}%
\def\tx@NCDiag{NCDiag }
\def\ncdiag{\pst@object{ncdiag}}
\def\ncdiag@i{\check@arrow{\ncdiag@ii}}
\def\ncdiag@ii#1#2{%
  \nc@object{Open}{#1}{#2}{1.5}{\ncangles@iii \psk@lineAngle\space \tx@NCDiag}}
\def\pcdiag{\pst@object{pcdiag}}
\def\pcdiag@i{\pc@object\ncdiag@ii}
\def\tx@NCDiagg{NCDiagg }
\def\ncdiagg{\pst@object{ncdiagg}}
\def\ncdiagg@i{\check@arrow{\ncdiagg@ii}}
\def\ncdiagg@ii#1#2{%
  \nc@object{Open}{#1}{#2}{.5}{\ncangles@iii \psk@lineAngle\space \tx@NCDiagg}}
\def\pcdiagg{\pst@object{pcdiagg}}
\def\pcdiagg@i{\pc@object\ncdiagg@ii}
\def\tx@NCLoop{NCLoop }
\define@key[psset]{pst-node}{loopsize}{\pst@getlength{#1}\psk@loopsize}
\psset[pst-node]{loopsize=1cm}
\def\ncloop{\pst@object{ncloop}}
\def\ncloop@i{\check@arrow{\ncloop@ii}}
\def\ncloop@ii#1#2{%
\nc@object{Open}{#1}{#2}{2.5}%
{\ncangles@iii /loopsize \psk@loopsize def \tx@NCLoop}}
\def\pcloop{\pst@object{pcloop}}
\def\pcloop@i{\pc@object\ncloop@ii}
\def\tx@NCCircle{NCCircle }
\def\nccircle{\pst@object{nccircle}}
\def\nccircle@i{\check@arrow{\nccircle@ii}}
\def\nccircle@ii#1#2{%
\pssetlength\pst@dima{#2}%
\nc@object{Open}{#1}{#1}{.5}{%
/AngleA \psk@angleA def
/r \pst@number\pst@dima def
\tx@NCCircle \psarc@v end}}
\def\tx@NCBox{NCBox }
\def\ncbox{\pst@object{ncbox}}
\def\ncbox@i{\check@arrow{\ncbox@ii}}
\def\ncbox@ii#1#2{%
\def\pst@linetype{2}%
\nc@object{Closed}{#1}{#2}{.5}{%
tx@Dict begin \psline@iii pop end
\psk@boxheight \psk@boxdepth
\tx@NCBox}}
\def\pcbox{\pst@object{pcbox}}
\def\pcbox@i{\pc@object\ncbox@ii}
\def\tx@NCArcBox{NCArcBox }
\define@key[psset]{pst-node}{boxheight}[0.4cm]{\pst@getlength{#1}\psk@boxheight}
\define@key[psset]{pst-node}{boxdepth}[0.4cm]{\pst@getlength{#1}\psk@boxdepth}
\define@key[psset]{pst-node}{boxsize}[0.4cm]{%
  \pst@getlength{#1}\psk@boxheight%
  \let\psk@boxdepth\psk@boxheight}
\psset[pst-node]{boxsize=0.4cm}
\def\ncarcbox{\pst@object{ncarcbox}}
\def\ncarcbox@i{\check@arrow{\ncarcbox@ii}}
\def\ncarcbox@ii#1#2{%
\def\pst@linetype{1}%
\nc@object{Closed}{#1}{#2}{.5}{%
\psk@arcangleA \psk@boxheight \psk@boxdepth \pst@number\pslinearc
\tx@NCArcBox}}
\def\pcarcbox{\pst@object{pcarcbox}}
\def\pcarcbox@i{\pc@object\ncarcbox@ii}
\def\tx@Tfan{Tfan }
\begingroup
\catcode`\:=13
\gdef\pst@activerot{\def:{\string:}}
\endgroup
\define@key[psset]{pst-node}{nrot}[0]{%
  \begingroup
  \pst@activerot
  \pst@expandafter{\@ifnextchar:{\psset@@nrot}{\psset@@rot}}{#1}\@nil
  \global\let\pst@tempg\psk@rot
  \endgroup
  \let\psk@nrot\pst@tempg}
\def\psset@@nrot:#1\@nil{%
  \psset@@rot#1\@nil
  \edef\psk@rot{NAngle \ifx\psk@rot\@empty\else\psk@rot add \fi}}
\psset[pst-node]{nrot=0}
\def\tx@LPutCoor{LPutCoor }
\def\tx@LPut{LPut }
\define@key[psset]{pst-node}{npos}[{}]{%
  \def\pst@tempa{#1}%
  \ifx\pst@tempa\@empty\def\psk@npos{\npos@default}\else\pst@checknum{#1}\psk@npos\fi}
\psset[pst-node]{npos=}
\def\ncput{\pst@object{ncput}}
\def\ncput@i{\pst@killglue\pst@makebox{\ncput@ii}}
\def\ncput@ii{%
  \begingroup%
  \use@par%
  \if@star\pst@starbox\fi%
  \pst@makesmall\pst@hbox%
  \pst@rotate\psk@nrot\pst@hbox%
  \ncput@iii%
  \endgroup%
  \pst@shortput}
\def\ncput@iii{%
  \leavevmode%
  \hbox{%
    \pst@Verb{
      \pst@nodedict
      /t \psk@npos def
      \tx@LPut
      end
      \tx@PutBegin}%
    \box\pst@hbox%
    \pst@Verb{\tx@PutEnd}}}
\def\naput{\pst@object{naput}}
\def\naput@i{\pst@killglue\pst@makebox{\naput@ii{NAngle 90 add}}}
\def\naput@ii#1{%
  \begingroup
  \use@par
  \if@star\pst@starbox\fi
  \def\psk@refangle{#1 }%
  \let\psk@rot\psk@nrot
  \pst@Verb{ 
    gsave  STV CP T /ps@refangle {#1 } def 
    /ps@rot { \psk@rot } def grestore }
  \uput@vii
  {exch pop add a \tx@PtoC h1 add exch w1 add exch }%
  {tx@Dict /NCLW known { NCLW add } if }%
  \ncput@iii
  \endgroup
  \pst@shortput}
\def\nbput{\pst@object{nbput}}
\def\nbput@i{\pst@killglue\pst@makebox{\naput@ii{NAngle 90 sub}}}
\define@key[psset]{pst-node}{tpos}[0.5]{%
  \pst@checknum{#1}\psk@tpos
  \ifdim\psk@tpos \p@<\z@
    \def\psk@tpos{.5}%
    \@pstrickserr{Bad `tpos' value: `#1'. Must be 0<tpos<1}\@ehpa
  \else
    \ifdim\psk@tpos \p@>\p@
      \def\psk@tpos{.5}%
      \@pstrickserr{Bad `tpos' value: `#1'. Must be 0<tpos<1}\@ehpa%
    \fi%
  \fi}
\psset[pst-node]{tpos=0.5}
\def\nlput{\pst@object{nlput}}
\def\nlput@i(#1)(#2)#3#4{%
  \begin@SpecialObj
  \psLDNode(#1)(#2){#3}{temp@lnput}
  \pcline[linestyle=none](#1)(temp@lnput)%
  \ncput[npos=1]{#4}%
  \end@SpecialObj}
\def\tvput{\pst@object{tvput}}
\def\tvput@i{\pst@makebox{\psput@tput{H}{1}}}
\def\tlput{\pst@object{tlput}}
\def\tlput@i{\pst@makebox{\psput@tput{H}{true}}}
\def\trput{\pst@object{trput}}
\def\trput@i{\pst@makebox{\psput@tput{H}{false}}}
\def\thput{\pst@object{thput}}
\def\thput@i{\pst@makebox{\psput@tput{V}{1}}}
\def\taput{\pst@object{taput}}
\def\taput@i{\pst@makebox{\psput@tput{V}{true}}}
\def\tbput{\pst@object{tbput}}
\def\tbput@i{\pst@makebox{\psput@tput{V}{false}}}
\def\tx@HPutAdjust{HPutAdjust }
\def\tx@VPutAdjust{VPutAdjust }
\def\psput@tput#1#2{%
  \begingroup
  \use@par
  \pst@tputmakesmall
  \leavevmode
  \hbox{%
    \pst@Verb{%
      \pst@nodedict
      /t \psk@tpos \pst@tposflip def
      tx@NodeDict /HPutPos known
        { #1PutPos }
        { CP /Y exch def /X exch def /NAngle 0 def /NCLW 0 def }
      ifelse
      /Sin NAngle sin def
      /Cos NAngle cos def
      /s \pst@number\pslabelsep NCLW add def
      /l \pst@number\pst@dima def
      /r \pst@number\pst@dimb def
      /h \pst@number\pst@dimc def
      /d \pst@number\pst@dimd def
      \ifnum1=0#2 \else
        /flag #2 def
        \csname tx@#1PutAdjust\endcsname
      \fi
      \tx@LPutCoor
      end
      \tx@PutBegin}%
    \box\pst@hbox
    \pst@Verb{\tx@PutEnd}}%
  \endgroup
  \pst@shortput}
\def\pst@tposflip{}
\def\pst@tputmakesmall{%
\pst@dima=\wd\pst@hbox
\divide\pst@dima 2
\pst@dimg=\psk@href\pst@dimg
\pst@dimb\pst@dima
\advance\pst@dima\pst@dimg 
\advance\pst@dimb-\pst@dimg 
\pst@dimd=\psk@vref\relax
\pst@dimc=\ht\pst@hbox
\advance\pst@dimc-\pst@dimd 
\advance\pst@dimd\dp\pst@hbox 
\setbox\pst@hbox=\hbox to\z@{%
\kern-\pst@dima\vbox to\z@{\vss\box\pst@hbox\vskip-\pst@dimd}\hss}}
\def\MakeShortNab#1#2{%
  \def\pst@shortput@nab{%
    \def\pst@tempg{\next}%
    \ifx#1\next
      \let\pst@tempg\naput
    \else
      \ifx#2\next
        \let\pst@tempg\nbput
      \else
        \ifx\@sptoken\next
          \let\pst@tempg\pst@shortput
        \fi
      \fi
    \fi
    \pst@tempg}}
\MakeShortNab{^}{_}
\def\MakeShortTablr#1#2#3#4{%
  \def\pst@shortput@tablr{%
    \def\pst@tempg{\next}%
    \ifx#1\next
      \let\pst@tempg\taput
    \else
      \ifx#2\next
        \let\pst@tempg\tbput
      \else
        \ifx#3\next
          \let\pst@tempg\tlput
        \else
          \ifx#4\next
            \let\pst@tempg\trput
          \else
            \ifx\@sptoken\next
              \let\pst@tempg\pst@shortput
            \fi
          \fi
        \fi
      \fi
    \fi
    \pst@tempg}}
\MakeShortTablr{^}{_}{<}{>}
\def\MakeShortTab#1#2{%
  \def\pst@shortput@tab{%
    \def\pst@tempg{\next}%
    \ifx#1\next
      \def\pst@tempg{%
        \@nameuse{%
          t\ifodd\psk@treemode\ifpstreeflip b\else a\fi
          \else\ifpstreeflip r\else l\fi\fi put}}%
    \else
      \ifx#2\next
        \def\pst@tempg{%
          \@nameuse{%
            t\ifodd\psk@treemode\ifpstreeflip a\else b\fi
            \else\ifpstreeflip l\else r\fi\fi put}}%
      \else
        \ifx\@sptoken\next
          \let\pst@tempg\pst@shortput
        \fi
      \fi
    \fi
    \pst@tempg}}
\MakeShortTab{^}{_}
\define@key[psset]{pst-node}{shortput}[none]{%
  \def\pst@tempg{#1}%
  \ifx\pst@tempg\@none
    \let\pst@shortput\ignorespaces
  \else
    \@ifundefined{pst@shortput@#1}%
     {\@pstrickserr{Bad short put: `#1'}\@ehpa}%
     {\edef\pst@shortput{\noexpand\afterassignment\expandafter\noexpand
      \csname pst@shortput@#1\endcsname\noexpand\let\noexpand\next}}%
  \fi}
\psset[pst-node]{shortput=none}
\def\lput{\def\pst@par{}\pst@ifstar{\@ifnextchar[{\lput@i}{\lput@ii}}}
\def\lput@i[#1]{\addto@par{ref=#1}\lput@ii}
\def\lput@ii{\@ifnextchar({\lput@iv}{\lput@iii}}
\def\lput@iii#1{\addto@par{nrot=#1}\@ifnextchar({\lput@iv}{\ncput@i}}
\def\lput@iv(#1){\addto@par{npos=#1}\ncput@i}
\def\mput{\def\pst@par{}\pst@ifstar{\@ifnextchar[{\mput@i}{\ncput@i}}}
\def\mput@i[#1]{\addto@par{ref=#1}\ncput@i}
\def\Lput{\def\pst@par{}\pst@ifstar{\@ifnextchar[{\Lput@ii}{\Lput@i}}}
\def\Lput@i#1{\addto@par{labelsep=#1}\Lput@ii}
\def\Lput@ii[#1]{\addto@par{ref={#1}}\@ifnextchar({\Lput@iv}{\Lput@iii}}
\def\Lput@iii#1{\addto@par{nrot={#1}}\@ifnextchar({\Lput@iv}{\Lput@v}}
\def\Lput@iv(#1){\addto@par{npos=#1}\Lput@v}
\def\Lput@v{\pst@killglue\pst@makebox{\Lput@vi}}
\def\Lput@vi{%
\begingroup
\use@par
\if@star\pst@starbox\fi
\Rput@vi
\pst@makesmall\pst@hbox
\ifx\psk@rot\@empty\else\pst@rotate{ps@rot }\pst@hbox\fi
\ncput@iii
\endgroup
\pst@shortput}
\def\Mput{\def\pst@par{}\pst@ifstar{\@ifnextchar[{\Mput@ii}{\Mput@i}}}
\def\Mput@i#1{\addto@par{labelsep=#1}\Mput@ii}
\def\Mput@ii[#1]{\addto@par{ref={#1}}\Lput@v}
\def\aput@#1{\def\pst@par{}\pst@ifstar{\@ifnextchar[{\aput@i#1}{\aput@ii#1}}}
\def\aput@i#1[#2]{\addto@par{labelsep=#2}\aput@ii#1}
\def\aput@ii#1{\@ifnextchar({\aput@iv#1}{\aput@iii#1}}
\def\aput@iii#1#2{\addto@par{nrot=#2}\@ifnextchar({\aput@iv#1}{#1}}
\def\aput@iv#1(#2){\addto@par{npos=#2}#1}
\def\aput{\aput@\naput@i}
\def\bput{\aput@\nbput@i}
\def\Aput{\def\pst@par{}\pst@ifstar{\@ifnextchar[{\Aput@i}{\naput@i}}}
\def\Aput@i[#1]{\addto@par{labelsep=#1}\naput@i}
\def\Bput{\def\pst@par{}\pst@ifstar{\@ifnextchar[{\Bput@i}{\nbput@i}}}
\def\Bput@i[#1]{\addto@par{labelsep=#1}\nbput@i}
\def\node@coor#1;#2\@nil{
  \pst@getnode{#1}\pst@tempg
  \edef\pst@coor{%
    \pst@nodedict
    tx@NodeDict \pst@tempg known
    \pslbrace \pst@tempg load \tx@GetCenter \psrbrace
    \pslbrace 0 0 \psrbrace ifelse
    end }}
\def\Node@coor[#1]#2;#3\@nil{
\begingroup
\psset{angle=0,#1}
\@ifnextchar\bgroup{\Node@@@coor}
                   {\Node@@coor}#2\@nil
\endgroup
\let\pst@coor\pst@tempg}
\def\Node@@coor#1\@nil{%
\pst@getnode{#1}\pst@tempg
\xdef\pst@tempg{%
\pst@nodedict
tx@NodeDict \pst@tempg known
  { \psk@nodesepA \psk@angleA 
    \pst@tempg load \psk@nodeseptypeA \tx@GetEdge
    \psk@offsetA \psk@angleA \tx@AddOffset
    \pst@tempg load \tx@GetCenter
    3 -1 roll add 3 1 roll add exch }
  { CP } ifelse end }}
\def\Node@@@coor#1{
\pst@@getcoor{#1}%
\def\psk@angleA{%
  \pst@tempg load \tx@GetCenter \pst@coor
  3 -1 roll sub 3 1 roll sub neg \tx@Atan \psk@angleB add
  }%
\Node@@coor}
\def\nput{\pst@object{nput}}
\def\nput@i#1#2{\pst@killglue\pst@makebox{\nput@ii{#1}{#2}}}
\def\nput@ii#1#2{%
  \begingroup
  \use@par
  \if@star\pst@starbox\fi%
  \psset[pstricks]{refangle=#1}%
  \let\psk@angleA\psk@refangle
  \edef\psk@nodesepA{\pst@number\pslabelsep}%
  \def\psk@nodeseptypeA{0 }%
  \pslabelsep\z@
  \uput@vi
  \Node@@coor#2\@nil
  \let\pst@coor\pst@tempg
  \leavevmode
  \psput@special\pst@hbox
  \endgroup
  \ignorespaces}
\newcount\psrow
\newcount\pscol
\newcount\psmatrixcnt
\newskip\psrowsep
\newskip\pscolsep
\define@key[psset]{pst-node}{colsep}[1.5cm]{\pssetlength\pscolsep{#1}}
\define@key[psset]{pst-node}{rowsep}[1.5cm]{\pssetlength\psrowsep{#1}}
\psset[pst-node]{colsep=1.5cm}
\psset[pst-node]{rowsep=1.5cm}
\newif\ifpsmatrix
\ifx\mscount\@undefined\let\mscount\@multicnt\fi
\def\psmatrix{\begingroup{\ifnum0=`}\fi 
  \@ifnextchar[{\psmatrix@i}{\ifnum0=`{\fi}{}\psmatrix@ii}}
\def\psmatrix@i[#1]{%
  \ifnum0=`{\fi}{}%
  \psset{#1}%
  \psmatrix@ii}
\def\psmatrix@ii{%
  \KillGlue
  \edef\psm@beginmath{%
    \ifmmode$\m@th\ifinner\textstyle\else\displaystyle\fi\fi}%
  \edef\psm@endmath{\ifmmode$\fi}%
  \let\\\psm@cr
  \advance\psmatrixcnt by \@ne
  \def\psm@thenode{M-\the\psmatrixcnt-\the\psrow-\the\pscol}%
  \tabskip\z@
  \psrow=\@ne
  \pscol\z@
  \psset{shortput=tablr}%
  \leavevmode
  \vbox\bgroup\halign\bgroup&%
  \begingroup
  \global\advance\pscol by \@ne
  \csname psrowhook\romannumeral\psrow\endcsname
  \csname pscolhook\romannumeral\pscol\endcsname
  \psm@beginnode##\psm@endnode\endgroup
  \cr}
\def\endpsmatrix{%
  \crcr\egroup\unskip\egroup
  \endgroup}
\def\psm@cr{{\ifnum0=`}\fi\ps@ifnextchar[{\psm@@cr}{\psm@@@cr{}}}
\def\psm@@cr[#1]{\psm@@@cr{\vskip#1\relax}}
\def\psm@@@cr#1{%
  \ifnum0=`{\fi}{}\cr
  \noalign{%
  \global\advance\psrow 1
  \global\pscol\z@
  \vskip\psrowsep
  #1}}
\def\psm@beginnode{%
  \@ifnextchar\psm@endnode
    {\let\psm@endnode@i\relax\setbox\pst@hbox=\hbox{}}%
    {\pst@object{psm@beginnode}}}
\def\psm@beginnode@i{%
  \setbox\pst@hbox=\hbox\bgroup
  \psm@beginmath
  \begingroup
  \ignorespaces}
\def\psm@endnode@i{%
  \unskip
  \endgroup
  \psm@endmath
  \egroup
  \use@par
  \@psttrue}
\def\psm@endnode{%
  \@pstfalse
  \psm@endnode@i
  \ifnum\pscol>1\relax \pshskip\pscolsep \fi
  \psk@mnodesize
  \hfil
  \Pst@nodealigntrue
  \if@pst\csname mnode@\psk@mnode\endcsname
  \else\csname mnode@\psk@emnode\endcsname\fi
  \psk@mcol
  \psk@@mnodesize}
\def\psspan#1{\global\mscount#1\relax\pstloop\ifnum\mscount>\@ne\sp@n\repeat}
\def\pstloop#1\repeat{\gdef\pstiterate{#1\relax\expandafter\pstiterate\fi}%
  \pstiterate
  \let\pstiterate\relax}
\define@key[psset]{pst-node}{name}[\relax]{\pst@getnode{#1}\psk@name}
\let\psk@name\relax
\define@key[psset]{pst-node}{mcol}[c]{%
  \ifx r#1\relax\let\psk@mcol\relax\else
    \ifx l#1\relax\let\psk@mcol\hfill\else
    \let\psk@mcol\hfil\fi\fi}
\psset[pst-node]{mcol=c}
\define@key[psset]{pst-node}{mnodesize}[-1pt]{%
  \pssetlength\pst@dimg{#1}%
  \ifdim\pst@dimg<\z@
    \let\psk@mnodesize\relax
    \let\psk@@mnodesize\relax
  \else
    \edef\psk@mnodesize{\noexpand\hbox to\number\pst@dimg sp\noexpand\bgroup}%
    \let\psk@@mnodesize\egroup
  \fi}
\psset[pst-node]{mnodesize=-1pt}
\def\mnode@R{\rnode@iii\Rnode@ii{\psm@thenode}}
\def\mnode@r{\rnode@iii\rnode@iv{\psm@thenode}}
\def\mnode@oval{\ovalnode@ii{\psm@thenode}}
\def\mnode@tri{\trinode@ii{\psm@thenode}}
\def\mnode@dia{\dianode@ii{\psm@thenode}}
\def\mnode@C{{\Pst@nodealigntrue\cnode@ii(\z@,\z@){\psk@radius}{\psm@thenode}}}
\def\mnode@f{{\Pst@nodealigntrue\fnode@ii(\z@,\z@){\psm@thenode}}}
\def\mnode@circle{\circlenode@ii{\psm@thenode}}
\def\mnode@Circle{\Circlenode@ii{\psm@thenode}}
\def\mnode@p{\pnode(\z@,\z@){\psm@thenode}}
\def\mnode@dot{\dotnode@ii(\z@,\z@){\psm@thenode}}
\def\mnode@none{\box\pst@hbox}
\define@key[psset]{pst-node}{mnode}[R]{%
  \@ifundefined{mnode@#1}%
    {\@pstrickserr{\string\psmatrix\space node `#1' not defined.}\@ehpa}%
    {\edef\psk@mnode{#1}}}
\define@key[psset]{pst-node}{emnode}[none]{%
  \@ifundefined{mnode@#1}%
    {\@pstrickserr{\string\psmatrix\space node `#1' not defined.}\@ehpa}%
    {\edef\psk@emnode{#1}}}
\psset[pst-node]{mnode=R,emnode=none}
\def\nccoil{\pst@object{nccoil}}
\def\nccoil@i{\check@arrow{\nccoil@ii}}
\def\nccoil@ii#1#2{\nc@object{Open}{#1}{#2}{.5}{
  \tx@NCCoor
  tx@Dict begin
  4 2 roll
  \psk@coilwidth \pscoilheight
  \psk@coilarmA \psk@coilarmB
  \psk@coilaspect \psk@coilinc
  \pst@coildict \tx@Coil end
  end}%
}
\def\nczigzag{\pst@object{nczigzag}}
\def\nczigzag@i{\check@arrow{\nczigzag@ii}}
\def\nczigzag@ii#1#2{\nc@object{Open}{#1}{#2}{.5}{
  \tx@NCCoor
  tx@Dict begin
  4 2 roll
  \pscoilheight
  \psk@coilwidth
  \psk@coilarmA
  \psk@coilarmB
  \pst@coildict \tx@ZigZag end
  \psline@iii
  \tx@Line
  end}%
}
\def\psGetNodeCenter#1{ tx@NodeDict begin /N@#1 load GetCenter end 
  \pst@number\psyunit div /#1.y exch def 	
  \pst@number\psxunit div /#1.x exch def }	
\def\psGetEdgeA#1#2{
  tx@NodeDict begin \psk@offsetA \psk@offsetB neg 
    \psk@nodesepA \psk@nodesepB 0 0 
    /N@#1 /N@#2 InitNC { NCCoor } if pop pop \tx@UserCoor end}
\def\psGetEdgeB#1#2{
  tx@NodeDict begin \psk@offsetA \psk@offsetB neg 
    \psk@nodesepA \psk@nodesepB 0 0 
    /N@#1 /N@#2 InitNC { NCCoor } if 4 2 roll pop pop \tx@UserCoor end}
\def\ncbarr{\pst@object{ncbarr}}
\def\ncbarr@i#1#2{%
  \begingroup
  \use@par%
  \psLNode(#1)(#2){0.5}{barr@tempNode}%
  \pst@dimc=\psk@angleA pt
  \pst@dimd=180pt
  \ifdim\pst@dimc=\z@\else\ifdim\pst@dimc=\pst@dimd\else\psset{angleA=0}\fi\fi
  \ncbar[arrows=-]{#1}{barr@tempNode}
  \ifdim\psk@angleA pt=\z@\relax
    \ncbar[angleA=180,angleB=180]{barr@tempNode}{#2}
  \else\ncbar[angleA=0,angleB=0]{barr@tempNode}{#2}\fi%
  \endgroup%
}
\def\psLNode(#1)(#2)#3#4{%
  \pst@getcoor{#1}\pst@tempA%
  \pst@getcoor{#2}\pst@tempB%
  \pnode(!
    \pst@tempA /YA exch \pst@number\psyunit div def
    /XA exch \pst@number\psxunit div def
    \pst@tempB /YB exch \pst@number\psyunit div def
    /XB exch \pst@number\psxunit div def
    /dx XB XA sub def
    /dy YB YA sub def
    XA dx #3\space mul add YA dy #3\space mul add){#4}}
\def\psLCNode(#1)#2(#3)#4#5{%
  \pst@getcoor{#1}\pst@tempA%
  \pst@getcoor{#3}\pst@tempB%
  \pnode(!
    \pst@tempA /YA exch \pst@number\psyunit div def
    /XA exch \pst@number\psxunit div def
    \pst@tempB /YB exch \pst@number\psyunit div def
    /XB exch \pst@number\psxunit div def
    XA #2\space mul XB #4\space mul add
    YA #2\space mul YB #4\space mul add){#5}}
\def\psLDNode(#1)(#2)#3#4{%
  \pst@getcoor{#1}\pst@tempA%
  \pst@getcoor{#2}\pst@tempB%
  \pssetlength\pst@dimb{#3}%
  \pnode(!%
    \pst@tempA /YA exch \pst@number\psyunit div def
    /XA exch \pst@number\psxunit div def
    \pst@tempB /YB exch \pst@number\psyunit div def
    /XB exch \pst@number\psxunit div def
    /dx XB XA sub def
    /dy YB YA sub def
    /angle dy dx Atan def
    /linelength \pst@number\pst@dimb \pst@number\psunit div def
    XA linelength angle cos mul add YA linelength angle sin mul add ){#4}%
}
\def\psRelNode{\pst@object{psRelNode}}
\def\psRelNode@i(#1)(#2)#3#4{{
  \use@par
  \pst@getcoor{#1}\pst@tempA%
  \pst@getcoor{#2}\pst@tempB%
  \pnode(!
    \pst@tempA /YA exch \pst@number\psyunit div def
    /XA exch \pst@number\psxunit div def
    \pst@tempB /YB exch \pst@number\psyunit div def
    /XB exch \pst@number\psxunit div def
    /AlphaStrich \psk@angleA\space def
    /unit \pst@number\psyunit \pst@number\psxunit div def 
    /dx XB XA sub  def
    /dy YB YA sub \ifPst@trueAngle\space unit mul \fi\space def
    /laenge dy dup mul dx dup mul add sqrt #3 mul def
    /Alpha dy dx atan def 
    /beta Alpha AlphaStrich add def
    laenge beta cos mul XA add
    laenge beta sin mul \ifPst@trueAngle\space unit div \fi\space YA add ){#4}%
}\ignorespaces}
\define@cmdkeys[psset]{pstricks-add}[PSTPSPNk@]{
  blName,bcName,brName,
  clName,ccName,crName,
  tlName,tcName,trName}[]{}%
\psset[pstricks-add]{%
  blName=PSPbl,bcName=PSPbc,brName=PSPbr,
  clName=PSPcl,ccName=PSPcc,crName=PSPcr,
  tlName=PSPtl,tcName=PSPtc,trName=PSPtr}
\def\psDefPSPNodes{\def\pst@par{}\pst@object{psDefPSPNodes}}
\def\psDefPSPNodes@i{%
  \pst@killglue
  \begingroup
  \use@par
  \expandafter\psDefPSPNodes@ii\pic@coor}
\def\psDefPSPNodes@ii(#1)(#2)(#3){%
    \pnode(#1){PSPN@temp}\pnode([angle=45]PSPN@temp){\PSTPSPNk@blName}
    \pnode(#3){PSPN@temp}\pnode([angle=-135]PSPN@temp){\PSTPSPNk@trName}
    \pnode(\PSTPSPNk@blName|\PSTPSPNk@trName){\PSTPSPNk@tlName}
    \pnode(\PSTPSPNk@trName|\PSTPSPNk@blName){\PSTPSPNk@brName}
    \ncline[linestyle=none]{\PSTPSPNk@blName}{\PSTPSPNk@tlName}
    \ncput[npos=.5]{\pnode{\PSTPSPNk@clName}}
    \ncline[linestyle=none]{\PSTPSPNk@blName}{\PSTPSPNk@brName}
    \ncput[npos=.5]{\pnode{\PSTPSPNk@bcName}}
    \pnode(\PSTPSPNk@brName|\PSTPSPNk@clName){\PSTPSPNk@crName}
    \pnode(\PSTPSPNk@bcName|\PSTPSPNk@trName){\PSTPSPNk@tcName}
    \pnode(\PSTPSPNk@bcName|\PSTPSPNk@clName){\PSTPSPNk@ccName}
  \endgroup
  \ignorespaces}
\def\psDefBoxNodes#1#2{\rnode[tl]{#1:tl}{\rnode[Bl]{#1:Bl}{\rnode[tr]{#1:tr}{%
\rnode[bl]{#1:bl}{\rnode[Br]{#1:Br}{\rnode[br]{#1:br}{#2}}}}}}%
\pnode(!\psGetNodeCenter{#1:bl}
          \psGetNodeCenter{#1:tl} 
          #1:bl.x #1:tl.x add 2 div #1:bl.y #1:tl.y add 2 div ){#1:Cl}%
\pnode(!\psGetNodeCenter{#1:tr}
          \psGetNodeCenter{#1:br} 
          #1:tr.x #1:br.x add 2 div #1:tr.y #1:br.y add 2 div ){#1:Cr}%
\pnode(!\psGetNodeCenter{#1:Cl}
          \psGetNodeCenter{#1:Cr} 
          #1:Cl.x #1:Cr.x add 2 div #1:Cl.y #1:Cr.y add 2 div ){#1:C}%
\pnode(!\psGetNodeCenter{#1:Br}
          \psGetNodeCenter{#1:Bl} 
          #1:Br.x #1:Bl.x add 2 div #1:Br.y #1:Bl.y add 2 div ){#1:BC}%
\pnode(!\psGetNodeCenter{#1:tr}
          \psGetNodeCenter{#1:tl} 
          #1:tr.x #1:tl.x add 2 div #1:tr.y #1:tl.y add 2 div ){#1:tC}%
\pnode(!\psGetNodeCenter{#1:br}
          \psGetNodeCenter{#1:bl} 
          #1:br.x #1:bl.x add 2 div #1:br.y #1:bl.y add 2 div ){#1:bC}}%
\newcount\pst@args
\newcount\num@pts
\newcount\pst@argcnt
\def\PST@root{}
\let\pst@next\relax
\def\my@tempA{}
\def\my@tempB{}
\def\my@tempC{}
\def\my@tempD{}
\def\my@next{}
\newif\if@paren%
\newif\if@equal%
\newif\if@colon%
\newif\ifshow
\def\plussign{+}\def\minussign{-}
\def\defaultvalue#1#2{
  \ifdefined#1\ifx#1\@empty\xdef#1{#2}\fi\else\xdef#1{#2}\fi}%
\def\testAlg#1|#2\@nil{%
\ifx\relax#2\relax%
   \let\my@next\psparnode\xdef\my@tempD{}%
\else%
   \let\my@next\algparnode\xdef\my@tempD{A}
\fi}%
\def\trim #1{\expandafter\trim@\expandafter{#1 }#1}%
\def\trim@ #1{\trim@@ @#1 @ #1 @ @@}%
\def\trim@@ #1@ #2@ #3@@{\trim@@@\empty #2 @}%
\def\unbrace#1{#1}%
\unbrace{\def\trim@@@ #1 } #2@#3{\expandafter\def%
  \expandafter #3\expandafter {#1}}%
\def\hasparen#1(#2\@nil{
  \ifx\relax#2\relax \@parenfalse \else \@parentrue\fi}%
\def\hasequal#1=#2\@nil{
  \ifx\relax#2\relax \@equalfalse \else \@equaltrue\fi
  \hascolon#2:\@nil}%
\def\hascolon#1:#2\@nil{
\ifx\relax#2\relax \@colonfalse \else \@colontrue\fi}%
\def\equalwhat#1=#2:#3\@nil{{#2}{#3}}%
\def\parsenodexn#1(#2)#3\@nil{%
  \def\coeffA{#1}\edef\nodeA{#2}%
  \trim\coeffA%
  \ifx\nodeA\@empty\else%
    \pnode(#2){@@TMP}%
    \ifx\coeffA\@empty\def\coeffA{1}\else%
      \ifx\coeffA\plussign\def\coeffA{1}\else\ifx\coeffA\minussign\def\coeffA{-1}\fi\fi\fi%
  \edef\cmd{\noexpand\psLCNode(@TMP\the\pst@argcnt){1}(@@TMP){\coeffA}{@TMP}}%
  \cmd%
  \advance\pst@argcnt by \@ne%
  \pnode(@TMP){@TMP\the\pst@argcnt}%
  \parsenodexn#3\@nil%
  \fi}%
\def\normalvec(#1)#2{%
  \psRelNodeVar(0,0)(#1)(0,1){#2}}%
\def\curvepnode#1#2#3{%
  \edef\my@tempA{#2}
  \expandafter\testAlg\my@tempA|\@nil\my@next {#1}{#2}{#3}}
\def\psparnode#1#2#3{%
  \pnode(!/t #1 def #2){#3}%
  \pnode(!/t #1 .001 sub def #2 
          /t #1 .001 add def 
           #2 3 -1 roll sub 3 1 roll sub neg 
           2 copy Pyth dup 3 1 roll div 3 1 roll div ){#3tang}}
\def\algparnode#1#2#3{%
  \pstVerb{tx@Dict begin /Func (#2) AlgParser cvx def end }
  \pnode(!/t #1 def Func){#3}
  \pnode(!/t #1 .001 sub def Func 
          /t #1 .001 add def 
          Func 3 -1 roll sub 3 1 roll sub neg 
          2 copy Pyth dup 3 1 roll div 3 1 roll div ){#3tang}
}%
\def\nodex#1{%
\expandafter\hasparen#1(\@nil%
\if@paren
  \pnode(0,0){@TMP0}%
  \pst@argcnt=0%
  \expandafter\parsenodexn#1()\@nil%
\else%
  \def\my@tempC{#1}%
  \ifx\my@tempC\@empty\pnode(0,0){@TMP}\else\pnode(#1){@TMP}\fi%
\fi}
\def\nodexn#1#2{%
\expandafter\hasparen#1(\@nil
\if@paren
  \pnode(0,0){@TMP0}%
  \pst@argcnt=0%
  \parsenodexn#1()\@nil%
  \pnode(@TMP){#2}%
\else%
  \def\my@tempC{#1}%
  \ifx\my@tempC\@empty\pnode(0,0){#2}\else\pnode(#1){#2}\fi%
\fi}
\def\psxline{\pst@object{psxline}}%
\def\psxline@i{\@ifnextchar({\psxline@iii}{\psxline@ii}}%
\def\psxline@ii#1{%
\addto@par{arrows=#1}%
\psxline@iii}%
\def\psxline@iii(#1)#2#3{{
\pst@killglue%
\use@par%
\nodexn{#2}{@TMP@a}%
\AplusB(#1)(@TMP@a){@TMP@A}%
\nodexn{#3}{@TMP@a}%
\AplusB(#1)(@TMP@a){@TMP@B}%
\psline(@TMP@A)(@TMP@B)%
}%
\ignorespaces}%
\def\curvepnodes{\pst@object{curvepnodes}}
\def\curvepnodes@i#1#2#3#4{{
  \pst@killglue
  \use@par
  \edef\my@tempA{#3}
  \expandafter\testAlg\my@tempA|\@nil %
  \pstVerb{%
	tx@Dict begin 
	/t0 #1 def
	/t1 #2 def  
	 t1 t0 sub end \psk@plotpoints div /dt exch def }%
  \pst@cntc=\psk@plotpoints\relax
  \advance\pst@cntc by \@ne\relax 
  \ifx\my@tempD\@empty\pstVerb{tx@Dict begin /Func (#3) cvx def end }
  \else\pstVerb{tx@Dict begin /Func (#3 ) AlgParser cvx def end }%
  \fi%
    \multido{\i=0+1}{\pst@cntc}{%
      \pnode(! /t #1 dt \i\space mul add def Func ){#4\i}}
    \expandafter\xdef \csname #4nodecount\endcsname {\psk@plotpoints}%
    \ifnum\Pst@Debug>0 \typeout{Created nodes #40 .. #4\psk@plotpoints}\fi%
}\ignorespaces}%
\def\fnpnode{\pst@object{fnpnode}}
\def\fnpnode@i#1#2#3{{
  \pst@killglue
  \use@par
  \ifPst@algebraic\pnode(*#1 {#2}){#3}\else\pnode(! /x #1 def x #2){#3}\fi
}\ignorespaces}%
\def\fnpnodes{\pst@object{fnpnodes}}
\def\fnpnodes@i#1#2#3#4{{
\pst@killglue
\use@par
\pst@dima=#1pt \pst@dimb=#2pt \advance\pst@dimb -\pst@dima%
\pst@cnta=\psk@plotpoints \relax 
\def\PST@root{#4}
\divide\pst@dimb by \pst@cnta
\pst@cntc=\pst@cnta %
\advance\pst@cntc by 1 \relax 
\ifPst@algebraic 
  \multido{\i=0+1}{\pst@cntc}{\pnode(*{\pst@number\pst@dima} {#3}){#4\i}
  \advance\pst@dima \pst@dimb}%
\else
    \multido{\i=0+1}{\pst@cntc}{\pnode(!/x \pst@number\pst@dima\space def x #3){#4\i}%
  \advance\pst@dima \pst@dimb}%
\fi%
  \expandafter\xdef \csname \PST@root nodecount\endcsname {\the\pst@cnta}%
  \ifnum\Pst@Debug>0 \typeout{Created nodes #40 .. #4\the\pst@cnta}\fi%
}\ignorespaces}%
\def\AtoB(#1)(#2)#3{\psLCNodeVar(#1)(#2)(-1,1){#3}}
\def\AplusB(#1)(#2)#3{\psLCNodeVar(#1)(#2)(1,1){#3}}
\def\midAB(#1)(#2)#3{\psLCNodeVar(#1)(#2)(.5,.5){#3}}
\def\psnline{\pst@object{psnline}}
\def\psnline@i{\pst@getarrows{\psnline@ii}}
\def\psnline@ii(#1,#2)#3{{%
\pst@killglue%
\use@par%
\pst@cnta=#2 \relax\advance\pst@cnta by 1
\edef\@tmp{}%
\multido{\i=#1+1}{\pst@cnta}{\xdef\@tmp{\@tmp(#3\i)}}%
\expandafter\psline\@tmp}%
\ignorespaces}%
\def\psnpolygon{\pst@object{psnpolygon}}
\def\psnpolygon@i{\pst@getarrows{\psnpolygon@ii}}
\def\psnpolygon@ii(#1,#2)#3{{%
\pst@killglue%
\use@par%
\pst@cnta=#2 \relax\advance\pst@cnta by 1
\edef\@tmp{}%
\multido{\i=#1+1}{\pst@cnta}{\xdef\@tmp{\@tmp(#3\i)}}%
\expandafter\pspolygon\@tmp}%
\ignorespaces}%
\def\psncurve{\pst@object{psncurve}}
\def\psncurve@i{\pst@getarrows{\psncurve@ii}}
\def\psncurve@ii(#1,#2)#3{{%
\pst@killglue%
\use@par%
\pst@cnta=#2 \relax\advance\pst@cnta by 1
\edef\@tmp{}%
\multido{\i=#1+1}{\pst@cnta}{\xdef\@tmp{\@tmp(#3\i)}}%
\expandafter\pscurve\@tmp}%
\ignorespaces}%
\def\psnccurve{\pst@object{psnccurve}}
\def\psnccurve@i{\pst@getarrows{\psnccurve@ii}}
\def\psnccurve@ii(#1,#2)#3{{%
\pst@killglue%
\use@par%
\pst@cnta=#2 \relax\advance\pst@cnta by 1
\xdef\@tmp{}%
\multido{\i=#1+1}{\pst@cnta}{\xdef\@tmp{\@tmp(#3\i)}}%
\expandafter\psccurve\@tmp}%
\ignorespaces}%
\def\shownode(#1){
  \pst@killglue%
  \pstVerb{%
    gsave tx@Dict begin %
    tx@NodeDict /N@#1 known { 
      /tmpar [(Node #1: ) <28> () (, ) () <29>] def %
      /str 12 string def 
      STV CP T \psGetNodeCenter{#1}\space 
      tmpar 2 #1.x str cvs put 
      /str 12 string def 
      tmpar 4 #1.y str cvs put 
      tmpar concatstringarray = }%
    {
      (Node #1: (NOT KNOWN)) = %
    } ifelse %
    end grestore }%
  \ignorespaces}%
\def\pnodes@ii#1{\getnodelist{#1}{}}
\def\getnodelist#1#2{%
\pst@args=0 \relax%
\def\PST@root{#1}%
\def\pst@next{#2}
\getnext@Node}%
\def\getnext@Node{\@ifnextchar({\getnext@Node@i}%
  {\advance\pst@args by \m@ne \expandafter\xdef \csname \PST@root nodecount\endcsname {\the\pst@args}
  \ifnum\Pst@Debug>0 \typeout{Created nodes \PST@root0 .. \PST@root\the\pst@args}\fi%
  \pst@next}%
}%
\def\getnext@Node@i(#1){%
\pnode(#1){\PST@root\the\pst@args}%
\advance\pst@args by \@ne\relax%
\getnext@Node}%
\def\psLCNodeVar(#1)(#2)(#3)#4{%
\pst@getcoor{#1}\my@tempA%
\pst@getcoor{#2}\my@tempB%
\pnode(#3){tmpLCn@de}%
\pnode(!%
  \my@tempA /YA exch \pst@number\psyunit div def
  /XA exch \pst@number\psxunit div def
  \my@tempB /YB exch \pst@number\psyunit div def
  /XB exch \pst@number\psxunit div def 
  \psGetNodeCenter{tmpLCn@de}\space
  XA tmpLCn@de.x mul XB tmpLCn@de.y mul add
  YA tmpLCn@de.x mul YB tmpLCn@de.y mul add){tmpLCn@deA}%
\pnode(tmpLCn@deA){#4}%
}%
\def\psRelNodeVar{\pst@object{psRelNodeVar}}
\def\psRelNodeVar@i(#1)(#2)(#3)#4{{
  \use@par
  \pst@getcoor{#1}\my@tempA%
  \pst@getcoor{#2}\my@tempB%
   \pnode(#3){tmpn@de}%
\pnode(!
  /unit \pst@number\psyunit \pst@number\psxunit div def 
    \my@tempA /YA exch \pst@number\psyunit div def
    /XA exch \pst@number\psxunit div def
    \my@tempB /YB exch \pst@number\psyunit div YA sub 
    \ifPst@trueAngle\space unit mul \fi\space def
    /XB exch \pst@number\psxunit div XA sub def
    \psGetNodeCenter{tmpn@de}
    XB tmpn@de.x mul YB tmpn@de.y mul sub
    YB tmpn@de.x mul XB tmpn@de.y mul add
    \ifPst@trueAngle\space unit div \fi\space 
   YA add exch XA add exch 
    ){#4}%
}}
\def\psRelLineVar{\pst@object{psRelLineVar}}
\def\psRelLineVar@i{\@ifnextchar({\psRelLineVar@iii}{\psRelLineVar@ii}}
\def\psRelLineVar@ii#1{%
  \addto@par{arrows=#1}%
  \psRelLineVar@iii}
\def\psRelLineVar@iii(#1)(#2)(#3)#4{{%
  \pst@killglue
  \use@par
  \psRelNodeVar(#1)(#2)(#3){#4}%
  \psline(#1)(#4)%
}\ignorespaces}
\def\rhombus#1(#2)(#3)#4#5{
\AtoB(#2)(#3){node@P}
\pnode(! 
/tmp \psGetNodeCenter{node@P} node@P.x node@P.y 
Pyth 2 div def 
/ang tmp #1\space div Acos def 
#1\space tmp 2 mul div 
dup ang cos mul exch ang sin mul ){node@A1}
\pnode(! \psGetNodeCenter{node@A1} node@A1.x node@A1.y neg ){node@A2}
\psRelNodeVar(#2)(#3)(node@A1){#4}%
\psRelNodeVar(#2)(#3)(node@A2){#5}%
}%
\def\psrline{\pst@object{psrline}}
\def\psrline@i{\@ifnextchar({\psrline@iii}{\psrline@ii}}%
\def\psrline@ii#1{%
\addto@par{arrows=#1}%
\psrline@iii}%
\def\psrline@iii{%
\getnodelist{@tmpnode}{\psrline@iv}%
}%
\def\psrline@iv{%
   \ifnum\pst@args<0\else
      \pnode(@tmpnode0){@tmpnodeB0}%
      \multido{\iA=1+1,\iB=0+1}{\pst@args}{%
      \AplusB(@tmpnodeB\iB)(@tmpnode\iA){@tmpnodeB\iA}}%
      \psrline@v%
   \fi%
}%
\def\psrline@v{{
  \pst@killglue%
  \use@par%
  \xdef\tmp{(@tmpnodeB0)}%
  \multido{\i=1+1}{\pst@args}%
{\xdef\tmp{\tmp(@tmpnodeB\i)}}%
\expandafter\psline\tmp%
}\ignorespaces}%
\def\polyIntersections#1#2(#3)(#4){%
\def\nodenameA{#1}\def\nodenameB{#2}%
\pnode(#3){P@A}\pnode(#4){P@B}%
\@ifnextchar({\polyIntersections@next}{\polyIntersections@ii}%
}%
\def\polyIntersections@ii#1#2{%
\def\root@node{#1}\num@pts=#2 \relax%
\polyIntersections@iii}%
\def\polyIntersections@next{
\def\root@node{P@}\getnodelist{P@}{\num@pts=\pst@args \relax\polyIntersections@iii}%
}%
\def\polyIntersections@iii{
\pst@cnta=\num@pts \relax\advance\pst@cnta by 1 \relax%
\pstVerb{%
 /xarray \the\pst@cnta\space array def
 /yarray \the\pst@cnta\space array def  tx@Dict begin }%
\multido{\i=0+1}{\the\pst@cnta}{\pstVerb{ \psGetNodeCenter{\root@node\i} xarray \i\space \root@node\i.x put yarray \i\space \root@node\i.y put }}%
\pstVerb{ /tposmin 100 def /tnegmax -100 def 
\psGetNodeCenter{P@B} \psGetNodeCenter{P@A} 
/dx P@B.x P@A.x sub def 
/dy P@B.y P@A.y sub def 
/lenAB dx dy Pyth def
/oldx xarray 0 get def /oldy yarray 0 get def 
1 1 \the\num@pts\space {/k exch def /newx xarray k get def /newy yarray k get def 
/ddx newx oldx sub def /ddy newy oldy sub def 
/det ddy dx mul ddx dy mul sub def
det abs lenAB ddx ddy Pyth mul .001 mul gt 
{/ac oldx P@A.x sub def /bd oldy P@A.y sub def 
 /tt  ac ddy mul bd ddx mul sub det div def 
 /ss ac  dy mul bd dx mul sub det div def 
ss 0 ge 
   {ss 1 le 
        {tt 0 lt {tt tnegmax gt {/tnegmax tt def} if } {tt tposmin lt {/tposmin tt def} if } ifelse }
    if } 
if }
 if 
 /oldx newx def /oldy newy def} for end }%
\pnode(! \psGetNodeCenter{P@A} \psGetNodeCenter{P@B} P@B.x P@A.x sub  tposmin mul P@A.x add  P@B.y P@A.y sub tposmin  mul P@A.y add ){\nodenameA}%
\pnode(! \psGetNodeCenter{P@A} \psGetNodeCenter{P@B} P@B.x P@A.x sub tnegmax mul P@A.x add P@B.y P@A.y sub tnegmax mul P@A.y add){\nodenameB}%
}%
\def\actualscale#1 #2 scale{
#1}
\def\psGetCenter#1{ tx@NodeDict begin /N@#1 load GetCenter end }
\def\ArrowNotch{\pst@object{ArrowNotch}}
\def\ArrowNotch@i#1#2#3#4{{%
\pst@killglue%
\use@par%
\def\inc{-1}%
\ifx#3<\def\inc{1}\fi
\pstVerb{ 
    1 \psk@arrowinset\space sub \psk@arrowlength\space \psk@arrowsize\space  
    \pst@number\pslinewidth \space mul add  mul mul 
    \expandafter\actualscale\psk@arrowscale \space  mul 
    /hh exch def /hh1 hh .05 sub def }
\def\root@node{#1}\num@pts=\csname\root@node nodecount\endcsname %
\pst@cntb=\num@pts \advance\pst@cntb by \@ne
\pst@cnta=\num@pts \advance\pst@cnta by \thr@@
\pst@cntc=#2 \relax
\ifnum\pst@cntc>\num@pts \pnode(0,0){#4}\else
\pstVerb{%
/PythSq { dup mul exch dup mul add } def
/PtSub {					
  3 -1 roll 		
  sub neg		
  3 1 roll 		
  sub			
  exch                     
} def
  /xarray \the\pst@cnta\space array def
  /yarray \the\pst@cnta\space array def  
  tx@Dict begin }
\multido{\i=0+1,\ib=1+1}{\the\pst@cntb}{\pnode(! \psGetCenter{\root@node\i}\space  
yarray \ib\space 3 -1 roll put xarray \ib\space 3 -1 roll put 0 0 ){@tmp}}
\pnode(! xarray 1 get dup yarray 1 get dup 3 1 roll 
xarray 2 get yarray 2 get PtSub  
2 copy Pyth hh div 2 div dup 
3 1 roll 
div 3 1 roll div 
3 1 roll 
add 3 1 roll add 
 xarray 0 3 -1 roll put yarray 0 3 -1 roll put 
 xarray length 2 sub /topnum exch def 
 xarray topnum get dup yarray topnum get dup 3 1 roll 
topnum 1 sub /topnum exch def xarray topnum get yarray topnum get 
3 -1 roll sub  neg 3 1 roll sub exch 
2 copy Pyth hh div 2 div dup 
3 1 roll div 3 1 roll div 
3 -1 roll add 3 1 roll 
topnum 2 add /topnum exch def xarray topnum 3 -1 roll put yarray topnum 3 -1 roll put 
 /oldcindex \the\pst@cntc\space 1 add def 
 xarray oldcindex get /xc exch def yarray oldcindex get /yc exch def
/inc \inc\space def 
/cindex oldcindex def 
{cindex inc add /cindex exch def xarray cindex get xc sub yarray cindex get yc sub Pyth dup hh1 gt 
{ exit } if } loop 
 hh1 .1 add lt { xarray cindex get yarray cindex get } 
{ xarray cindex inc sub get dup yarray cindex inc sub get dup 4 -1 roll exch 
xarray cindex get yarray cindex get PtSub /dy1 exch def /dx1 exch def dx1 dy1 PythSq /Aterm exch def 
 2 copy xc yc PtSub 
 2 copy 2 copy 3 -1 roll mul 3 1 roll mul add hh dup mul sub 
 Aterm div /Cterm exch def  
 dx1 dy1 
 4 1 roll mul 3 1 roll mul add Aterm div /Bterm exch def 
 Bterm abs neg dup dup mul Cterm sub sqrt add dup /tval exch def
 dup dx1 dy1 4 1 roll mul 3 1 roll mul  
 PtSub } ifelse 
 \pst@number\psyunit div exch \pst@number\psxunit div exch  
){#4}\fi%
\pstVerb { end } 
}\ignorespaces}%
\def\saveDataAsNodes#1#2{
  \psLoopIndex=0\relax
  \typeout{Open file #1}%
  \openin7=#1
  \loop
    \read7 to \@Data
    \ifeof7\else
      \ifx\@Data\@empty
      \else
        \pnode(!\@Data){#2\the\psLoopIndex}%
        \typeout{#2\the\psLoopIndex -> \@Data}%
	\advance\psLoopIndex by 1
        \let\@oldData\@Data
      \fi
  \repeat
  \closein7
  \advance\psLoopIndex by -1
  \pnode(!\@oldData){#2Last}%
}
\catcode`\@=\TheAtCode\relax
 
\csname PSTcoilsLoaded\endcsname
\let\PSTcoilsLoaded 
\ifx\PSTricksLoaded   \else\input pstricks.tex\fi
\ifx\PSTnodeLoaded    \else\fi
\ifx\PSTXKeyLoaded    \else\input pst-xkey \fi
\def\fileversion{1.07}
\def\filedate{2015/05/13}
\edef\TheAtCode{\the\catcode`\@}
\catcode`\@=11
\pst@addfams{pst-coil}
\pstheader{pst-coil.pro}
\edef\pst@theheaders{\pst@theheaders,pst-coil.pro}
\def\pst@CoilDict{tx@CoilDict begin }
\def\tx@CoilLoop  {\pst@CoilDict CoilLoop   end }
\def\tx@Coil      {\pst@CoilDict Coil       end }
\def\tx@AltCoil   {\pst@CoilDict AltCoil    end }
\def\tx@ZigZag    {\pst@CoilDict ZigZag     end }
\def\tx@ZigZagCirc{\pst@CoilDict ZigZagCirc end }
\def\tx@Sin       {\pst@CoilDict Sin        end }
\define@key[psset]{pst-coil}{coilwidth}[1cm]{\pst@getlength{#1}\psk@coilwidth}
\define@key[psset]{pst-coil}{coilheight}[1]{\pst@checknum{#1}\pscoilheight}
\define@key[psset]{pst-coil}{coilarmA}[0.5cm]{\pst@getlength{#1}\psk@coilarmA}
\define@key[psset]{pst-coil}{coilarmB}[0.5cm]{\pst@getlength{#1}\psk@coilarmB}
\define@key[psset]{pst-coil}{coilarm}[0.5cm]{%
  \pst@getlength{#1}\psk@coilarmA%
  \let\psk@coilarmB\psk@coilarmA}
\define@key[psset]{pst-coil}{coilaspect}[45]{\pst@getangle{#1}\psk@coilaspect}
\define@key[psset]{pst-coil}{coilinc}[10]{\pst@getangle{#1}\psk@coilinc}
\psset[pst-coil]{coilaspect=45,coilarm=.5cm,coilheight=1,coilwidth=1cm,coilinc=10}
\def\pscoil{\def\pst@par{}\pst@object{pscoil}}
\def\pscoil@i{\pst@getarrows\pscoil@ii}
\def\pscoil@ii(#1){\@ifnextchar({\pscoil@iii{1}(#1)}{\pscoil@iii{\z@}(0,0)(#1)}}
\def\pscoil@iii#1(#2)(#3){%
  \begin@OpenObj
  \pst@getcoor{#2}\pst@tempa
  \pst@getcoor{#3}\pst@tempb
  \pst@optcp{#1}\pst@tempa
  \addto@pscode{%
    \pst@tempa \pst@tempb
    \psk@coilwidth \pscoilheight
    \psk@coilarmA \psk@coilarmB
    \psk@coilaspect \psk@coilinc
    \tx@Coil }%
    \showpointsfalse
  \end@OpenObj}
\def\psCoil{\def\pst@par{}\pst@object{psCoil}}
\def\psCoil@i#1#2{%
  \begin@AltOpenObj
  \showpointsfalse
  \pst@getangle{#1}\pst@tempa
  \pst@getangle{#2}\pst@tempb
  \addto@pscode{%
    \pst@tempa
    \pst@tempb
    \psk@coilwidth
    \pscoilheight
    \psk@coilaspect
    \psk@coilinc
    \tx@AltCoil  
    \@nameuse{psls@\pslinestyle} }%
  \end@OpenObj}
\define@key[psset]{pst-coil}{bow}[0]{%
  \pst@getlength{#1}\psk@bow%
  \pst@dimm=\psk@bow pt%
  \pst@absdim{\pst@dimm}{\pst@dimn}%
  \ifdim\pst@dimn<1pt \def\psk@bow{0}\fi}%
\psset[pst-coil]{bow=0}
\def\pszigzag{\def\pst@par{}\pst@object{pszigzag}}
\def\pszigzag@i{\pst@getarrows\pszigzag@ii}
\def\pszigzag@ii(#1){\@ifnextchar({\pszigzag@iii{1}(#1)}{\pszigzag@iii{\z@}(0,0)(#1)}}
\def\pszigzag@iii#1(#2)(#3){%
  \addbefore@par{bow=0}%
  \begin@OpenObj%
  \pst@getcoor{#2}\pst@tempA%
  \pst@getcoor{#3}\pst@tempB%
  \pst@optcp{#1}\pst@tempA%
  \addto@pscode{%
    \pst@tempA
    \pst@tempB
    \pscoilheight
    \psk@coilwidth
    \psk@coilarmA
    \psk@coilarmB 
    \ifdim\psk@bow pt=\z@ \tx@ZigZag \else \psk@bow\space \tx@ZigZagCirc \fi
    \psline@iii
    \tx@Line }%
  \end@OpenObj}
\def\nccoil{\pst@object{nccoil}}
\def\nccoil@i{\check@arrow{\nccoil@ii}}
\def\nccoil@ii#1#2{\nc@object{Open}{#1}{#2}{.5}{%
  \tx@NCCoor
  tx@Dict begin
  4 2 roll
  \psk@coilwidth \pscoilheight
  \psk@coilarmA \psk@coilarmB
  \psk@coilaspect \psk@coilinc
  \tx@Coil 
  end }}
\def\pccoil{\def\pst@par{}\pst@object{pccoil}}
\def\pccoil@i{\pc@object\nccoil@ii}
\def\nczigzag{\pst@object{nczigzag}}
\def\nczigzag@i{\check@arrow{\nczigzag@ii}}
\def\nczigzag@ii#1#2{\nc@object{Open}{#1}{#2}{.5}{%
  \tx@NCCoor
  tx@Dict begin
  4 2 roll
  \pscoilheight
  \psk@coilwidth
  \psk@coilarmA
  \psk@coilarmB
  \ifdim\psk@bow pt=\z@\tx@ZigZag\else\psk@bow\space\tx@ZigZagCirc\fi 
  \psline@iii
  \tx@Line
  end }}
\def\pczigzag{\def\pst@par{}\pst@object{pczigzag}}
\def\pczigzag@i{\pc@object\nczigzag@ii}
\def\pst@checkUnit#1#2{\expandafter\pst@checkUnit@i#1!!#2}
\def\pst@checkUnit@i{\@ifnextchar*%
  {\def\pst@roundValue{0 }\pst@checkUnit@ii}%
  {\def\pst@roundValue{-1 }\pst@checkUnit@iii**}}
\def\pst@checkUnit@ii*{\@ifnextchar*%
  {\def\pst@roundValue{1 }\pst@checkUnit@iii*}%
  {\pst@checkUnit@iii**}}
\def\pst@checkUnit@iii**#1!!#2{%
  \edef\ps@next{#1}%
  \ifx\ps@next\@empty\let\pst@num\z@%
  \else\expandafter\pst@@checknum\ps@next..\@nil%
  \fi%
  \ifnum\pst@num=\z@\pst@getlength{#1}{#2}\def\pst@relativePeriod{false }%
  \else%
    \def\pst@relativePeriod{true }%
    \edef#2{\ifnum\pst@num=\tw@-\fi\the\pst@cntg.%
    \expandafter\@gobble\the\pst@cnth\space}%
  \fi}
\define@key[psset]{pst-coil}{periods}[1]{\pst@checkUnit{#1}{\psk@periods}}
\define@key[psset]{pst-coil}{amplitude}[1]{\def\psk@amplitude{#1 }}
\define@key[psset]{pst-coil}{ppoints}[360]{\def\psk@ppoints{#1 }}
\define@key[psset]{pst-coil}{function}[sin]{\def\psk@function{#1 }}
\psset[pst-coil]{periods=1,amplitude=1,ppoints=360,function=sin}
\def\pssin{\pst@object{pssin}}
\def\pssin@i{\pst@getarrows\pssin@ii}
\def\pssin@ii(#1){\@ifnextchar({\pssin@iii{1}(#1)}{\pssin@iii{\z@}(0,0)(#1)}}
\def\pssin@iii#1(#2)(#3){%
  \begin@OpenObj
  \pst@getcoor{#2}\pst@tempa
  \pst@getcoor{#3}\pst@tempb
  \pst@optcp{#1}\pst@tempa
  \addto@pscode{%
    \pst@tempa \pst@tempb
    \psk@periods 
    \pst@relativePeriod 
    \pst@roundValue
    \psk@amplitude \pst@number\psyunit mul
    \psk@coilarmA \psk@coilarmB 
    \psk@ppoints
    { \psk@function }
    \tx@Sin
  }%
  \showpointsfalse%
  \end@OpenObj}
\def\ncsin{\pst@object{ncsin}}
\def\ncsin@i{\check@arrow{\ncsin@ii}}
\def\ncsin@ii#1#2{\nc@object{Open}{#1}{#2}{.5}{%
  \tx@NCCoor
  tx@Dict begin
  4 2 roll
  \psk@periods 
  \pst@relativePeriod 
  \pst@roundValue
  \psk@amplitude \pst@number\psyunit mul
  \psk@coilarmA \psk@coilarmB 
  \psk@ppoints
  { \psk@function }
  \tx@Sin 
  end }}
\def\pcsin{\def\pst@par{}\pst@object{pcsin}}
\def\pcsin@i{\pc@object\ncsin@ii}
\catcode`\@=\TheAtCode\relax

\newcommand{\ShortFig}{_Short}

 \newcommand{\Gt}{\bar{G}}
 \newcommand{\GtO}{\bar{G}_0}
 \newcommand{\Gtoo}{\bar{G}_\infty}
 \newcommand{\Mt}{\bar{M}}
 \newcommand{\MtO}{\bar{M}_0}
 \newcommand{\Mtoo}{\bar{M}_\infty}
 \newcommand{\varphit}{\bar{\varphi}}
 \newcommand{\varphitO}{\bar{\varphi}_0}
 \newcommand{\varphitoo}{\bar{\varphi}_\infty}
 \newcommand{\PtO}{\bar{P}_0}
 \newcommand{\Ptoo}{\bar{P}_\infty}

\renewcommand{\mod}[1]{{\rm mod}(#1)}
\newcommand{\punto}{\pscircle*{0.1}}

\renewcommand{\a}{\tilde{a}}
\renewcommand{\b}{\tilde{b}}
\newcommand{\G}{\tilde{G}}
\newcommand{\s}{\tilde{s}}
\newcommand{\p}{\tilde{s}}
\newcommand{\oB}{\tilde{B}}
\newcommand{\oDelta}{\tilde{\Delta}}
\newcommand{\onabla}{\tilde{\nabla}}
\newcommand{\ovarphi}{\tilde{\varphi}}
\newcommand{\oomega}{\tilde{\omega}}

\newcommand{\ds}{\displaystyle}
\newcommand{\ts}{\textstyle}
\definecolor{dark_green}{rgb}{0.2,0.4,0.2}
\newrgbcolor{mygreen}{0.0 0.5 0.0}

 \newtheorem{Theo}{\bf Theorem}
 \newtheorem{Lemm}{\bf Lemma}
 \newtheorem{Prop}{\bf Property}
 \newtheorem{Def}{\bf Definition}

\journal{}
\date{}

\begin{document}

\begin{frontmatter}

\title{On the Analysis of Qualitative Nyquist Plots}

\author{Davide Tebaldi\corref{mycorrespondingauthor} \orcidlink{0000-0003-1432-0489}}
\cortext[mycorrespondingauthor]{Corresponding author}
\ead{davide.tebaldi@unimore.it}

\author{Roberto Zanasi \orcidlink{0000-0001-5507-825X}}
\ead{roberto.zanasi@unimore.it}

\affiliation{organization={University of Modena and Reggio Emilia},
            addressline={Via Pietro Vivarelli 10 - int. 1},
            city={Modena},
            postcode={41125},
            state={},
            country={Italy}}

\begin{abstract}
A powerful tool in control and systems engineering is represented by Nyquist plots, for which a qualitative representation often gives a clearer visualization of the frequency response function that is typically not given by computer programs, especially if portions of the Nyquist plot extend to infinity.
This letter addresses the graphical analysis of the frequency response function, with the objective of enhancing the procedure for the qualitative construction of Nyquist plots. Several results supported by analytical proofs are derived for what concerns the low and high frequency behavior, which enable to improve the qualitative construction of Nyquist plots in the vicinity of the initial and final points.
\end{abstract}


\end{frontmatter}

\section{Introduction}
\label{sec:introduction}
Nyquist plots are one of the possible graphical representations of the frequency response function~\cite{LTI_Syst_1} having
a large variety of different applications in the field of control theory, including the Nyquist stability criterion~\cite{Nyquist_9,Nyquist_2,Nyquist_2_bisss,Nyquist_1} for linear systems
and the stability analysis of nonlinear systems using criteria such as the Circle criterion~\cite{Nyquist_3}, the Popov criterion~\cite{Nyquist_3}, or the describing function method~\cite{Nyquist_4}. Furthermore, Nyquist plots are also employed to perform the passivity analysis of physical systems~\cite{Nyquist_5,Nyquist_6},
and to design lead/lag networks using the concept of admissible domain~\cite{Nyquist_8}.
Nyquist plots also find application in several other fields, including the medical one
as in~\cite{Altro_2}, where they are employed within an algorithm that recognizes different types of swallowing events, which is used for screening and treatment of dysphagia. In~\cite{Altro_4}, Nyquist plots are used for the thermal characterization of electronic packages, while a method based on the use of Nyquist plots to relax linear matrix inequalities conservatism for robust mechatronics synthesis is proposed in \cite{Altro_5}.

For what concerns the construction of Nyquist plots, an interesting feature is described in \cite{Nyquist_10}, showing that many Nyquist plots can be associated with characteristic curves on the complex plane.
However, whenever a detailed and punctual representation of Nyquist plots is desired, their
construction is typically addressed making use of computer aided control system design software, such as the Control System Toolbox available in the MATLAB/Simulink environment~\cite{CST_Ref}, the LabVIEW software~\cite{LabVIEW_Octave_Ref}, and others. However, the reading of the resulting Nyquist plots may not be easy, because of the large range in magnitude that can result when spanning a wide range of frequency. Following the same principle employed for Bode plots, some works employ the logarithmic scale for the magnitude representation \cite{Altroo_6}
which, however, still require the use of computer aided control system design software and to arbitrary set the minimum magnitude value to be represented.

A qualitative representation
is often sufficient to extract the desired features of the Nyquist plot. The typical approaches for constructing Nyquist plots
require a large amount of calculations \cite{LTI_Syst_1,Nyquist_12,Nyquist_13} and are essentially based on the analysis of the frequency response function at low and high frequencies.
In \cite{questo_2}, a revised version of the method for plotting qualitative Nyquist plots was proposed, based on the calculation of two novel parameters $\Delta_\tau$ and $\Delta_p$ allowing to improve the Nyquist plot representation at low and high frequencies, with the specific objective of discriminating whether the Nyquist plot is leading or lagging compared to the initial and final phases, respectively. This procedure allows to enhance the comprehension of the plots given by computer aided control system design software.
However, the proposed procedure failed whenever $\Delta_\tau=0$ or $\Delta_p=0$. Furthermore, these two parameters only encompass information about the phase, without giving information about the modulus of the Nyquist plot in the vicinity of the initial and final point.

In the present letter, we extend the work in \cite{questo_2} by providing the following new contributions: 1) a much deeper frequency response analysis at low frequency; 2) the generalization of the low-frequency response analysis to the high-frequency case, providing dual results;
3) the generalization of the previous parameters $\Delta_\tau$ and $\Delta_p$, overcoming the limitation occurring whenever they are equal to zero and also providing information about the modulus, as well as the phase, of the Nyquist
plot in the vicinity of the initial and final point; 4) several results supported by analytical proofs allowing to discriminate the qualitative behavior of the frequency response function in the vicinity of the initial and final points when constructing the qualitative Nyquist plot.

The remainder of this letter is structured as follows. Sec.~\ref{sec:MathTools} introduces the qualitative graphical analysis of the frequency response function, which is deepened in Sec.~\ref{Sec_Freq_Res_s_0} and Sec.~\ref{Sec_Freq_Res_s_oo} by deriving several results for the low and high frequency response, respectively. 
To validate the proposed results,
two case studies are presented in Sec.~\ref{Case_studies_sect}, while the conclusions are given in Sec.~\ref{Conclusion_sect}.

\section{Qualitative graphical analysis of the frequency response in the complex plane}
 \label{sec:MathTools}

\begin{figure}[tb]
 \includegraphics[clip,width=0.95\columnwidth]{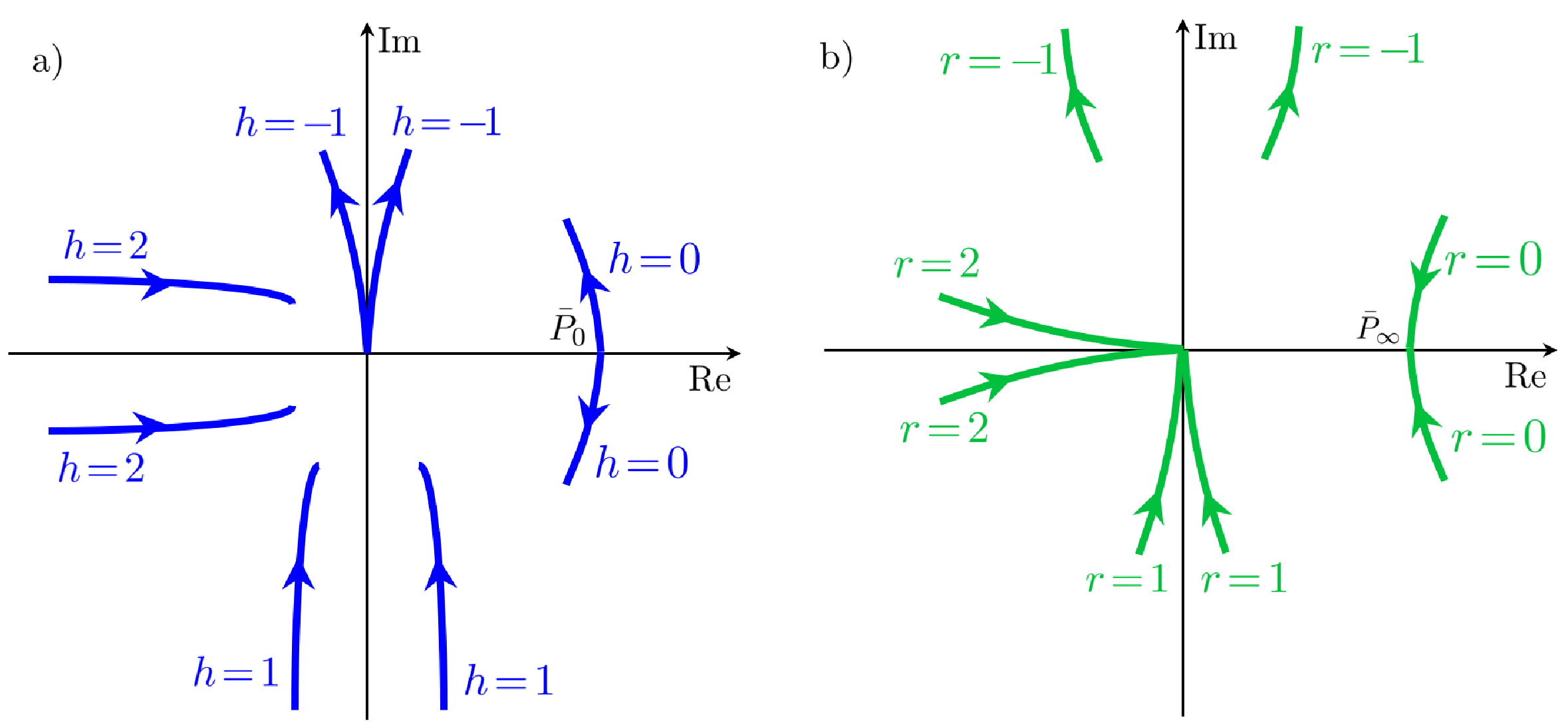}
    \caption{\label{fig:InitEndPolarPlot} (a) Nyquist plots in the low frequency range  for different values of the number $h$ of poles at the origin and (b) Nyquist plots in the high frequency range     for different values of the relative degree $r$.}
    \vspace{-2mm}
\end{figure}

Consider the transfer function $\Gt(s)$ expressed as follows:
\begin{equation}
\Gt(s)=\frac{K}{s^h} \, G(s),
 \label{eq:TransferFunction}
\end{equation}
where the function $G(s)$ has the following structure:
\begin{equation}
 G(s) =
 \frac{a_m s^m+\ldots+a_3 s^3+a_2 s^2+a_1 s+a_0}
       {b_n s^n+\ldots+b_3 s^3+b_2 s^2+b_1 s+b_0}
 \label{Gs_base}
\end{equation}
and is characterized by non-null coefficients
$a_m$, $b_n$,  $a_0$ and $b_0$,
without loss of generality.
The initial and final points of the Nyquist plot of function $\Gt(s)$ can be obtained by considering the  functions $\GtO(s)$ and $\Gtoo(s)$, that
approximate $\Gt(s)$ for $s\simeq 0^{+}$ and $s\simeq \infty $,
respectively. The approximating function $\GtO(s)$ is obtained from
$\Gt(s)$ in (\ref{eq:TransferFunction}) by neglecting all the terms in $s$
except for the zeros and poles at the origin:
\begin{equation}
 \GtO(s)
 \!=\!\! \lim_{s\simeq 0^+} \!\Gt(s)
  \!=\! \frac{K}{s^h}G_0,
 \hspace{3mm}
 \mbox{where}
 \hspace{3mm}
  G_0 \!= \!\frac{a_0}{b_0}
 \label{eq:KDefinition}
\end{equation}
and  where $h$ is the number of poles of the transfer  function $\Gt(s)$ at the origin.
The approximating function $\Gtoo(s)$ is deduced from $\Gt(s)$ in
(\ref{eq:TransferFunction}) by considering only the terms in $s$ having the highest degree:
\begin{equation}
 \Gtoo(s)
=\lim_{s\simeq \infty} \Gt(s)
= \frac{K}{s^r}G_\infty,
 \hspace{2mm}
 \mbox{where}
 \hspace{2mm}
G_\infty = \frac{a_m}{b_n}
 \label{eq:K1Definition}
\end{equation}
and where $r=h+n-m$ is the relative degree of function $\Gt(s)$.
The initial point $\PtO = \MtO\, e^{j\varphitO}$ and the final point $\Ptoo =
\Mtoo\, e^{j\varphitoo}$ of the Nyquist plot can be computed by considering the
magnitude and the phase of the frequency response functions $\GtO(j\omega)$
and $\Gtoo(j\omega)$ for $\omega\!\simeq\! 0^+$ and $\omega
\!\simeq\! \infty$:
\begin{equation}
  \label{M_varphi_0_infty}
\begin{array}{@{}c@{\;}c@{}}
\begin{array}{@{}r@{\,}c@{\,}l@{}}
 \MtO &=&  \ds \lim_{\omega\simeq 0^+}\frac{|K|}{\omega^h}M_0 =
 \left\{
 \begin{array}{@{}l@{\;\;\;}l}
 \infty &\mbox{ if } h>0\\[0mm]
 |K|M_0 &\mbox{ if } h=0\\[0mm]
 0 &\mbox{ if } h<0
 \end{array}\right.
 \end{array},\;\;\;\;  M_0= |G_0|,
 \\[9mm]
  \varphitO =  \arg \left( K \right)-h \ds \frac{\pi}{2}+\varphi_0, \;\;\;\;\;\;
\varphi_0 = \mbox{arg}(G_0),
  \\[5mm]
  \begin{array}{@{}r@{}c@{}l@{}}
 \Mtoo &=& \ds\lim_{\omega\simeq \infty}\frac{|K|}{\omega^r}M_\infty =
 \left\{
 \begin{array}{@{}l@{\,}l}
 0 &\mbox{ if } r>0\\[0mm]
 |K|M_\infty &\mbox{ if } r=0\\[0mm]
 \infty &\mbox{ if } r<0
\end{array}\right.
\end{array},\;\;\;\;  M_{\infty}= |G_{\infty}|,
\\[9mm]
 \varphitoo=\arg \left(K\right)-r \ds \frac{\pi}{2}+\varphi_\infty, \;\;\;\;\;
\varphi_\infty = \mbox{arg}(G_\infty).
\end{array}
\end{equation}
From \eqref{M_varphi_0_infty}, it follows that the Nyquist plot starts from the origin when $\omega\simeq 0^+$ if $h<0$,
from a point on the real axis if $h=0$, or  from infinity if $h>0$, as graphically shown
in Fig.~\ref{fig:InitEndPolarPlot}(a). Dually, from \eqref{M_varphi_0_infty} it follows that the Nyquist plot ends at
infinity when $\omega\simeq \infty$ 
if $r<0$, at a point on the real axis if $r=0$, or at the origin if $r>0$, as graphically shown in Fig.~\ref{fig:InitEndPolarPlot}(b).
The initial and final directions of the Nyquist plot are determined by
the parameters $\varphitO$ and $\varphitoo$. However, as shown
in Fig.~\ref{fig:InitEndPolarPlot}, the 
behavior of the Nyquist curve
is
not univocal either when $\omega\simeq 0^+$ or $\omega \simeq \infty$,
because the system $\Gt(s)$ may exhibit a phase lead or a
phase lag with respect to the initial and final phases $\varphitO$ and $\varphitoo$, respectively. This aspect is investigated in detail in the following sections.

Note that the frequency response of function $\Gt(s)$ for $s\simeq 0$ and $s\simeq \infty$ is strictly related to the frequency response of function $G(s)$ through \eqref{eq:TransferFunction}-\eqref{M_varphi_0_infty}. In the following sections, the frequency response of function $G(s)$ is carried out for $s\simeq 0$ and $s\simeq \infty$ to study the behavior in the vicinity of the initial and final points $P_{0} = M_{0}\, e^{j\varphi_{0}}$ and $P_{\infty} =
M_{\infty}\, e^{j\varphi_{\infty}}$. The obtained results can be directly extended to the frequency response of function $\Gt(s)$ by using the relations \eqref{M_varphi_0_infty} between the initial points $\PtO$,\! $P_{0}$ and between the final points $\Ptoo$,\! $P_{\!\infty}$.

 \section{Frequency response of $G(s)$ for $s\simeq 0$} \label{Sec_Freq_Res_s_0}

 The Taylor series expansion for $s=0$ of function $G(s)$ in (\ref{Gs_base}) has the following form:
\begin{equation}
  \label{Gs}
 G(s)
  = \sum_{k=0}^\infty G_k \,s^k, \hspace{8mm}\mbox{where}\hspace{8mm}
    G_k\!=\!
 \left.\frac{1}{k!}\frac{d^kG(s)}{ds^k}\right|_{s=0}.
 \end{equation}


 \noindent
From (\ref{Gs}) it follows that the frequency response function $G(j\omega)$ of function $G(s)$ can be expressed as follows:
 \begin{equation}
  \label{Gjw}
\begin{array}{@{}r@{\;}c@{\;}l@{}}
 G(j\omega)
  &=& \ds
  \!\sum_{k=0}^\infty G_k \,(j\omega)^k=
  \ds
  \underbrace{\!\sum_{n=0}^\infty\! G_{2 n}(-1)^n \omega^{2n} }_{G_{ev}(\omega)}
  \!+ j\omega\!
  \underbrace{\left[
  \sum_{n=0}^\infty G_{2 n+1} \,(-1)^n \omega^{2n}
  \!\right]\!}_{G_{od}(\omega)},\!

\end{array}
 \end{equation}
  where $G_{ev}(\omega)$ and $\omega \,G_{od}(\omega)$ are the real and imaginary parts of the frequency response function $G(j\omega)$. Note that the functions $G_{ev}(\omega)$ and $G_{odd}(\omega)$ are only composed of
  the even and odd coefficients $G_k$ of the Taylor series expansion (\ref{Gs}), respectively.

\subsection{Sign of the phase $\varphi_0(\omega)$   when $\omega\simeq 0^+$}
 The phase $\varphi_0(\omega)$ of function $G(j\omega)$ when $\omega\simeq 0^+$ can be expressed from \eqref{Gjw} as follows:
 \begin{equation}
  \label{varphiw}
\begin{array}{@{}r@{\;}c@{\;}l@{}}
 \varphi_0(\omega)
  &=& \ds
 \arctan\left[\frac{\omega \,G_{od}(\omega) }{G_{ev}(\omega) }\right] 
 \\[4mm]
  &=& \ds
 \arctan\left[\frac{\omega ( G_1 \!-\! G_3 \omega^2  \!+\! G_5  \omega^2 \!-\! G_7 \omega^6 \!+\!\ldots )}
 {G_0 \!-\! G_2 \omega^2  \!+\! G_4  \omega^2 \!-\! G_6 \omega^6 +\ldots}\right].
 \end{array}
 \end{equation}

\noindent
\begin{Def} \label{Def_Delta_tau}
{\it  Let parameter $\Delta_{\tau_{k}}$ be defined as follows:
\begin{equation}
  \label{Delta_tau_i}
 \Delta_{\tau_{k}}
 =
 \ds (-1)^{\left\lfloor\frac{k}{2}\right\rfloor}\frac{G_{k}}{G_0},
\end{equation}
 where $k>0$ is a positive integer number,
 and function $\left\lfloor x\right\rfloor$ is the floor of $x$.
}
\end{Def}
\vspace{2mm}

\noindent
\begin{Prop} \label{Prop_Delta_tau}
{\it Let $k$ be an odd integer index $k\in\{1,\,3,\,5,\,\ldots,\,\infty\}$.
If $G_1=G_3=G_5=\cdots=G_{k-2}=0$, then the phase $\varphi_0(\omega)$ of function $G(j\omega)$ when $\omega\simeq 0^+$ can be approximated as follows:
 \begin{equation}
  \label{varphiwi}
 \varphi_0(\omega)
 \!\simeq\!
 (-1)^\frac{k-1}{2}\frac{G_{k}}{G_0} \,\omega^{k}
 = \Delta_{\tau_{k}} \,\omega^{k}.
 \end{equation}
Relation (\ref{varphiwi}) holds since $\frac{k-1}{2}={\left\lfloor\frac{k}{2}\right\rfloor}$ for odd 
$k$.
}
\end{Prop}
\vspace{2mm}
{\it Proof.} The expression of $\varphi_0(\omega)$ in \eqref{varphiwi} follows directly from \eqref{Delta_tau_i} and \eqref{varphiw} when $\omega \simeq 0^+$ and $G_1\!=\!G_3\!=\!G_5\!=\!\cdots\!=\!G_{k-2}\!=\!0$.

\vspace{2mm}

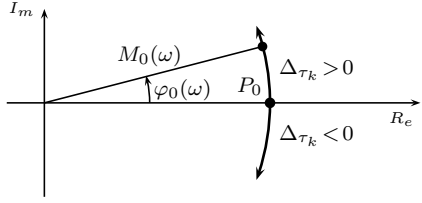
\begin{figure}[tb]
\begin{center}
 \setlength{\unitlength}{5.0mm}
 \psset{unit=\unitlength}
 \SpecialCoor
 \begin{pspicture}(-1,-2)(10,2.75)
 \psline[linewidth=0.6pt]{->}(-1,0)(10,0)
 \rput[t](9.5,-0.25){\tiny $R_e$}
 \psline[linewidth=0.6pt]{->}(0,-2.5)(0,2.5)
 \rput[r](-0.25,2.5){\tiny $I_m$}
 \psarc[linewidth=1.0pt]{*->}(0,0){6}{0}{20}
 \psarc[linewidth=1.0pt]{<-}(0,0){6}{-20}{0}
 \rput[l](6.3;8){\scriptsize $\Delta_{\tau_k}\!>\!0$}
 \rput[l](6.3;-8){\scriptsize $\Delta_{\tau_k}\!<\!0$}
 \rput[rb](5.8;2){\scriptsize $P_0$}
  \psline[linewidth=0.6pt]{-}(0,0)(5.8,1.5)
  \pscircle[fillstyle=solid,fillcolor=black,linecolor=black](5.8,1.5){.1}
 \rput(2.8,1.25){\scriptsize $M_0(\omega)$}
 \psarc[linecolor=black,linewidth=0.6pt]{->}(0,0){2.8}{0}{15}
 \rput(3.68,0.36){\scriptsize $\varphi_0(\omega)$}
%
 %
 \end{pspicture}
\end{center}
 \caption{Nyquist plot of $\!G(j\omega)\!$ exiting the initial point $P_0$ by moving clockwise or counter-clockwise depending on the sign of parameter $\!\Delta_{\tau_k}$.\!}\label{partenze_G0_zero}
\end{figure}

Note that the sign of the parameter $\Delta_{\tau_{k}}$ used in (\ref{varphiwi}) determines if the Nyquist plot of the frequency response function $G(j\omega)$ exits the initial point $P_0=M_0\, e^{j\varphi_0}$ defined in \eqref{M_varphi_0_infty}
by moving clockwise or counter-clockwise in the complex plane, as shown in Fig.~\ref{partenze_G0_zero}. Equivalently,
the sign of the parameter $\Delta_{\tau_{k}}$ used in (\ref{varphiwi}) determines if the Nyquist plot of the frequency response function $\Gt(j\omega)$ exits the initial point $\PtO=\MtO\, e^{j\varphitO}$ defined in \eqref{M_varphi_0_infty}
by moving clockwise or counter-clockwise in the complex plane.

%

\vspace{2mm}
\noindent
\begin{Def} \label{Def_1}
{\it With reference to the transfer function $G(s)$ defined in (\ref{Gs_base}), let $\Delta_{ij}$ denote the following determinant:
 \begin{equation}
  \label{Delta_ij_def}
\Delta_{ij} =
\left|\begin{array}{cc}
a_i & a_j \\
b_i & b_j \\
\end{array}\right| = a_ib_j-a_jb_i,
 \end{equation}
for $i,j\in\{0,\,1,\,2,\,\ldots,\,n\}$. }
\\[2mm]
\end{Def}
\vspace{2mm}

\begin{Prop} \label{Prop_Gi_odd} {\it
All the odd coefficients $G_{k}$ of the series expansion (\ref{Gs}), for  $k\in\{1,\,3,\,5,\,\ldots,\,\infty\}$,  can be computed recursively using the following formula:
 \begin{equation}
  \label{G_k_odd}
 G_{k} = \frac{1}{b_0^2}\left(\nabla_{k}
 -\sum_{h=1}^\frac{k-1}{2} (-1)^{h}
 G_{k-2h}B_{h}\right),
\end{equation}
where the coefficients $\nabla_{k}$ are function of the determinants $\Delta_{ij}$ defined in (\ref{Delta_ij_def}):
\begin{equation}
  \label{nabla_k}
 \nabla_{k} = \sum_{j=0}^{\frac{k-1}{2}} (-1)^j\Delta_{k-j,j},
\end{equation}
and $B_{h}$ are polynomials of the coefficients $b_k$ in \eqref{Gs_base}:
\begin{equation}
  \label{B_h}
B_{h}=b_h^2 +2\sum_{j=1}^h (-1)^j b_{h-j} b_{h+j}.
\end{equation}
}
\end{Prop}
\vspace{3mm}
Property~\ref{Prop_Gi_odd} can be verified symbolically in Matlab:  by using the {\tt Symbolic Math Toolbox}, it can be verified that the coefficients $G_k$ in \eqref{G_k_odd}, for odd values of $k$, are equal to the coefficients $G_k$ defined in \eqref{Gs}.
\\[1mm]
From \eqref{Delta_ij_def} and \eqref{nabla_k}, it can be observed that the coefficients $\nabla_{k}$ are function only of the parameters $a_i$ and $b_i$ of $G(s)$ in (\ref{Gs_base}) characterized by indexes $i\leq k$.

\vspace{2mm}
\begin{Theo} \label{Prop_napla_all} {\it
The sign of the phase $\varphi_0(\omega)$ of function $G(j\omega)$ when $\omega\simeq 0^+$ can be determined as follows:
 \begin{equation}
  \label{Sign_varphi_w}
 \mbox{sign}(\varphi_0(\omega))
  = \mbox{sign}(\Delta_{\tau_{k}}),
\end{equation}
where $\Delta_{\tau_{k}}$ is the first non-zero element of the series of odd coefficients $\Delta_{\tau_{k}}$ for increasing values of the odd index  $k\in\{1,\,3,\,5,\,\ldots,\,\infty\}$. Furthermore, the coefficients $\Delta_{\tau_{k}}$ in (\ref{Sign_varphi_w}) can also be expressed as follows:
 \begin{equation}
  \label{New_Delta_tau}
\Delta_{\tau_{k}}
 = (-1)^\frac{k-1}{2}\frac{\nabla_{k}}{a_0 b_0}.
\end{equation}
}
\end{Theo}
\vspace{1mm}
{\it Proof.} The first equality $\mbox{sign}(\varphi(\omega))  = \mbox{sign}(\Delta_{\tau_{k}})$ in (\ref{Sign_varphi_w}) comes directly from  (\ref{varphiwi}) in Property~\ref{Prop_Delta_tau} and from the Definition~\ref{Def_Delta_tau} of $\Delta_{\tau_k}$ in (\ref{Delta_tau_i}).
Then, from (\ref{G_k_odd}) it follows that when  $G_1=G_3=G_5=\cdots=G_{k-2}=0$, the coefficients $G_{k}$ can be expressed as follows:
 \begin{equation}
  \label{G_2ip1}
 G_{k} = \frac{\nabla_{k}}{b_0^2}.
\end{equation}
Therefore, substituting (\ref{G_2ip1})  in (\ref{Delta_tau_i}) and recalling the static gain $G_0=a_0/b_0$ from \eqref{M_varphi_0_infty}, yield the expression of $\Delta_{\tau_k}$ in \eqref{New_Delta_tau}.
\vspace{2mm}

\subsection{Behavior of the Modulus $M_0(\omega)$ when $\omega\simeq 0^+$}

\begin{Theo} \label{Prop_Mod_Gjw} {\it
The modulus
$M_0(\omega)$
of function $G(j\omega)$ when $\omega\simeq 0^+$ increases or decreases as follows:
 \begin{equation}
  \label{Modulus_Gjw}
\begin{array}{cc}
M_0(\omega) \;\mbox{increases if} \;\Delta_{\tau_k}\!>\!0, \\[2mm]
M_0(\omega) \;\mbox{decreases if}\; \Delta_{\tau_k}\!<\!0, \end{array}
 \end{equation}
where $\Delta_{\tau_{k}}$ is the first non-zero element of the series of even coefficients $\Delta_{\tau_{k}}$ for increasing values of the even index  $k\in\{0,\,2,\,4,\,\ldots,\,\infty\}$.}
\end{Theo}
\vspace{1mm}
{\it Proof.}
When $\omega\simeq 0^+$, from  \eqref{Gjw} it follows that the modulus of $G(j\omega)$ can be approximated as follows:
 \begin{equation}
  \label{Mod_Gjw}
 \begin{array}{@{}r@{\,}c@{\,}l}
 M_0(\omega) &\simeq&
 |G_{\!ev}(\omega)|
 \!\simeq\!
 |G_{\!0}| \left|1 - \frac{G_2}{G_0} \omega^2  + \frac{G_4}{G_0}  \omega^4 \!\!-\! \frac{G_6}{G_0} \omega^6 +\ldots\right|\,
\end{array}\!,
 \end{equation}
  where $G_{ev}(\omega)$ is the real part of  $G(j\omega)$. Note that the  modulus $M_0(\omega)$ of point $G(j\omega)$ when $\omega\simeq 0^+$
contains only the even coefficients $G_k$, for  $k\in\{0,\,2,\,4,\,\ldots,\,\infty\}$, of the series expansion (\ref{Gs}). From (\ref{Mod_Gjw}), it can be observed that when $\omega\simeq 0^+$ the  modulus $|G(j\omega)|$ increases or decreases depending on the sign of the
coefficients  $- \frac{G_2}{G_0}$, $\frac{G_4}{G_0}$, $-\frac{G_6}{G_0}$,  etc. Replacing the Definition~\ref{Def_Delta_tau} of $\Delta_{\tau_k}$ from (\ref{Delta_tau_i}) in \eqref{Mod_Gjw} yields \eqref{Modulus_Gjw}.
\vspace{1mm}

\noindent The recursive formula given in (\ref{G_k_odd}) can only be used for odd values of index $k$. The coefficients $G_k$ for the calculation of the even coefficients $\Delta_{\tau_k}$ in \eqref{Modulus_Gjw}
 can instead be computed by using the recursive formula (\ref{G_k_all}) given in the following Property~\ref{Prop_Gi_all}.
\vspace{2mm}

\begin{Prop} \label{Prop_Gi_all} {\it
All the coefficients $G_k$ present in the series expansion (\ref{Gs}), for $k\geq 0$,  can also be computed recursively using the following formula:
 \begin{equation}
  \label{G_k_all}
 G_k = \frac{1}{b_0^2}\left(\Delta_{k0}-b_0\sum_{j=1}^{k-1} G_{k-j}b_j\right).
\end{equation}
}
\end{Prop}
\vspace{2mm}
Property~\ref{Prop_Gi_all} can be verified symbolically in Matlab:  by using the {\tt Symbolic Math Toolbox}, it can be verified that the coefficients $G_k$ in \eqref{G_k_all},  for both even and odd values of $k$, are equal to the coefficients $G_k$ defined in \eqref{Gs}.
\vspace{2mm}


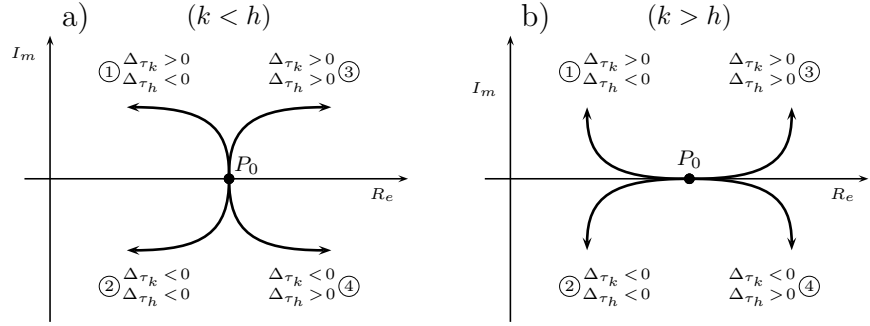
\begin{figure}[tb]
\begin{center}
 \setlength{\unitlength}{6.8mm}
 \psset{unit=\unitlength}
 \psset{yunit=0.7\unitlength}
 \SpecialCoor
 \begin{pspicture}(1,-3.5)(17,5.68)
\rput(2,4.5){a)}
\rput(11,4.5){b)}
\rput(5,4.5){\footnotesize $(k<h)$}
\rput(14,4.5){\footnotesize $(k>h)$}
%
 \psline[linewidth=0.6pt]{->}(1,0)(8.5,0)
 \rput[t](8,-0.25){\tiny $R_e$}
 \psline[linewidth=0.6pt]{->}(1.5,-4)(1.5,4)
 \rput[r](1.25,3.5){\tiny $I_m$}
  \psbezier[linewidth=1.0pt]{*->}(5,0)(5,1.5)(4.5,2)(3,2)
  \psbezier[linewidth=1.0pt]{*->}(5,0)(5,1.5)(5.5,2)(7,2)
  \psbezier[linewidth=1.0pt]{->}(5,0)(5,-1.5)(4.5,-2)(3,-2)
  \psbezier[linewidth=1.0pt]{->}(5,0)(5,-1.5)(5.5,-2)(7,-2)
 \rput[r](4.5,3){\tiny  \cput[linewidth=0.3pt,framesep=1pt](0,0.1){\tiny 1}
 $\begin{array}{r}
       \Delta_{\tau_k}\!>\!0 \\ \Delta_{\tau_h}\!<\!0
 \end{array}$}
 \rput[r](4.5,-3){\tiny \cput[linewidth=0.3pt,framesep=1pt](0,0.1){\tiny 2}
 $\begin{array}{r}
       \Delta_{\tau_k}\!<\!0 \\ \Delta_{\tau_h}\!<\!0
 \end{array}$}
 \rput[l](5.5,3){\tiny
 $\begin{array}{l}
       \Delta_{\tau_k}\!>\!0 \\ \Delta_{\tau_h}\!>\!0
 \end{array}$ \cput[linewidth=0.3pt,framesep=1pt](0,0.1){\tiny 3}
 }
 \rput[l](5.5,-3){\tiny
 $\begin{array}{l}
       \Delta_{\tau_k}\!<\!0 \\ \Delta_{\tau_h}\!>\!0
 \end{array}$ \cput[linewidth=0.3pt,framesep=1pt](0,0.1){\tiny 4}
 }
  \rput(5,0){\punto}
 \rput[lb](5.1;1){\scriptsize $P_0$}
 %
 %
\rput(9,0){
  \psline[linewidth=0.6pt]{->}(1,0)(8.5,0)
 \rput[t](8,-0.25){\tiny $R_e$}
 \psline[linewidth=0.6pt]{->}(1.5,-4)(1.5,4)
 \rput[r](1.25,2.5){\tiny $I_m$}
  \psbezier[linewidth=1.0pt]{*->}(5,0)(3.5,0)(3,0.5)(3,2)
  \psbezier[linewidth=1.0pt]{*->}(5,0)(6.5,0)(7,0.5)(7,2)
  \psbezier[linewidth=1.0pt]{->}(5,0)(3.5,0)(3,-0.5)(3,-2)
  \psbezier[linewidth=1.0pt]{->}(5,0)(6.5,0)(7,-0.5)(7,-2)
 \rput[r](4.5,3){\tiny  \cput[linewidth=0.3pt,framesep=1pt](0,0.1){\tiny 1}
 $\begin{array}{r}
       \Delta_{\tau_k}\!>\!0 \\ \Delta_{\tau_h}\!<\!0
 \end{array}$}
 \rput[r](4.5,-3){\tiny \cput[linewidth=0.3pt,framesep=1pt](0,0.1){\tiny 2}
 $\begin{array}{r}
       \Delta_{\tau_k}\!<\!0 \\ \Delta_{\tau_h}\!<\!0
 \end{array}$}
 \rput[l](5.5,3){\tiny
 $\begin{array}{l}
       \Delta_{\tau_k}\!>\!0 \\ \Delta_{\tau_h}\!>\!0
 \end{array}$ \cput[linewidth=0.3pt,framesep=1pt](0,0.1){\tiny 3}
 }
 \rput[l](5.5,-3){\tiny
 $\begin{array}{l}
       \Delta_{\tau_k}\!<\!0 \\ \Delta_{\tau_h}\!>\!0
 \end{array}$ \cput[linewidth=0.3pt,framesep=1pt](0,0.1){\tiny 4}
 }
  \rput(5,0){\punto}
 \rput[b](5,0.3){\scriptsize $P_0$}
 }
 %
 %
 \end{pspicture}
\end{center}
\caption{Different types of Nyquist plots when exiting the initial point $P_0$.}\label{partenze_G0}
\end{figure}

\subsection{Exiting the initial point $P_0$  when $\omega\simeq 0^+$}
\label{Getting_out}

Theorem~\ref{Prop_napla_all} and Theorem~\ref{Prop_Mod_Gjw} have shown that the odd coefficients $G_1$,  $G_3$,  $G_5$,  $\ldots$ and the even coefficients $G_2$,  $G_4$,  $G_6$,  $\ldots$
in \eqref{Gs}
can be used to determine the sign of the phase $\varphi_0(\omega)$ and whether the modulus $M_0(\omega)$ increases or the decreases when $\omega\simeq 0^+$.
The combined results of Theorem~\ref{Prop_napla_all} and Theorem~\ref{Prop_Mod_Gjw} allow to determine how the frequency response function $G(j\omega)$ exits the initial point $P_0$ when $\omega\simeq 0^+$.
Specifically, the following two Lemmas originate from Theorem~\ref{Prop_napla_all}.

\vspace{1mm}\noindent \begin{Lemm}\label{Th1_Lm1}{\it
When $\omega\simeq 0^+$ and the first non-zero odd parameter $\Delta_{\tau_k}$ has an index $k<h$, where $h$ is the index of the first not-zero even parameter $\Delta_{\tau_h}$, the function $G(j\omega)$ exits the initial point $P_0$ by
moving in the complex plane along a direction {\rm perpendicular} to the real axis. The  possible four Nyquist curves that can be obtained  are shown in Fig.~\ref{partenze_G0}(a).}
\end{Lemm}
\vspace{1mm}
{\it Proof.} From (\ref{Gjw}) and \eqref{Delta_tau_i}, if $\omega\simeq 0^+$,  $\Delta_{\tau_k}\not=0$ is the first non-zero odd parameter, and $\Delta_{\tau_h}\not=0$ is the first non-zero even parameter, it follows that function $G(j\omega)$ can be expressed as:
\begin{equation}
  \label{Gjw_caso_uno}
\begin{array}{@{}r@{\,}c@{\,}l@{}}
 G(j\omega)
  &\simeq & \ds
    G_0 - G_h \omega^h  + j    G_k \omega^k
  \simeq
    G_0( 1+ \Delta_{\tau_h} \omega^h  + j    \Delta_{\tau_k} \omega^k).
    \\[1mm]
\end{array}
 \end{equation}
From \eqref{Gjw_caso_uno} it follows that, when $\omega\simeq 0^+$, the function $G(j\omega)$ exits the initial point $P_0$ by drawing a Nyquist curve that moves along a direction which is {\it perpendicular} to the real axis because the infinitesimal $\Delta_{\tau_k} \, \omega^k$ is larger than the infinitesimal $\Delta_{\tau_h} \omega^h$, since $k<h$. The four possible  Nyquist curves shown in Fig.~\ref{partenze_G0}(a) depend on the positive or negative values of the two parameters $\Delta_{\tau_k}$ and $\Delta_{\tau_h}$.

\vspace{1mm}

\vspace{1mm}\noindent \begin{Lemm}\label{Th1_Lm2}{\it
When $\omega\simeq 0^+$ and the first non-zero odd parameter $\Delta_{\tau_k}$ has an index $k>h$, where $h$ is the index of the first not-zero even parameter $\Delta_{\tau_h}$, the function $G(j\omega)$ exits the initial point $P_0$ by drawing a Nyquist curve that moves along a direction {\rm parallel} to the real axis. The Nyquist curves that can be obtained in this case  are shown in Fig.~\ref{partenze_G0}(b).
}
\end{Lemm}
\vspace{1mm}
{\it Proof.} From (\ref{Gjw}) and \eqref{Delta_tau_i}, when $\omega\simeq 0^+$,  $\Delta_{\tau_k}\not=0$ is the first non-zero odd parameter, and $\Delta_{\tau_h}\not=0$ is the first non-zero even parameter, it follows that function $G(j\omega)$ can be expressed as in \eqref{Gjw_caso_uno}. In this case, from \eqref{Gjw_caso_uno} it follows that, when $\omega\simeq 0^+$, the function $G(j\omega)$ exits the initial point $P_0$ by drawing a Nyquist curve that moves along a direction which is {\it parallel} to the real axis because the infinitesimal $\Delta_{\tau_k} \, \omega^k$ is smaller than the infinitesimal $\Delta_{\tau_h} \omega^h$, since $k>h$.  In this case, the four possible Nyquist curves are shown in Fig.~\ref{partenze_G0}(b).

 \section{Frequency response of  $G(s)$ at $s\simeq \infty$}\label{Sec_Freq_Res_s_oo}

Substituting $\s=\frac{1}{s}$ in (\ref{eq:TransferFunction}) yields:
\begin{equation}
  \label{Gt_di_p}
  \Gt(s)
  = \ds
   \frac{K}{s^r}\, \G(s),  \hspace{8mm} \mbox{where} \hspace{8mm} r=h+n-m
 \end{equation}
 is the relative degree of function $\Gt(s)$ and function $\G(s)$ is:
\begin{equation}
  \label{G_di_p}
\begin{array}{@{}r@{\,}c@{\,}l@{}}
  \G(s)
  &=& \ds
  \frac{(a_0\s^m+ a_{1}\s^{m-1} + \ldots+ a_{m-2}\s^2  + a_{m-1}\s+ a_m )}
       {(b_0\s^n+ b_{1}\s^{n-1} + \ldots+ b_{n-2}\s^2  + b_{n-1}\s+ b_n )}.
 \end{array}
 \end{equation}
Function $\G(s)$ has a similar structure as function $G(s)$ in (\ref{Gs_base}), in which the coefficients $a_i$ and $b_i$ are in reversed order. 
\vspace{2mm}
\noindent
\begin{Def} \label{Def_a_b_bar}
{\it
 Let $\a_i$ and $\b_j$ denote the ``dual version'' of coefficients $a_i$ and $b_j$, defined as follows:
 \begin{equation}
 \label{Prop_a_b_bar}
\a_i = a_{m-i}
\hspace{16mm}\mbox{and}\hspace{16mm}
\b_j = b_{n-j},
 \end{equation}
 for  $i\!\in\!\{1,2,\ldots,m\}$ and  $j\!\in\! \{1,2,\ldots,n\}$. From (\ref{Prop_a_b_bar}), it follows:
\begin{equation}
  \label{prop_a_b_conj}
a_i = \a_{m-i}
\hspace{16mm}\mbox{and}\hspace{16mm}
b_j = \b_{n-j}.
 \end{equation}
}
\end{Def}
\vspace{2mm}
Substituting (\ref{prop_a_b_conj}) in (\ref{G_di_p}) yields:
\begin{equation}
  \label{G_di_p_dual_form}
  \G(s)
 =
   \frac{(\a_m\,\s^m+ \a_{m-1}\,\s^{m-1} + \ldots+ \a_{2}\,\s^2  + \a_{1}\,\s+ \a_0)}
       {(\b_n\,\s^n+ \b_{n-1}\,\s^{n-1} + \ldots+ \b_{2}\,\s^2  + \b_{1}\,\s+ \b_0 )},
 \end{equation}
 having now the same structure as function $G(s)$ in (\ref{Gs_base}) but using
 the dual coefficients $\a_i$ and $\b_j$ and the 
 new variable $\s$.

The frequency response of function $G(s)$ at $s\simeq \infty$ can be studied using the Taylor series expansion of
$\G(s)$
at $\s = 0$:
 \begin{equation}
  \label{Gs_bar}
 \G(s)
  = \sum_{k=0}^\infty   \G_k \,\s^k =
 \G_0
 + \G_1 \s
 + \G_2 \s^2
  + \ldots,
 \end{equation}
where the coefficients $\G_k$ in (\ref{Gs_bar}) are the dual version of the coefficients $G_k$ in \eqref{Gs} using (\ref{Prop_a_b_bar}).
Equivalently, also the coefficients $\Delta_{\tau_k}$ in (\ref{Delta_tau_i}),  $\Delta_{ij}$ in (\ref{Delta_ij_def}), $\nabla_k$ in (\ref{nabla_k}) and  $B_h$ in (\ref{B_h})  can be dualized into  $\oDelta_{\tau_k}$,  $\oDelta_{ij}$, $\onabla_k$ and  $\oB_h$ using (\ref{Prop_a_b_bar}). In particular, the dual determinant $\oDelta_{ij}$ is defined as follows:
 \begin{equation}
  \label{dual_Delta_ij_def}
\oDelta_{ij} =
\left|\begin{array}{cc}
a_{m-i} & a_{m-j} \\
b_{n-i} & b_{n-j} \\
\end{array}\right| = a_{m-i} b_{n-j}-a_{m-j} b_{n-i}.
 \end{equation}
 Note that $\G_0=\frac{a_m}{b_n}$ in \eqref{Gs_bar} is equal to $G_\infty$ defined in \eqref{eq:K1Definition}.
\vspace{2mm}
\begin{Theo} \label{Prop_napla_all_oo} {\it
(Dual of Theorem~\ref{Prop_napla_all}) The sign of the phase $\varphi_\infty(\omega)$ of function 
$G(j\omega)$
when $\omega\simeq \infty$ can be determined as follows:
 \begin{equation}
  \label{Sign_varphi_w_oo}
 \mbox{\rm sign}(\varphi_\infty(\omega))
  = \mbox{\rm sign}(-\oDelta_{\tau_{k}}),
\end{equation}
where $\oDelta_{\tau_{k}}$ is the first non-zero element of the series of odd coefficients $\oDelta_{\tau_{k}}$, for increasing values of the odd index  $k\in\{1,\,3,\,5,\,\ldots,\,\infty\}$.
Furthermore, the coefficient $\oDelta_{\tau_{k}}$ in (\ref{Sign_varphi_w_oo}) can also be expressed as follows:
 \begin{equation}
  \label{New_Delta_tau_oo}
\oDelta_{\tau_{k}}
 = (-1)^{\left\lfloor\frac{k}{2}\right\rfloor}\frac{\onabla_{k}}{a_m b_n}.
\end{equation}
}
\end{Theo}
{\it Proof.} The frequency response $\G(j\omega)$ of function $\G(s)$ in    \eqref{Gs_bar} when $\omega\simeq \infty$ can be expressed as follows:
 \begin{equation}
  \label{Gjw_oo}
\begin{array}{@{\!\!}r@{\,}c@{\,}l@{}}
 \G(j\omega)
  &=& \ds
 \sum_{k=0}^\infty \G_k \,\frac{1}{(j\omega)^k}
  = \sum_{k=0}^\infty \G_k \,\frac{(-j)^k}{\omega^k}
  \\[8mm]
  &=& \ds
    \G_0 \!-\!  \frac{\G_2}{\omega^{2}}  \!+\!   \frac{\G_4}{\omega^{4}}
    +\ldots
  -\frac{j}{\omega}
  \left[
    \G_1 \!-\!  \frac{\G_3}{\omega^{2}}  \!+\!   \frac{\G_5}{\omega^{4}}
    \!+\!\ldots
  \right].
\end{array}
 \end{equation}
When $\omega\simeq \infty$ and $\G_1=\G_3=\G_5=\cdots=\G_{k-2}=0$, the phase $\varphi_\infty(\omega)$ of $\G(j\omega)$ in (\ref{Gjw_oo}) and, equivalently, of $G(j\omega)$, can be approximated as follows:
 \begin{equation}
  \label{varphiwi_oo}
 \varphi_\infty(\omega)
   \!\simeq\!
  \arctan\!\left[-\frac{ (-1)^{\left\lfloor\frac{k}{2}\right\rfloor} \G_{k}  }
  {\G_0 \,\omega^{k}}\right]
 \!\simeq\!
 \frac{-\oDelta_{\tau_{k}}}{\omega^{k}},
 \end{equation}
 from which equality (\ref{Sign_varphi_w_oo}) follows. Finally, the expression (\ref{New_Delta_tau_oo}) is the dual version of
(\ref{New_Delta_tau}) in Theorem~\ref{Prop_napla_all}.
\vspace{2mm}

\begin{Theo} \label{Prop_Mod_Gjw_oo} {\it
(Dual of Theorem~\ref{Prop_Mod_Gjw}) The modulus $M_\infty(\omega)$ of function $G(j\omega)$ when $\omega\simeq \infty$ increases or decreases as follows:
 \begin{equation}
  \label{Modulus_Gjw_oo}
\begin{array}{cc}
M_\infty(\omega) \;\;\;\mbox{\rm increases if} \;\;\;\oDelta_{\tau_k}\!>\!0, \\[2mm]
M_\infty(\omega) \;\;\;\mbox{\rm decreases if}\;\;\; \oDelta_{\tau_k}\!<\!0, \end{array}
 \end{equation}
where $\oDelta_{\tau_{k}}$ is the first non-zero element of the series of even coefficients $\oDelta_{\tau_{k}}$ for increasing values of the even index  $k\in\{0,\,2,\,4,\,\ldots,\,\infty\}$.}

\end{Theo}
\vspace{1mm}
{\it Proof.}
 When $\omega\simeq \infty$,
  from (\ref{Gjw_oo}) it follows that $|\G(j\omega)|$ and, equivalently, $M_\infty(\omega)$ can be expressed as:
  \begin{equation}
  \label{Mod_Gjw_oo}  \ts
M_{\infty}(\omega)= |\G(j\omega)|
 \simeq
 |\G_0| \left|1 - \frac{\G_2}{\G_0\omega^2}   + \frac{\G_4}{\G_0 \omega^4}   - \frac{\G_6}{\G_0\omega^6}  +\ldots\right|,
 \end{equation}
 which is the dual of $M_0(\omega)$ in (\ref{Mod_Gjw}). Therefore, the same proof as Theorem~\ref{Prop_Mod_Gjw} applies here using the dual coefficients
 $\oDelta_{\tau_k}$.

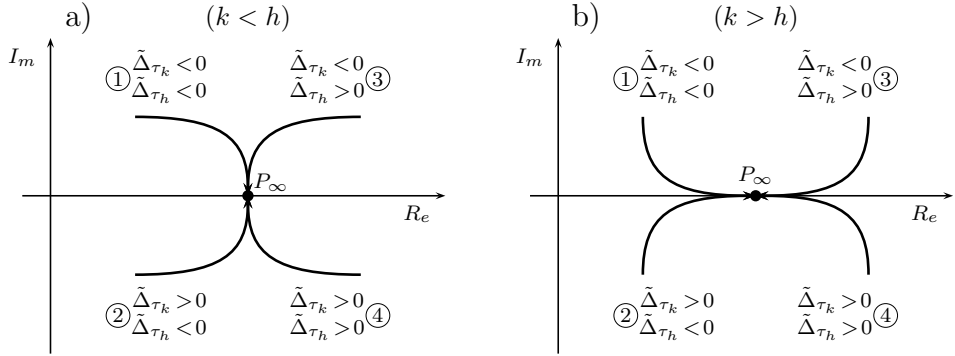
\begin{figure}[tb]
\begin{center}
 \setlength{\unitlength}{7.5mm}
 \psset{unit=\unitlength}
 \psset{yunit=0.7\unitlength}
 \SpecialCoor
 \begin{pspicture}(1,-3.5)(17,5.68)
\rput(2,4.5){a)}
\rput(11,4.5){b)}
\rput(5,4.5){\footnotesize $(k<h)$}
\rput(14,4.5){\footnotesize $(k>h)$}
%
\psline[linewidth=0.6pt]{->}(1,0)(8.5,0)
 \rput[t](8,-0.25){\scriptsize $R_e$}
 \psline[linewidth=0.6pt]{->}(1.5,-4)(1.5,4)
 \rput[r](1.25,3.5){\scriptsize $I_m$}
  \psbezier[linewidth=1.0pt]{<-}(5,0)(5,1.5)(4.5,2)(3,2)
  \psbezier[linewidth=1.0pt]{<-}(5,0)(5,1.5)(5.5,2)(7,2)
  \psbezier[linewidth=1.0pt]{<-}(5,0)(5,-1.5)(4.5,-2)(3,-2)
  \psbezier[linewidth=1.0pt]{<-}(5,0)(5,-1.5)(5.5,-2)(7,-2)
 \rput[r](4.5,3){\scriptsize  \cput[linewidth=0.3pt,framesep=1pt](0,0.1){\scriptsize 1}
 $\begin{array}{r}
       \oDelta_{\tau_k}\!<\!0 \\ \oDelta_{\tau_h}\!<\!0
 \end{array}$}
 \rput[r](4.5,-3){\scriptsize \cput[linewidth=0.3pt,framesep=1pt](0,0.1){\scriptsize 2}
 $\begin{array}{r}
       \oDelta_{\tau_k}\!>\!0 \\ \oDelta_{\tau_h}\!<\!0
 \end{array}$}
 \rput[l](5.5,3){\scriptsize
 $\begin{array}{l}
       \oDelta_{\tau_k}\!<\!0 \\ \oDelta_{\tau_h}\!>\!0
 \end{array}$ \cput[linewidth=0.3pt,framesep=1pt](0,0.1){\scriptsize 3}
 }
 \rput[l](5.5,-3){\scriptsize
 $\begin{array}{l}
       \oDelta_{\tau_k}\!>\!0 \\ \oDelta_{\tau_h}\!>\!0
 \end{array}$ \cput[linewidth=0.3pt,framesep=1pt](0,0.1){\scriptsize 4}
 }
  \rput(5,0){\punto}
 \rput[lb](5.1;1){\scriptsize $P_\infty$}
 %
 \rput(9,0){
\psline[linewidth=0.6pt]{->}(1,0)(8.5,0)
 \rput[t](8,-0.25){\scriptsize $R_e$}
 \psline[linewidth=0.6pt]{->}(1.5,-4)(1.5,4)
 \rput[r](1.25,3.5){\scriptsize $I_m$}
  \psbezier[linewidth=1.0pt]{<-}(5,0)(3.5,0)(3,0.5)(3,2)
  \psbezier[linewidth=1.0pt]{<-}(5,0)(6.5,0)(7,0.5)(7,2)
  \psbezier[linewidth=1.0pt]{<-}(5,0)(3.5,0)(3,-0.5)(3,-2)
  \psbezier[linewidth=1.0pt]{<-}(5,0)(6.5,0)(7,-0.5)(7,-2)
 \rput[r](4.5,3){\scriptsize  \cput[linewidth=0.3pt,framesep=1pt](0,0.1){\scriptsize 1}
 $\begin{array}{r}
       \oDelta_{\tau_k}\!<\!0 \\ \oDelta_{\tau_h}\!<\!0
 \end{array}$}
 \rput[r](4.5,-3){\scriptsize \cput[linewidth=0.3pt,framesep=1pt](0,0.1){\scriptsize 2}
 $\begin{array}{r}
       \oDelta_{\tau_k}\!>\!0 \\ \oDelta_{\tau_h}\!<\!0
 \end{array}$}
 \rput[l](5.5,3){\scriptsize
 $\begin{array}{l}
       \oDelta_{\tau_k}\!<\!0 \\ \oDelta_{\tau_h}\!>\!0
 \end{array}$ \cput[linewidth=0.3pt,framesep=1pt](0,0.1){\scriptsize 3}
 }
 \rput[l](5.5,-3){\scriptsize
 $\begin{array}{l}
       \oDelta_{\tau_k}\!>\!0 \\ \oDelta_{\tau_h}\!>\!0
 \end{array}$ \cput[linewidth=0.3pt,framesep=1pt](0,0.1){\scriptsize 4}
 }
  \rput(5,0){\punto}
 \rput[b](5,0.3){\scriptsize $P_\infty$}
 }
 %
 %
 \end{pspicture}
\end{center}
\caption{$\!$Different types of Nyquist plots when entering the final point $P_{\!\infty}$.}\label{arrivi_Goo}
\end{figure}

\subsection{Entering the final point $P_{\infty}$  when $\omega\simeq \infty$}
\label{Getting_in}

Following the same outline of Sec.~\ref{Getting_out}, the dual of Lemma~\ref{Th1_Lm1} and Lemma~\ref{Th1_Lm2} originate from Theorem~\ref{Prop_napla_all_oo}.
\vspace{1mm}\noindent \begin{Lemm}\label{Th3_Lm1}{\it
(Dual of Lemma~\ref{Th1_Lm1}) When $\omega\simeq \infty$ and the first non-zero odd parameter $\oDelta_{\tau_k}$ has an index $k<h$, where $h$ is the index of the first not-zero even parameter $\oDelta_{\tau_h}$, the function $G(j\omega)$ enters the final point $P_\infty$ by drawing a Nyquist curve that moves along a direction {\rm perpendicular} to the real axis. }
\end{Lemm}
\vspace{1mm}
\noindent From Lemma \ref{Th3_Lm1} and Theorem \ref{Prop_Mod_Gjw_oo} it follows that, when $\omega\simeq \infty$ and $k<h$, the function $G(j\omega)$ enters the final point $P_\infty$ by drawing one of the four Nyquist curves shown in Fig.~\ref{arrivi_Goo}(a).
\vspace{1mm}\noindent \begin{Lemm}\label{Th3_Lm2}{\it (Dual of Lemma~\ref{Th1_Lm2})
When $\omega\simeq \infty$ and the first non-zero odd parameter $\oDelta_{\tau_k}$ has an index $k>h$, where $h$ is the index of the first not-zero even parameter $\oDelta_{\tau_h}$, the function $G(j\omega)$ enters the final point $P_\infty$ by drawing a Nyquist curve that moves along a direction {\rm parallel} to the real axis. }
\end{Lemm}
\vspace{1mm}
\noindent From Lemma \ref{Th3_Lm2} and Theorem \ref{Prop_Mod_Gjw} it follows that, when $\omega\simeq \infty$ and $k>h$, the function $G(j\omega)$ enters the final point $P_\infty$ by drawing one of the four Nyquist curves shown in Fig.~\ref{arrivi_Goo}(b).

 \section{Case studies}\label{Case_studies_sect}

 \begin{figure}[tb]
\centering
\setlength{\unitlength}{5.0mm}
 \psset{unit=\unitlength}
 \psset{yunit=0.7\unitlength}
 \SpecialCoor
 \begin{pspicture}(1,-12)(20,8.5)
 %
\psfrag{Im}[][][0.9][0]{Im}
\psfrag{Re}[][][0.9][0]{Re}
\psfrag{Poo}[b][b][0.9][0]{$\Ptoo$}
\psfrag{P0}[bl][tl][0.8][0]{\rput(0.5,1.25){$\PtO$}}
 \psfrag{ph0}[][][0.9]{}
\psfrag{phoo}[][][0.9][0]{}
 \rput(10,0){ \includegraphics[clip,width=0.75\columnwidth]{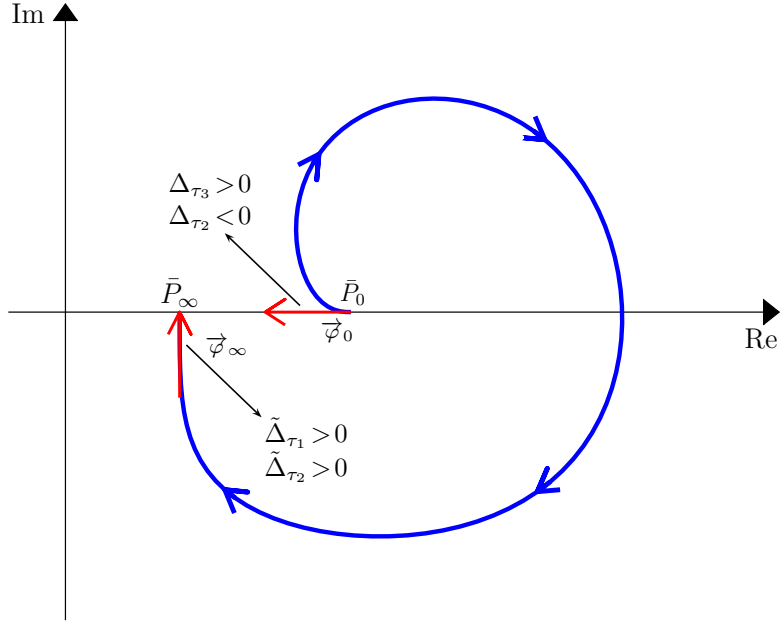}
}

\rput(8.5,-0.75){\scriptsize $\overrightarrow{
\varphi}_0$}
\rput(5.56,-1.3){\scriptsize $\overrightarrow{
\varphi}_\infty$}
 \psline[linewidth=0.6pt]{->}(7.5,0.25)(5.5,3)
 \rput(5.25,4.2){$\footnotesize
\begin{array}{c}
\Delta_{\tau_3}\!>\!0\\
 \Delta_{\tau_2}\!<\!0\\
\end{array}$
 }
 \psline[linewidth=0.6pt]{->}(4.48,-1.25)(6.48,-4)
 \rput(7.8,-5.3){$\footnotesize
\begin{array}{c}
\oDelta_{\tau_1}\!>\!0\\
 \oDelta_{\tau_2}\!>\!0\\
\end{array}$
 }
 %
 \end{pspicture}
    \caption{\label{Plot_Nyquist_Fig_1} Qualitative Nyquist plot of $G(j\omega)$ associated with $G(s)$ in \eqref{Gs_es_1}.}
\end{figure}

 \subsection{First case study}
 Reference is made to the following transfer function:
\begin{equation}\label{Gs_es_1}
\Gt(s)=
 \frac{2 s^3+6 s^2+2 s+1}{4 s^3+5 s^2+2 s+1}.
\end{equation}
The initial and final points $\PtO$ and $\Ptoo$ of the Nyquist plot of $\Gt(j\omega)$ are characterized by: $\MtO=1$, $\Mtoo=0.5$, $\varphitO=\varphitoo=0$.
 Applying Definition~\ref{Def_Delta_tau} and Properties~\ref{Prop_Gi_odd} and~\ref{Prop_Gi_all} yields:
\[
\begin{array}{c} \ds
 \Delta_{\tau_1}
 = \frac{G_1}{G_0}
 =0,
 \hspace{.5mm}
 \Delta_{\tau_2}
 = -\frac{G_2}{G_0}
 =-1,
 \hspace{.5mm}
 \Delta_{\tau_3}
 = -\frac{G_3}{G_0}
 =4, \\[6mm]
 \oDelta_{\tau_1} \ds
 = \frac{\G_1}{\G_0}
 =\frac{7}{4},
 \hspace{6.5mm}
 \oDelta_{\tau_2}
 = -\frac{\G_2}{\G_0}
 =\frac{27}{16}.
\end{array}
 \]
In order to obtain the qualitative Nyquist plot of the frequency response function $\Gt(j\omega)$, the procedure described in~\cite{questo_2} would fail in this case, since the parameter called $\Delta_\tau$, which is $\Delta_{\tau_1}$ in this letter, is equal to zero. Therefore, it would not be possible to discriminate the behavior of the phase $\varphi_0(\omega)$ when $\omega\simeq 0^+$, namely whether the Nyquist plot starts by leading or lagging compared to the initial phase $\varphitO$. The detailed frequency analysis performed in this letter allows to still be able to discriminate the behavior of the phase $\varphi_0(\omega)$ when $\omega\simeq 0^+$ thanks to Theorem~\ref{Prop_napla_all}, and also allows to discriminate the behavior of the modulus $M_0(\omega)$ when $\omega\simeq 0^+$ thanks to Theorem~\ref{Prop_Mod_Gjw}.
Applying Theorem~\ref{Prop_napla_all} and Theorem~\ref{Prop_Mod_Gjw}, since $\Delta_{\tau_3}\!>\!0$, $\Delta_{\tau_2}\!<\!0$ (case $k\!>\!h$), the Nyquist plot exits the initial point $\PtO$ as the curve
\;\cput[linewidth=0.3pt,framesep=1pt](0,0.1){\tiny 1}\; in Fig.~\ref{partenze_G0}(b), as shown in Fig.~\ref{Plot_Nyquist_Fig_1}.
 Applying Theorem~\ref{Prop_napla_all_oo} and Theorem~\ref{Prop_Mod_Gjw_oo}, since $\tilde{\Delta}_{\tau_1}\!>\!0$, $\tilde{\Delta}_{\tau_2}\!>\!0$ (case $k\!<\!h$), the Nyquist plot enters the final point $\Ptoo$ as the curve
 \;\cput[linewidth=0.3pt,framesep=1pt](0,0.1){\tiny 4}\; in Fig.~\ref{arrivi_Goo}(a), as shown in Fig.~\ref{Plot_Nyquist_Fig_1}. The qualitative behavior of the remaining part of the Nyquist plot when $\omega\neq 0^+$ and $\omega\neq \infty$ can be determined by calculating the intersections with the real axis using the Routh–Hurwitz criterion.
 The red arrows in Fig.~\ref{Plot_Nyquist_Fig_1} denote the following  vectors:
 $$
 \overrightarrow{
\varphi}_0=\left.\frac{d \Gt(j\omega)}{d\omega}\right|_{\omega \simeq 0^+}
\hspace{8mm}\mbox{and} \hspace{8mm}
\overrightarrow{
\varphi}_\infty=\left.\frac{d \Gt(j\omega)}{d\omega}\right|_{\omega \simeq \infty}
$$
indicating the direction tangential to the Nyquist plot at the initial and final points.

\begin{figure}[t]
\centering
\setlength{\unitlength}{5.0mm}
 \psset{unit=\unitlength}
 \psset{yunit=0.7\unitlength}
 \SpecialCoor
 \begin{pspicture}(1,-9.75)(20,10)
 %

\psfrag{DT}[b][b][1.0][0]{\scriptsize$\Delta_{\tau_1}\!<\!0$}
\psfrag{Dp}[rt][t][0.9][0]{ }
\psfrag{Im}[rt][t][0.9][0]{Im}
\psfrag{Re}[tr][b][0.9][0]{Re}
\psfrag{sa}[b][b][0.9][0]{$\sigma_a$}
 \psfrag{phi0}[r][c][0.9][0]{$\overrightarrow{\varphi}_{0}\;\;\;\;\;\;$}
\psfrag{phioo}[b][r][0.9]{$\!\!\!\!\!\!\!\!\!\!\!\overrightarrow{\varphi}_{\infty}$}
  \rput(10,0){
  \includegraphics[clip,width=0.75\columnwidth]
{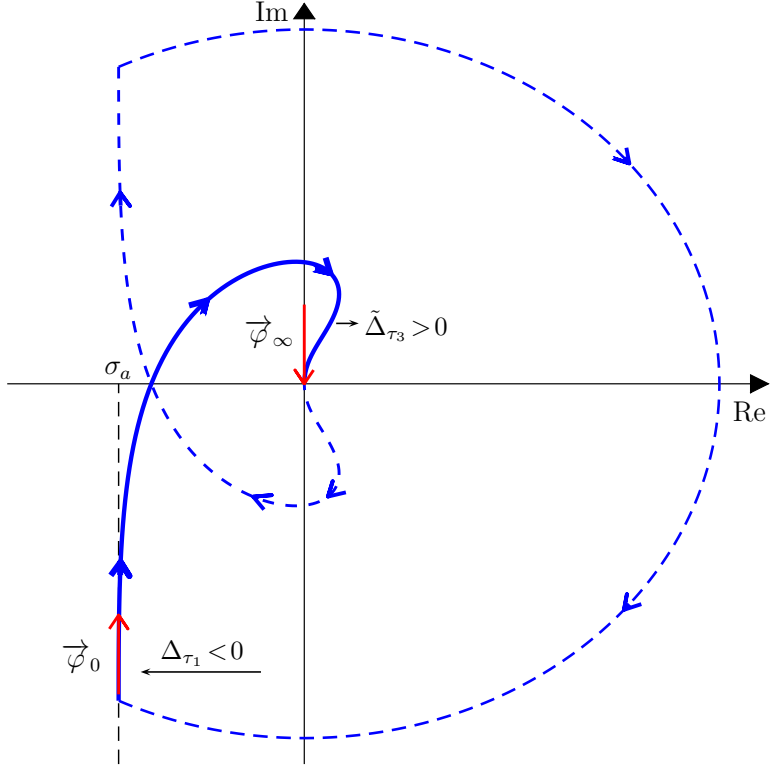}
}
  \psline[linewidth=0.6pt]{->}(6.5,-11.0)(3.3,-11.0)
  \rput(5.05,-10.3){\footnotesize$
 \Delta_{\tau_1}\!<\!0$
  }
 \psline[linewidth=0.6pt]{->}(8.5,2.25)(9.1,2.25)
 \rput[lt](9.25,2.75){\footnotesize$\oDelta_{\tau_3}\!>\!0$}
 %
 \end{pspicture}
 \vspace{15.8mm}
    \caption{Qualitative Nyquist plot of $G(j\omega)$ associated with $G(s)$ in \eqref{Gs_es_2}.}\label{Plot_Nyquist_Fig_2}
\end{figure}

 \subsection{Second case study}
Reference is made to the following transfer function:\begin{equation}\label{Gs_es_2}
\Gt(s)=
 \frac{s^2+12 s+35}
      {s(s^4+12 s^3+30 s^2+28 s+ 9)}.
\end{equation}
 Applying Definition~\ref{Def_Delta_tau} and Properties~\ref{Prop_Gi_odd} and~\ref{Prop_Gi_all} yields $\Delta_{\tau_1}=-2.76$, $\oDelta_{\tau_1}=0$,
 $\oDelta_{\tau_3}=88$. Applying Theorem~\ref{Prop_napla_all}, since $\Delta_{\tau_1}\!<\!0$, the Nyquist plot starts by lagging with respect to the initial phase $\varphitO$, as shown in Fig.~\ref{Plot_Nyquist_Fig_2}.
The procedure described in~\cite{questo_2} would fail to discriminate the phase of $\Gt(j\omega)$
when $\omega\simeq \infty$ since the parameter called $\Delta_p$, which is $\oDelta_{\tau_1}$ in this letter, is equal to zero.
The detailed frequency analysis performed in this letter allows to still be able to discriminate the behavior
thanks to Theorem~\ref{Prop_napla_all_oo}:
since $\tilde{\Delta}_{\tau_1}\!=\!0$ and $\tilde{\Delta}_{\tau_3}\!>\!0$, the Nyquist plot ends by lagging with respect to the final phase $\varphitoo$.
 The qualitative behavior of the remaining part of the Nyquist curve when $\omega\neq 0^+$ and $\omega\neq \infty$ can be determined by calculating the intersections with the real axis using the Routh–Hurwitz criterion and abscissa $\sigma_a$ of the asymptote shown in Fig.~\ref{Plot_Nyquist_Fig_2}.

\begin{figure}[t]
 %
 %
\centering
\psfrag{Diagramma di Nyquist}[b][b][1.0][0]{}
\psfrag{Imag}[][][0.9][0]{Im}
\psfrag{Real}[b][b][0.9][0]{Re}
\psfrag{Sa}[b][b][0.9][0]{$\sigma_a$}
\includegraphics[clip,width=0.85\columnwidth]
{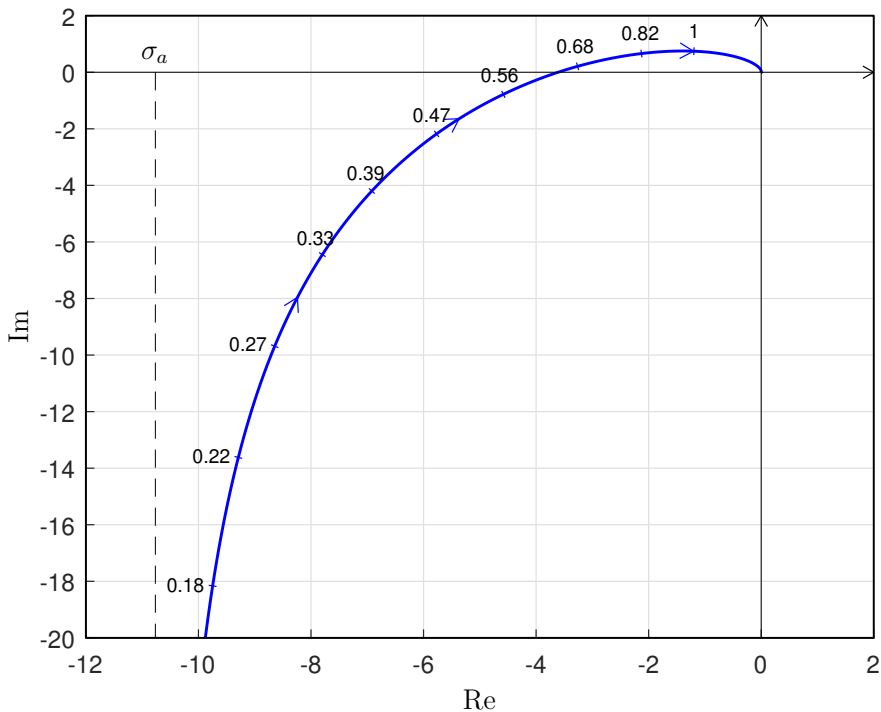}
 \vspace{-2mm}
    \caption{\label{Plot_Nyquist_Fig_3} Actual Nyquist plot of $G(j\omega)$ associated with $G(s)$ in \eqref{Gs_es_2}.}
\end{figure}

 The actual Nyquist plot is shown in Fig.~\ref{Plot_Nyquist_Fig_3}. By comparing Fig.~\ref{Plot_Nyquist_Fig_2} and Fig.~\ref{Plot_Nyquist_Fig_3}, it can be observed that qualitative representations of Nyquist plots are much more effective in providing a global view of the frequency response function both at low and high frequency. Furthermore, Fig.~\ref{Plot_Nyquist_Fig_2} shows that qualitative Nyquist plots make it easier to construct the complete diagram in order to perform a stability analysis.

\section{Conclusion}\label{Conclusion_sect}

This letter has addressed the graphical analysis of the frequency response function, with the objective of improving the procedure for the qualitative construction of Nyquist plots. Several analytical results have been derived and proven, concerning the graphical behavior of the frequency response function at low and high frequency. Two case studies have been considered, showing how the obtained results allow to discriminate the qualitative
behavior of the frequency response on the Nyquist plot in the vicinity
of the initial and final points.

\setcounter{secnumdepth}{0}
\section{Funding}
The work was partly supported by the University of Modena and Reggio Emilia
through the action FARD (Finanziamento Ateneo Ricerca Dipartimentale) 2023/2024, and funded under the National Recovery and Resilience Plan (NRRP), Mission 04 Component 2 Investment 1.5 - NextGenerationEU, Call for tender n. 3277 dated 30/12/2021
Award Number:  0001052 dated 23/06/2022.


\begin{thebibliography}{00}
 \bibitem{LTI_Syst_1} G. F. Franklin, J. D. Powell, and A. Emami-Naeini, ``Feedback control of dynamic systems'', fourth edition, \emph{Prentice Hall PTR, Upper Saddle River}, NJ, USA, 2001.

 %
 \bibitem{Nyquist_9} A. G. J. MacFarlane and H. Nyquist, ``Frequency-Response Methods in Control Systems'', \emph{IEEE Comput. Soc. Press}, New York, NY, USA:  1979.


%
 \bibitem{Nyquist_2} Y.-L. Zhang, M. Zhu, D. Li, J.-M. Wang,
``Static boundary feedback stabilization of an anti-stable wave equation with both collocated and non-collocated measurements'',
\emph{Syst. Control Lett.},
vol. 154, 2021, 104967

 \bibitem{Nyquist_2_bisss} M. Bucolo, A. Buscarino, L. Fortuna and M. Frasca, ``Nyquist Plots for MIMO Systems Under Frequency Transformations'', \emph{IEEE Control Syst. Lett.}, vol. 6, pp. 169-174, 2022.

%
 \bibitem{Nyquist_1} A. Buscarino, L. Fortuna, M. Frasca,
``Nyquist plots under frequency transformations'',
\emph{Syst. Control Lett.},
vol. 125, 2019, pp. 16-21.
 
%
 \bibitem{Nyquist_3} H. Khalil, ``Nonlinear Systems'', \emph{Prentice Hall PTR}, 2002.

%
 \bibitem{Nyquist_4} K. J. Astrom and R.M. Murray, ``Feedback Systems: An Introduction for Scientists and
Engineers'', \emph{Princeton University Press}, Princeton, NJ, USA, 2008.

%
 \bibitem{Nyquist_5} G. Braglia, D. Tebaldi, and L. Biagiotti. ``Phase-free Dynamic Movement Primitives Applied to Kinesthetic Guidance in Robotic Co-manipulation Tasks'', \emph{arXiv preprint arXiv:2401.08238}, 2024.


%
 \bibitem{Nyquist_6} D. Tebaldi and R. Zanasi,
``A unified methodology for the power efficiency analysis of physical systems'', \emph{J. Franklin Inst.}, vol. 361, no. 1, pp. 283-300, 2024.


 %
 \bibitem{Nyquist_8} R. Zanasi, S. Cuoghi, L. Ntogramatzidis, ``Analytical Design of Lead-Lag Compensators on Nyquist and Nichols planes'',
\emph{IFAC Proceedings Volumes}, vol. 44, no. 1, pp. 7666-7671, 2011.


\bibitem{Altro_2} F. Zhang et al., ``Swallowing Events Recognition Method Based on Complex Impedance Pharyngography and Nyquist Plots'', \emph{IEEE Sensors J.}, vol. 22, no. 18, pp. 18076-18084, Sept. 2022.


\bibitem{Altro_4} P. Kawka, G. De Mey and B. Vermeersch, ``Thermal Characterization of Electronic Packages Using the Nyquist Plot of the Thermal Impedance'', \emph{IEEE Trans. on Compon. and Packag. Technol.}, vol. 30, no. 4, pp. 660-665, Dec. 2007.

\bibitem{Altro_5} Y. Z. Tan and C. K. Pang, ``Relaxing LMI Conservatism Using Nyquist Plots and Its Application to Robust Mechatronics Synthesis'', \emph{IEEE Trans. Control Syst. Technol.}, vol. 25, no. 2, pp. 600-610, March 2017.

 %
 \bibitem{Nyquist_10} A. Emami-Naeini, ``The shapes of Nyquist plots [Lecture Notes]'', \emph{IEEE Control Syst. Mag.}, vol. 29, no. 5, pp. 102-115, Oct. 2009.



\bibitem{CST_Ref}
Control System Toolbox Documentation:
{\small \url{https://it.mathworks.com/products/control.html}}. Last accessed: September 2024.

\bibitem{LabVIEW_Octave_Ref}
LabVIEW Documentation:
{\small \url{https://www.ni.com/en/shop/labview.html}}.  Last accessed: September 2024.


\bibitem{Altroo_6} T. Andresen, ``A logarithmic-amplitude polar diagram'', \emph{Model. Identif. Control}, 22
(2), 65–72, 2001.

 %
 \bibitem{Nyquist_12} B. C. Kuo and F. Golnaraghi, ``Automatic Control Systems'', ninth ed., \emph{John Wiley \& Sons},
Inc., New York, NY, USA, 2002.

 %
 \bibitem{Nyquist_13} R. C. Dorf and R. H. Bishop, ``Modern Control Systems'', eleventh ed., \emph{Prentice Hall PTR},
Upper Saddle River, NJ, USA, 2008.

\bibitem{questo_2} R. Zanasi, F. Grossi, and L. Biagiotti,
``Qualitative graphical representation of Nyquist plots'',
\emph{Syst. Control Lett.},
vol. 83, pp. 53-60, 2015.

\end{thebibliography}
\end{document}